\documentclass[
reprint,
superscriptaddress,
nofootinbib,
amsmath,amssymb,
aps,
prx,
showkeys
]{revtex4-2}

\usepackage{graphicx}
\usepackage{bm}
\usepackage{xspace}
\usepackage[table]{xcolor}
\usepackage{float}  
\usepackage{booktabs}  
\usepackage{wasysym}  
\usepackage[caption=false]{subfig}
\usepackage[
  colorlinks=true,
  linkcolor=blue,
  citecolor=blue,
  urlcolor=blue,
  breaklinks=true
]{hyperref}  
\newcommand{\etal}[0]{\textit{et al.}\xspace}
\newcommand{\figref}[2][]{Fig.\ #2#1\xspace}  
\newcommand{\eqqref}[1]{Eq.\ #1\xspace}
\newcommand{\sectref}[1]{Sect.\ #1\xspace}

\newcommand{\refref}[1]{Ref.\ #1\xspace}
\newcommand{\refrefs}[1]{Refs.\ #1\xspace}
\newcommand{\axref}[1]{Appendix {#1}\xspace}
\newcommand{\tabref}[1]{Tab.\ #1\xspace}
\newcommand{\vidref}[2]{\tabref{#1}/#2\xspace}

\newcommand{\vidTableIdColumnWidth}{.5cm}
\newcommand{\vidTableSnapshotColumnWidth}{2.6cm}
\newcommand{\vidTableDescriptionColumnWidth}{8cm}

\newcommand{\new}[1]{\textcolor{black}{#1}\xspace} 

\newcommand{\affilsimtech}{Stuttgart Center for Simulation Science,
Cluster of Excellence EXC 2075,\\ University of Stuttgart,
Universit\"atsstra{\ss}e 32, 70569 Stuttgart, Germany}
\newcommand{\affilwin}{WIN-Kolleg of the Young Academy $\vert$ Heidelberg Academy of Sciences and Humanities, Karlstraße 4, 69117 Heidelberg, Germany}

\begin{document}

\title{
Optimal information injection and transfer mechanisms\\for active matter reservoir computing
}

\author{Mario U. Gaimann}
\affiliation{\affilsimtech}%

\author{Miriam Klopotek}
\email{miriam.klopotek@simtech.uni-stuttgart.de}
\affiliation{\affilsimtech}%
\affiliation{\affilwin}

\date{February 27, 2026}

\begin{abstract}
Reservoir computing (RC) is a state-of-the-art machine learning method that makes use of the power of dynamical systems (the reservoir) for real-time inference.
When using biological complex systems as reservoir substrates, it serves as a testbed for basic questions about bio-inspired computation  -- of how self-organization generates proper spatiotemporal patterning.
Here, we use a simulation of an active matter system, driven by a chaotically moving input signal, as a reservoir.
So far, it has been unclear whether such complex systems possess the capacity to process information efficiently and independently of the method by which it was introduced.
We find that when switching from a repulsive to an attractive driving force, the system completely changes the way it computes, while the predictive performance landscapes remain nearly identical.
The nonlinearity of the driver's injection force improves computation by decoupling the single-agent dynamics from that of the driver.
Triggered are the (re-)growth, deformation, and active motion of smooth structural boundaries (interfaces), and the emergence of coherent gradients in speed -- features found in many soft materials and biological systems. 
The nonlinear driving force activates emergent regulatory mechanisms, which manifest enhanced morphological and dynamic diversity -- arguably improving fading memory, nonlinearity,  expressivity, and thus, performance. 
We further perform RC in a broad variety of non-equilibrium active matter phases that arise when tuning internal (repulsive) forces for information transfer.
Overall, we find that active matter agents forming liquid droplets are particularly well suited for RC.
The consistently convex shape of the predictive performance landscapes, together with the observed phenomenological richness, conveys robustness and adaptivity.
Our findings pave the way for bio-inspired unconventional computing methods exploiting collectivity and fusing physical dynamics across scales.
\end{abstract}


\maketitle

\begingroup
\let\clearpage\relax
\section{\label{sec:introduction} Introduction}

Over the course of millions of years, living systems have evolved strategies to survive in uncertain and fluctuating environments. To name some examples, neural circuits adapt to external stimuli via synaptic plasticity and receptor modulation \cite{Chaudhary2025, Citri2008, Bazzari2019}, bacteria adapt their sensing apparatus to follow changing concentration gradients of nutrients \cite{Li2024,Berg1975}, and cell collectives reorganize dynamically \cite{Weady2024, Cheung2025, Kabla2012}. These evolved mechanisms of adaptation \cite{Tu2018}, memory \cite{Crystal2013}, and learning \cite{Bialek2025} are often surprisingly robust and inspire human engineering in different fields. These encompass neuromorphic \cite{Maass2007,Indiveri2015,Markovic2020,Jaeger2021-unconventional_general,Mehonic2022} and morphological computing \cite{Hauser2012,Hauser2011,Joshi2004,Hauser2021,Mertan2024,Pfeifer2007,Mueller2017} including robotics, and  machine learning \cite{Bloembergen2015}. 
Prominent overarching topics include the plasticity-stability dilemma \cite{Verbeke2019}, robustness \cite{Daniels2016,Aldana2007,Kitano2004,Freiesleben2023}, 
memory or forgetting \cite{Boyd1985,Indiveri2015,Maass2002,Maass2007} and its tradeoff with nonlinearity \cite{Inubushi2017}, and computing ``near criticality'' \cite{Mora2011-biological_criticality}.

One key field that lies at these crossroads is reservoir computing (RC) \cite{Jaeger2010,Atiya2000, Maass2002-real-time_computing, Jaeger2001-short_term_memory, Lukosevicius2009, teVrugt2024, Yan2024}, an unconventional computing paradigm and state-of-the-art method for chaotic time-series forecasting \cite{Pathak2018_model_free,Liu2010,Gilpin2023,Shahi2022}. It describes how an arbitrary dynamical or physical \cite{Tanaka2019} system -- the \textit{reservoir} -- can be harnessed for computation. Reservoir computing works by transforming an input signal via the nonlinear dynamics of the reservoir into a high-dimensional, spatiotemporal response (patterning). These responses can then be mapped via a trained linear readout layer into a lower dimension to solve a range of computational problems:  
chaotic time series prediction \cite{Jaeger2010}, spatiotemporal pattern recognition and classification \cite{Jalalvand2015}, as well as real-time signal processing \cite{Zhong2021}, which are usually solved using digital devices.

The fact that RC directly taps into spatiotemporal pattern generation makes it a generic paradigm for physical \cite{Horsman2014} and morphological \cite{Hauser2012,Hauser2011,Joshi2004,Hauser2021} computing, allowing it to serve as a testbed for exploring the connections between adaptive systems and computing.

In biology, active matter systems \cite{teVrugt2025, Ramaswamy2010, Bowick2022-review_active_mat, Reichhardt2017} are well known for producing complex spatiotemporal patterns. Active matter systems consist of individual, self-propelled particles that consume energy at the microscopic scale, follow simple rules, and typically interact locally with their neighboring particles. Surprisingly, these simple rules give rise to emergent, collective behavior and global patterns, such as flocking \cite{Vicsek1995}.
They display adaptive qualities in response to external perturbations. Examples from nature, spanning multiple scales, include biological microfilms \cite{Davies1998} and starling flocks \cite{Cavagna2010}. Examples from engineering include self-organizing magnetic microrobot collectives \cite{Ceron2023, Wang2022b}, macroscopic kilobots \cite{el-Showk2025}, and swarms of UAV drones \cite{Verdoucq2022}.

Combining these two perspectives -- reservoir computing and active matter -- raises the question of which biologically inspired systems can serve as useful, functional computing substrates. This effort is part of a broader vision to develop \textit{intelligent matter}, which also encompasses controlling active particles using artificial intelligence, for instance \cite{Loewen2025, Cichos2020, Tovey2025}.
Biologically and bio-inspired reservoir computing \cite{Scardapane2017} has been performed using human soft tissue \cite{Kobayashi2025}, the \textit{E.\ Coli} bacterium \cite{Jones2007, Ahavi2024}, brain organoids \cite{Smirnova2023, Cai2023}, cultured neurons \cite{Iannello2025, Sumi2023}, biocompatible organic
electrochemical networks \cite{Cucchi2021} and even raw egg albumen \cite{Fortulan2025}.
More specifically, several active matter RC frameworks have been proposed recently, including laser-driven active gold nano-particle arrays \cite{Wang2024}, electrically modulated active hydrogen ions residing in an electroactive polymer gel \cite{Strong2022}, micro-particles in a medium driven by ultrasonic pressure waves \cite{Jeggle2025}, light-modulated micro-particles with a symmetry-broken refractive index profile \cite{Jeggle2025}, and simulations of a Reynolds boids model \cite{Lymburn2021}. 

In this paper, we observe different response mechanisms of simple active matter systems to timely varying, chaotic external input signals through particle-based active matter simulations. 
By embedding such an adaptive physical system within the information processing context of reservoir computing, we aim to elucidate the physical roots of high computational capabilities, particularly in terms of collectivity. 

Here, external control or feedback is not needed apart from training the readout; hence, the full bulk of the computation is offloaded to the material itself \cite{Hauser2012,Stepney2008}.
To test the ability to physically express various forms of information, we study different variants of information injection into the system, and subsequently, parameter tuning the processing or transfer.
As we will see, when moving from repulsive to attractive forces at the source, markedly different processes of (re-)emergence are triggered, including a shift towards growth, translation, and metamorphosis of coherent structural, as well as dynamical forms. 

In an elementary way, forces are the basic physical quantities by which information is transferred and propagated between and through systems.  In condensed matter, repulsive forces dominate the transfer of momentum and energy (densities). In active matter, these are also essential.  In combination with exchanges with the surrounding environment of the agents (via dissipative and non-Hamiltonian forces acting on them, collective behaviors emerge that unlock active matter RC as reported in Refs.~\cite{Gaimann2025,Lymburn2021, Uehara2024}.
Thus, it is a primary aim to try to assess the effect of these force components on the swarm dynamics and relate them to the resulting RC performance.

\new{This work builds on our previous study \cite{Gaimann2025}, where we showed that tuning the intrinsic agent relaxation dynamics robustly improves predictive performance in various scenarios. In the second part of this study, we focus on (i) how to optimally inject the input signal into the active matter system, and (ii) how to optimally propagate the input signal through the active matter system. This means studying the effect of different driver-agent interactions (``information injection'') and agent-agent interactions \new{(``information propagation'')} (see \sectref{\ref{sec:mm-interactions}} and \sectref{\ref{sec:new_forces}})}. The physical simulations are used as an information processing system (the \textit{reservoir}) in a common reservoir computing framework described in \sectref{\ref{sec:rc_setup}}. 

We analyze active matter trajectories using observables inspired by collective behavior research and the statistical mechanics of complex systems out of equilibrium in \sectref{\ref{sec:mm-observables}}. Simulation and analysis details are described in \sectref{\ref{sec:mm-simulation-details}}.

In \sectref{\ref{sec:attractive-driver}}, we introduce a novel way of injecting information into an active matter system for reservoir computing using an attractive (instead of a repulsive) driver. We compare two different attraction mechanisms -- linear versus inverse -- and connect spatiotemporal swarm patterns and their dynamics with their predictive performance in the reservoir computing framework. We enrich our analysis with driver-agent cross-correlations, connected velocity correlations, and other observables to explain why the inverse attraction mechanism works better than the linear one.
In \sectref{\ref{subsec:driver-repulsion}} we revisit the previously employed driver repulsion interaction and show that strong and large spatial swarm responses improve predictive performance for an optimal (near critically damped) speed-controller setting, which regulates the exchanges of energy with the agents' environment.
The information \new{propagation through} the swarm -- enabled \new{via} collectivity through agent-agent repulsion interactions -- is investigated in \sectref{\ref{sec:repulsion}}. It reveals the optimality of static and following droplets for repulsive and attractive drivers, respectively.
In \sectref{\ref{sec:n_agents}} we discuss the influence of the number of agents, along with their spatial distribution through agent-agent repulsion interactions, on predictive performance. This leads to reservoir computing with new active matter states, such as pressurized liquids or amorphous solids. We show that the highest performance achieved to date with reservoir computing with active matter is delivered by a driver immersed in and repelling a highly susceptible liquid droplet.
\new{We follow the approach of previous studies \cite{Lymburn2021,Gaimann2025} and restrict our analysis in the main text to a standard benchmark, the open-loop Lorenz-63 n-steps-ahead prediction task, and use simple performance metrics, specifically the Pearson correlation coefficient of the $x$-coordinate. In the appendix, we show additional performance measures (\axref{\ref{sec:additional-performance-measures}}), statistics (\axref{\ref{ax:supplementary_statistics}}), short-term memory capacity measurements (\axref{\ref{sec:short-term-mc}}), separability (kernel rank) measurements (\axref{\ref{sec:kernel_rank}}), NARMA-10 benchmarks (\axref{\ref{ax:narma10}}), and tables for control parameters and performance metrics (\axref{\ref{ax:tables}}). Comprehensive benchmarking and task-specific optimization are beyond the scope of this work and are left for future studies.}
Lastly, in \sectref{\ref{sec:discussion}} we discuss the advantages and disadvantages of attractive and repulsive dynamic mechanisms in the active matter reservoir computing framework.

\endgroup

\begingroup
\let\clearpage\relax
\section{\label{sec:materials-methods} Methods}
\subsection{\label{sec:mm-interactions} Driven Active Matter Model Design}

\begin{figure*}
    \centering
    \includegraphics{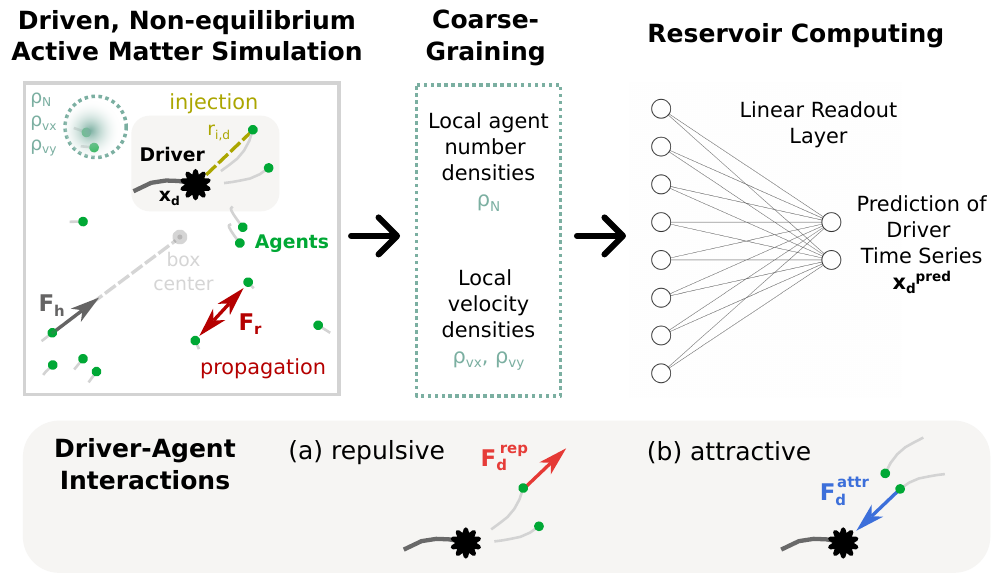}
    \caption{
        \textbf{Concept of reservoir computing with a driven, non-equilibrium active matter simulation.} The reservoir consists of a collection of self-propelled particles (green dots). External information in the form of an external time series enters the reservoir through a special particle (driver, black spiked ball).  All agents are subject to several forces specified in \sectref{\ref{sec:mm-interactions}} and \ref{sec:new_forces}. In this paper, we focus on the agent-agent repulsion force $\mathbf{F}_{r}$ \new{(for information propagation)} and the driver-agent interaction $\mathbf{F}_{d}$ \new{(for information injection)}, which may be repulsive or attractive \new{(see gray box)}. Other interactions include an attraction to the center of the simulation box $\mathbf{F}_{h}$, a speed-regulating force $\mathbf{F}_{sc}$ (not shown here), and a force clamp (not shown here). The response of the agents to the external driving signal is measured in a coarse-grained fashion through Gaussian observation kernels placed on the simulation canvas (symbolized through the \new{turquoise}, dotted circle) (see \sectref{\ref{sec:rc_setup}}). Observations are local densities of agent count and velocity. Swarm patterns generated by certain input time series patterns can then be mapped to future states of the input time series via a linear readout layer, which is trained using Ridge regression.
    }
    \label{fig:concept}
\end{figure*}

To simulate a driven swarm, we follow a basic active matter reservoir computing setup used in \refref{\cite{Gaimann2025}} and visualized in \figref{\ref{fig:concept}}. Agents have time-dependent positions $\mathbf{x}_i (t)$ and velocities $\mathbf{v}_i (t)$; the driver position is denoted as $\mathbf{x}_d (t)$. The driven swarm serving as the reservoir is described by the following set of forces:
\begin{align}
    \mathbf{F}_{r,i} &= \sum_{j=1}^{N_r} \frac{\mathbf{x}_i-\mathbf{x}_j}{\left\|\mathbf{x}_i-\mathbf{x}_j\right\|^2} \label{eq:repulsion_force} \\
    \mathbf{F}_{h,i} &= \mathbf{x}_{h}-\mathbf{x}_i \label{eq:homing_force} \\
    \mathbf{F}_{{sc},i} &= -\mathbf{v}_i \frac{\left(\left\|\mathbf{v}_i\right\|-s\right)}{s} \label{eq:speed-controller} \\
    \mathbf{F}_{d,i}^{\text{rep}} &= \theta\left(r_d-\left\|\mathbf{x}_i-\mathbf{x}_d\right\|\right) \frac{\mathbf{x}_i-\mathbf{x}_d}{\left\|\mathbf{x}_i-\mathbf{x}_d\right\|^2} \label{eq:driver_repulsion}
\end{align}
An agent $i$ experiences four types of forces: individual, driver, local, and global forces.
Individually, each agent regulates its speed to achieve a constant target agent speed $s$. This speed-controlling force can thus accelerate or decelerate the agent (\eqqref{\ref{eq:speed-controller}}).
Agents seek to avoid the driving particle and experience a repulsive force when within a radial distance $r_d$ around the driver position, described by the Heaviside step function ($\theta ( r_d- \left\|\mathbf{x}_i-\mathbf{x}_d\right\|) = 1$ if $\left\|\mathbf{x}_i-\mathbf{x}_d\right\| \leq r_d$, and $0$ otherwise) (\eqqref{\ref{eq:driver_repulsion}}).
Locally, each agent interacts with neighboring agents $j=1,\dots,N_r$ based on radial neighborhoods with radius $r_r$. Here, agents avoid collisions through a local repulsive force (\eqqref{\ref{eq:repulsion_force}}). 
Globally, agents are attracted to the center of the simulation box $\mathbf{x}_h$ (\eqqref{\ref{eq:homing_force}}).

All force terms are scaled with a force strength parameter $K$, summed up (\eqqref{\ref{eq:total_force}}), and then processed by a hyperbolic tangent force wrapper (\eqqref{\ref{eq:force_wrapper}}) which is inspired by the limited power output that an active matter agent can generate: 
\begin{align}
    \mathbf{F}_i(t) &= 
    K_r \mathbf{F}_{r,i} + K_{sc} \mathbf{F}_{sc,i} + K_h \mathbf{F}_{h,i} + K_d \mathbf{F}_{d,i}\,, \label{eq:total_force} \\
    \mathbf{F}_i(t) &\mapsto \alpha \tanh \left(\beta \mathbf{F}_i(t)\right). \label{eq:force_wrapper}
\end{align}
Here, $\alpha$ is the asymptotically maximum force an agent can experience, and $\beta$ tunes the slope of the wrapper. 

The trajectory of the driver particle itself is fixed according to a determined driving protocol; it acts as a non-reciprocal (non-mutually acting) external force and does not experience a force from the agents. By default, we choose the $x$ and $y$ coordinates of the chaotic Lorenz-63 attractor \cite{Lorenz1963} as driving protocol, and fit it to a fraction of the total simulation box.

We evolve the driven swarm then by integrating agent forces using the Euler-forward scheme \cite{Griffiths2010},  moving the driver one step forward, and repeating these steps until the final simulation time is reached.

\subsection{New Forces for Information Injection and Transmission}\label{sec:new_forces}

We study two different types of attractive driving forces. The first one is the \textit{inverse driver attraction force}
\begin{equation}
    \mathbf{F}_{d,i}^{\text{attr,inv}} = - \mathbf{F}_{d,i}^{\text{rep}}\label{eq:driver_attraction_inverse}
\end{equation}
that is obtained by changing the sign of the previously presented driver repulsion force (\eqqref{\ref{eq:driver_repulsion}}).

The \textit{linear driver attraction force} 
\begin{equation}
    \mathbf{F}_{d,i}^{\text{attr,lin}} = \mathbf{x}_{d}-\mathbf{x}_i
    \label{eq:driver_attraction_linear}
\end{equation}
is defined analogously to the previously introduced static homing force (\eqqref{\ref{eq:homing_force}}), but is now attached to the dynamically moving (chaotic) driving particle. Hence, the homing position $\mathbf{x}_{h}$ is replaced with the time-dependent driver position $\mathbf{x}_{d}(t)$.

To study the effects of different agent-agent interactions, we replace the agent-agent repulsive force (\eqqref{\ref{eq:repulsion_force}}) in the total force sum (\eqqref{\ref{eq:total_force}}) with a short-range repulsive force
\begin{equation} \label{eq:short_range_repulsion}
    \mathbf{F}_{srr,i}= \sum_{j=1}^{N_{r}} \exp\left(-\frac{\left\|\mathbf{x}_i-\mathbf{x}_j\right\|}{\lambda}\right)\frac{\mathbf{x}_i-\mathbf{x}_j}{\left\|\mathbf{x}_i-\mathbf{x}_j\right\|}\,,
\end{equation}
where $\lambda$ is the characteristic decay length.

Another choice is replacing \eqqref{\ref{eq:repulsion_force}} with a long-range repulsive force
\begin{equation} \label{eq:long_range_repulsion}
    \mathbf{F}_{lrr,i}=\sum_{j=1}^{N_{r}} \frac{\mathbf{x}_i-\mathbf{x}_j}{\left\|\mathbf{x}_i-\mathbf{x}_j\right\|^{m+1}}\,,
\end{equation}
where $m$ is the power law decay exponent.

\subsection{Reservoir Computing Setup}\label{sec:rc_setup}

To perform reservoir computing with driven swarm systems described in the previous section, we again follow the setup used in \refref{\cite{Gaimann2025}}.
We use a coarse-grained state description of our active matter reservoir based on local densities of agent counts and agent $x$ and $y$ velocities. These are obtained using Gaussian observation kernels to capture local densities. Each kernel records
\begin{equation}
    \begin{aligned}
    & r_{1}(t)=\sum_{i=1}^{N} \psi\left(\mathbf{x}_i(t)\right)\,, \\
    & r_{2}(t)=\sum_{i=1}^{N} \psi\left(\mathbf{x}_i(t)\right) v_{x i}(t)\,, \text{and} \\
    & r_{3}(t)=\sum_{i=1}^{N} \psi\left(\mathbf{x}_i(t)\right) v_{y i}(t)\,,
    \end{aligned}
\end{equation}
which correspond to the coarse-grained densities of agent count, agent velocity in $x$ and $y$ directions. 
The factors
\begin{equation}
    \psi_m(\mathbf{x}_i) =e^{-\frac{\left(\mathbf{x}_i-\mathbf{c}_m\right)^2}{2 w_m}}\,.
    \label{eq:gaussian-kernel}
\end{equation}
determine the contribution of each agent $i$ to the coarse-grained observation, based on the distance of an agent to the position $\mathbf{c}_m$ of the $m$th kernel and kernel width $w_m$. 

We place $M$ Gaussian observation kernels -- with Gaussian distributed kernel widths uniformly randomly on the square simulation canvas.

The observation matrix $\mathbf{X} \in \mathbb{R}^{3M \times T}$ then collects for each time step $t$ of $T$ total simulation steps $3M$ observations. The driver time series is given as $\mathbf{Y}_{\mathrm{target}} = (\mathbf{x}_d(t_1), ..., \mathbf{x}_d(t_T)) \in \mathbb{R}^{2 \times T}$.
We then connect observations at a time $t$ with future states of the driver at a time $t + \Delta t_{\mathrm{pred}}$ via a linear readout layer $\bm{W}_{\mathrm{out}} \in \mathbb{R}^{2 \times \new{(3M+1)}}$:
\begin{equation} \label{eq:rc-prediction}
    \mathbf{W}_{\mathrm{out}} \mathbf{X}^\mathrm{train} = \mathbf{Y}_{\mathrm{target}}^{\mathrm{train}}
\end{equation}
$\mathbf{W}_{\mathrm{out}}$ can be obtained using Ridge regression and then used to predict the future state of a testing time series with the same underlying dynamics but different initial conditions.
To measure the predictive performance $P$ of our reservoir, we compute the \new{previously employed \cite{Lymburn2021}} Pearson correlation coefficient between the $x$ coordinate of the target time series and the predicted time series
\begin{equation}
    P =\frac{\langle x_{\mathrm{target}}(t)\ x_{\mathrm{pred}}(t)\rangle}{\sqrt{\left\langle(x_{\mathrm{target}}(t))^2\right\rangle\left\langle(x_{\mathrm{pred}}(t))^2\right\rangle}}\,,
    \label{eq:correlation-coefficient}
\end{equation}
\new{both centered around zero by subtracting half of the simulation box size.
We also characterize RC performance by measuring the Pearson correlation coefficient for the $y$ dimension as well as other common metrics using both the first two dimensions (see also \axref{\ref{sec:additional-performance-measures}}):
the Normalized Root Mean Squared Error by standard deviation (NRMSE), the Normalized Mean Squared Error (NMSE) by variance, and the Symmetric Mean Absolute Percentage Error (sMAPE). We use the \texttt{sktime} library \cite{Kiraly2025, Loening2019} to compute these measures.}
\subsection{\label{sec:mm-observables} Observables for Driven Nonequilibrium Soft Matter Systems}

To assess the connection between agent dynamics and driver dynamics we compute \textit{driver-agent cross-correlations} as lagged Pearson correlations between an agent time series $x = (x_1, x_2, \dots, x_T)$ and a driver time series $y = (y_1, y_2, \dots, y_T)$ with lags $\tau \in [-50; 50]$ as
\begin{widetext}
    \begin{equation}
        \rho_{xy}(\tau) \;=\; 
    \frac{\sum_{i=1}^{M} \big(x_{i+\max(\tau,0)} - \overline{x_\tau}\big)\,\big(y_{i+\max(-\tau,0)} - \overline{y_\tau}\big)}
    {\sqrt{\sum_{i=1}^{M} \big(x_{i+\max(\tau,0)} - \overline{x_\tau}\big)^2}\;\sqrt{\sum_{i=1}^{M} \big(y_{i+\max(-\tau,0)} - \overline{y_\tau}\big)^2}}\,,
    \end{equation}
\end{widetext}
where $M = N - |\tau|$ time steps are taken into account in the sums, and $\overline{x_\tau}, \overline{y_\tau}$ and $\overline{x_\tau}, \overline{y_\tau}$ are the sample means of the lag-reduced time series. Specifically, we use the x position and the speed time series.

To quantify the agent structure around the driver, we define a radial distribution function (RDF) around the driver -- the \textit{driver-centric RDF} -- as
\begin{equation}
    g(r) = \frac{N_\text{obs}(r)}{\rho \, 2\pi r \, \Delta r},\,
\end{equation}
where $N_\text{obs}(r)$ is the number of agents counted in the shell $[r, r + \Delta r]$ around the driver, and the denominator describes the expected number of agents in a shell of width $\Delta r$ and area $2\pi r \Delta r$ if the agents were uniformly distributed in the simulation box with a density of $\rho = N/l_{\text{box}}^2$.\\

We also quantify agent dynamics through the absolute velocity autocorrelation function
\begin{equation}
    \left| \mathrm{VAC}(\tau) \right| = \left| \frac{1}{N N_{\mathcal{T}}} \sum_{i=1}^N \sum_{j=1}^{N_{\mathcal{T}}} \mathbf{v}_i(t_{0,j}) \cdot \mathbf{v}_i(t_{0,j}+\tau) \right|
\end{equation}
averaged over all agents $N$ and all $N_{\mathcal{T}}$ time windows $\mathcal{T}_{t_{0,j}} = [t_{0,j}, t_{0,j} + \tau_\text{max})$ for a total number of simulation steps $T$ and a given time lag $\tau \in \mathcal{T}$. We note that the windows $\mathcal{T}_{t_{0,j}}$ do not overlap, and random waiting times drawn uniformly from $\mathcal{T}$ are used to separate each two windows.\\

In \refref{\cite{Gaimann2025}}, inspired by collective behavior research \cite{Attanasi2014_collective_behaviour, Cavagna2010}, it was shown that agent dynamics can be understood by studying correlations between velocity fluctuations of agents.
The velocity fluctuation of an agent $i$
\begin{equation}
    \delta \mathbf{v}_i=\mathbf{v}_i - \frac{1}{N} \sum_{i=1}^N \mathbf{v}_i  \,,
    \label{eq:velocity-fluctuation}
\end{equation}
describes the deviation of an individual agent's velocity from the center of mass velocity of the total swarm.
Normalizing these fluctuations by their standard deviation yields dimensionless velocity fluctuations $\delta \bm{\varphi}_i$. The spatial correlation that captures how the change of behavior (velocity) of an agent $i$ correlates with the change of behavior of a neighboring agent $j$ in a distance $r_{ij}$ is described by the radially binned \textit{connected velocity correlation function}
\begin{equation}
    \mathrm{CVC}(r)=\frac{\sum_{i \neq j}^N \delta \bm{\varphi}_i \cdot \delta \bm{\varphi}_j \ \delta\left(r-r_{i j}\right)}{\sum_{i \neq j}^N \delta\left(r-r_{i j}\right)} \,,
    \label{eq:connected_velocity_correlation}
\end{equation}
where $\delta(r-r_{ij})$ is the Dirac delta function ($\delta\left(r-r_{i j}\right) = 1 \text { if } r < r_{i j} < r + dr, \text{ and } 0 \text { otherwise}$).\\

The size of the driver exclusion zone forming around a repulsive driver is estimated by computing the shortest distance $r_{i,d}^{min}$ between the driver and the active matter system. This can be used to obtain a lower bound for the exclusion zone by computing the circle area $A = \pi (r_{i,d}^{min})^2$, averaged over $T = 1,000$ recorded time steps.\\

In the appendix in \figref{\ref{fig:attraction_observables_1}} and \ref{fig:attraction_observables_2}, we provide further supplementary observables to the main text (polarity, rotation, mean squared displacement, dynamical susceptibility, and the connected velocity correlation at the first local minimum) that have been previously analyzed for different parameter scans in \refref{\cite{Gaimann2025}}. 

\subsection{\label{sec:mm-simulation-details} Simulation, Reservoir Computing, and Analysis Details}

We choose our active matter reservoir computing parameters based on common settings used in previous studies \cite{Gaimann2025, Lymburn2021}. 
We simulate by default $N=200$ agents residing in a square box with length $l_{\text{box}} = 16.0$ and periodic boundary conditions. All agents are placed uniformly at random in the box with initial speeds $\left|\mathbf{v}_i\right| = 1.0$ and uniformly at random directions. The chaotic input (driver) trajectory is generated by integrating the coupled differential equations of the Lorenz-63 system \cite{Lorenz1963} ($\dot{x} =a(y-x); \dot{y} =x(b-z)-y; \dot{z} =x y-c z$) using default values ($a = 10; b = 28; c = 8/3$) and Euler integration with an integration time step $\Delta t = 0.02$ and initial conditions $(x_0,y_0,z_0) = (0.0, 1.0, 1.05)$ for training and $(0.0, 1.1, -1.5)$ for testing. We consider only the $x$ and $y$ coordinates here and scale the time series to fit in a box with length $l_{\text{box}}^{\text{driver}} = 8.0$, centered in the simulation box.

All swarm interactions are described in \sectref{\ref{sec:mm-interactions}} and \sectref{\ref{sec:new_forces}}. We use default values from \refref{\cite{Gaimann2025}}: a local repulsion force with $r_r = 1.0$ and $K_r = 2.0$; a global homing force with $K_h = 2.0$; and a sigmoid force clamp with $\alpha = 200.0$ and $\beta = 0.1$.
If not mentioned otherwise, we present results for active matter systems with near-critically damped speed-controller settings ($(K_{sc}, s) = (0.02069, 0.04833)$) used in \refref{\cite{Gaimann2025}}. These speed-controller settings were found to deliver optimal intrinsic dynamics and predictive performance given the above specifications and also across different prediction tasks (attractors) \cite{Gaimann2025}. A previous study used speed-controlling settings at a ``critical'' transition from a condensed droplet-like swarm to a gas-like swarm \cite{Lymburn2021} ($(K_{sc}, s) = (2.0, 10.0)$), which we also used for further comparisons in the appendix. (We note that the original publication states $K_{sc} = 20.0$, but dynamics matching the described phenomenology could only be reproduced with a value of $K_{sc} = 2.0$.)
We do not add the agent-agent alignment force employed in \refref{\cite{Lymburn2021}} to our model because it was found not to improve performance for agents with near-critically damped speed-controller settings \cite{Gaimann2025} and because leaving it out simplifies the model. 
All scans with a driver attraction interaction have no attractive force to the center of the simulation box (homing force) $\bm{F}_h$ to avoid conflicting attraction interactions and to simplify the model. The driver attraction interaction radius is not limited ($r_d \equiv \infty$) if not mentioned otherwise.

In \sectref{\ref{sec:attractive-driver}} we vary the speed-controller parameters $K_{sc}, s \in [10^{-5}; 10^{2}]$ for linear and inverse driver attraction (each with $K_d = 100.0$).
In \sectref{\ref{subsec:driver-repulsion}} we vary the driver repulsion strength $K_{d} \in [10^{-2}; 10^{6}]$ and the driver repulsion radius $r_{d} \in [10^{-1}; 8.0]$.
In \sectref{\ref{sec:repulsion}} we vary the agent-agent repulsion strength $K_r \in [10^{-3}; 10^{2}]$ and the repulsion radius $r_r \in [10^{-1}; 8.0]$ for a repulsive driver and an inversely attractive driver.
In \sectref{\ref{sec:n_agents}} we vary the number of agents quasi-logarithmically $N_a \in [10^0; 10^{3}]$ (18 unique integers), with added special values $N_a = \{100; 200\}$, and the agent-agent repulsion strength $K_r \in [10^{-1}; 10^{2}]$ for a repulsive driver and an inversely attractive driver.

We perform all scans using 20 $\times$ 20 logarithmically spaced parameter combinations and plot observables in contour plots inside heatmaps using linear interpolation (Gouraud shading). We show predictive performances obtained using test data (test initial conditions).

In each time step, we observe the swarm using Gaussian observation kernels (see \sectref{\ref{sec:rc_setup}}. In total, we place $M=200$ kernels uniformly at random in the simulation box with normally distributed widths $w_m$ with mean $\mu = 0.0$ and standard deviation $\sigma = 1/\sqrt{2}$ at the beginning of the simulation. 
We run two active matter simulations: a training run and a testing run, each with different initial conditions of the chaotic Lorenz-63 system. Each run contains $T = 50,000$ simulation time steps preceded by $1,000$ burn-in (equilibration) time steps. 
The connected velocity correlation function is recorded using a bin width of $\Delta r_{\text{bin}} = 1.0$. The driver-centric radial distribution function is computed using a bin width of $\Delta r_{\text{bin}} = 0.5$. Velocity auto-correlation functions are recorded with a maximum lag time of 75 integration time steps ($\tau_{\text{max}} = 1.5$).
We predict $\Delta t_{\text{pred}} = 25$ integration time steps ahead at each time step of the simulation ($\approx 0.45283$ Lyapunov times for the Lorenz-63 system) after training the readout layer using Ridge regression with a Ridge parameter of $\lambda_{\text{Ridge}} = 1.0$.

\new{We show results for a single reservoir computing run with the same initializations for agents and driver for all simulations. For all predictive performance figures in the main text, we also provide complementary means and standard deviations obtained from averaging over 100 different seeds (with both random agent and driver initial positions) in \axref{\ref{ax:supplementary_statistics}}. The chosen initial conditions for the single reservoir computing run used in the main text yield representative (not outlying) predictive performances.}

\endgroup

\begingroup
\let\clearpage\relax
\section{\label{sec:results} Results}
\subsection{\label{sec:attractive-driver} Information Injection with an Attractive Driver}

\begin{figure*}[htbp]
    \centering
    \includegraphics{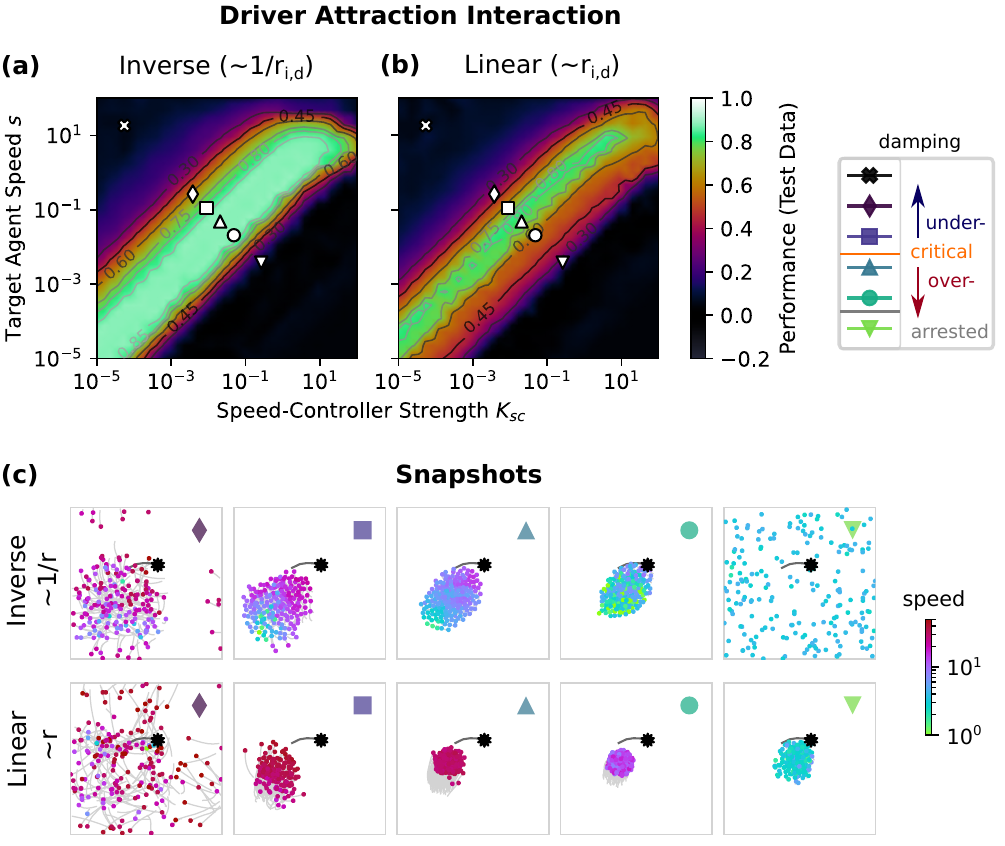}
    \caption{\textbf{Replacing the global attraction to the center of the simulation box and the local driver repulsion force with two different driver attraction forces.} (a,b) Predictive performance for a speed-controller parameter scan with (a) an inversely driver attraction force $\sim 1/r_{i,d}$ (where $r_{i,d}$ is the distance between the driver $d$ and an agent $i$, see \eqqref{\ref{eq:driver_attraction_inverse}}) and (b) a linearly attractive driver ($\sim r_{i,d}$), both with $K_d = 100.0$ (\eqqref{\ref{eq:driver_attraction_linear}}). (b) Predictive performance for a speed-controller parameter scan with a driver homing force $\sim r_{i,d}$  with $K_d = 100.0$. (c) Snapshots of the driven swarm for different parameter combinations (symbols) in the parameter scans for each driver attraction force. Refer to \tabref{\ref{tab:supplementary_videos_driver_attraction_inverse}} and \tabref{\ref{tab:supplementary_videos_driver_attraction_linear}} for corresponding videos. Parameter values $(K_{sc}, s)$ for symbols: cross: (0.00005, 18.32981); diamond: (0.00379, 0.26367); square: (0.00886, 0.11288); pyramid: (0.02069, 0.04833); circle: (0.04833, 0.02069); nabla: (0.26367, 0.00379).}\label{fig:driver_attraction_speed_controller}
\end{figure*}

To predict a chaotic time series using active matter as a reservoir, its information has to be injected into the active matter reservoir by physical means. This was accomplished so far using a non-reciprocal (in the sense of \emph{non-mutually acting} \cite{Fruchart2021}) interaction between a special, moving driver particle and the active matter system: a repulsive interaction from the driver to the agents \cite{Lymburn2021, Gaimann2025}. The driver moves along the trajectory of the temporally varying 2D input signal. An attractive interaction has not been explored and is the most basic physical counterpart. Whether this change yields better, similar, or worse performances is fully unclear up front, as are the effects on the highly nonequilibrium phenomenology resulting from adaptive self-organization.

In \figref{\ref{fig:driver_attraction_speed_controller}}, we show the predictive performances for swarm configurations obtained by varying the parameters of the speed-controlling force, for two types of attractive drivers: linearly and inversely attractive drivers (see \sectref{\ref{sec:new_forces}}). The speed-controlling force regulates intrinsic agent dynamics, which are critically instrumental for predictive performance \cite{Gaimann2025}.  \figref{\ref{fig:driver_attraction_speed_controller}} reveals that the inversely attractive driver heatmap shows a broader high-performance regime ($P > 0.80$) compared to the linearly attractive driver. The highest performance is obtained in the near-critically damped regime (pyramid symbol). It is significantly higher for the inversely attractive driver ($P = 0.88$;  
\new{$P_y = 0.80$, 
$\mathrm{NRMSE}= 0.55$, 
$\mathrm{NMSE}= 0.31$, 
$\mathrm{sMAPE}= 6.7\,\%$}) 
compared to the linear case ($P = 0.79$; 
\new{$P_y = 0.68$, 
$\mathrm{NRMSE}= 0.76$, 
$\mathrm{NMSE}= 0.57$, 
$\mathrm{sMAPE}= 11\,\%$
}). 
\new{Compared to a simple Echo State Network (ESN) with $N=600$ neurons (see \refref{\cite{Gaimann2025}}, Appendix D: $P= 0.833$, $P_y = 0.746$,  $\mathrm{NRMSE}=0.619$, $\mathrm{NMSE}=0.384$, $\mathrm{sMAPE}=7.23\,\%$), the $\mathrm{NRMSE}$ of this active matter RC system is about $11\,\%$ lower (see also \figref{\ref{fig:time-series-comparison}}).}     

In this regime, the near-critically damped agent dynamics in combination with inverse driver attraction force lead to speed gradients in the swarm, visible in the snapshot in \figref[c]{\ref{fig:driver_attraction_speed_controller}} (upper panel, pyramid symbol (\href{https://darus.uni-stuttgart.de/file.xhtml?fileId=401421}{Video} \vidref{\ref{tab:supplementary_videos_driver_attraction_inverse}}{4})). This can also be observed for adjacent parameter combinations (upper panel, square symbol (\href{https://darus.uni-stuttgart.de/file.xhtml?fileId=401420}{Video} \vidref{\ref{tab:supplementary_videos_driver_attraction_inverse}}{3}); circle symbol (\href{https://darus.uni-stuttgart.de/file.xhtml?fileId=401422}{Video} \vidref{\ref{tab:supplementary_videos_driver_attraction_inverse}}{5})). The shape that swarms take on varies strongly during the driving evolution: For small and less abrupt driver excursions, we observe a droplet shape, while after abrupt, large driver excursions, the swarm fans out and changes into a teardrop shape, with the tip pointing towards the driver.

This near-critically damped, inversely attracted swarm can be further described by intermediate polarity ($\Phi_P \approx 0.65$), very low rotation ($\Phi_R \approx 0.10$), low dynamical susceptibility ($\chi \approx 6.0$ compared to the repulsive driver case studied in \refref{\cite{Gaimann2025}}), and a low CVC at its first local minimum ($\approx -0.15$) (see also \figref[]{\ref{fig:attraction_observables_1} and \ref{fig:attraction_observables_2}}). 
Conversely, no coherent gradients in speed are visible for the linearly attractive driver, as all agents move with nearly uniform speed towards the driver (\figref[c]{\ref{fig:driver_attraction_speed_controller}}, lower panel, pyramid symbol (\href{https://darus.uni-stuttgart.de/file.xhtml?fileId=401433}{Video} \vidref{\ref{tab:supplementary_videos_driver_attraction_linear}}{4})). We think this is an essential difference. The shape of the droplet remains stable in this case. The linearly attracted swarm can also be characterized through high polarity ($\Phi_P \approx 0.90$), zero rotation, and zero dynamical susceptibility (see also \figref{\ref{fig:attraction_observables_1}} and \figref{\ref{fig:attraction_observables_2}}). Independent of the attraction mechanism, we observe for both cases low mean squared displacements of $< 15.0$.

When moving towards lower ratios $r = K_{sc} / s$ (upper left-hand side), the agents are more strongly accelerated towards the driver and accumulate high inertia due to the given speed-controller setting (square, diamond symbols). Since there is no driver repulsion in place, this leads to agents overshooting and orbiting around the driver position in both cases (see \href{https://darus.uni-stuttgart.de/file.xhtml?fileId=401420}{Video} \vidref{\ref{tab:supplementary_videos_driver_attraction_inverse}}{3} and \href{https://darus.uni-stuttgart.de/file.xhtml?fileId=401432}{Video} \vidref{\ref{tab:supplementary_videos_driver_attraction_linear}}{3}). This reduces predictive performance, likely because the reservoir state contains a growing share of individualistic (as opposed to collective) agent dynamics (``noise'') and a reduced share of coherent, thus informative, response dynamics caused by recent driver excitations (``signal''). 
For the cross symbol, agents accumulate very high speeds and behave gas-like at all times, causing undistinguishable spatiotemporal reservoir states for different driving patterns and a drop of predictive performance towards $0.0$. 

Moving towards the opposite side of the performance heatmap, the active matter system becomes overdamped.  Agents linearly attracted to the driver track large driver excursions worse because of their slower response, leading to a strong decrease in performance below $0.60$ (\figref[c]{\ref{fig:driver_attraction_speed_controller}}, lower panel, circle symbol (\href{https://darus.uni-stuttgart.de/file.xhtml?fileId=401434}{Video} \vidref{\ref{tab:supplementary_videos_driver_attraction_linear}}{5})). For the inverse attraction, the speed gradient becomes weaker while the performance stays very high above $0.85$ (\href{https://darus.uni-stuttgart.de/file.xhtml?fileId=401422}{Video} \vidref{\ref{tab:supplementary_videos_driver_attraction_inverse}}{5}). 
Further increasing the ratio $r$ leads to arrested agent dynamics (nabla symbol), which results from numerical integration with the chosen integration time step of $\Delta t = 0.02$ (see \refref{\cite{Gaimann2025}} for a more detailed discussion) and causes a sharp drop in performance towards $0.0$.

We confirm that the choice of the Ridge parameter used during readout layer training only marginally affects the predictive performance for the linearly attractive driver and cannot improve performance beyond $P \approx 0.80$
(see     \figref{\ref{fig:driver_attraction_lambda_ridge_200}} and \figref{\ref{fig:driver_attraction_lambda_ridge_scan}}). This suggests that the performance difference is rooted in the difference in the spatiotemporal response of the active matter system itself. It also underscores how the essence of computation at the final inference step is offloaded to the active matter system, in the original spirit of RC and morphological computing \cite{Hauser2012}.

Comparing the inversely attractive driver with the previously used (inversely) repulsive driver in \refref{\cite{Gaimann2025}} (see also \figref{\ref{fig:comparison_attractive_to_repulsive_driver}} in the appendix), we note that the performance in the near-critically damped regime is the same ($P = 0.88$, pyramid symbol). The band of high performances ($P > 0.85$) is slightly broader in the attractive case and extends to the circle symbol, which shows a slightly lower performance of $0.83$ in the repulsive case. Moving towards the broad underdamped regime (diamond, cross symbols), the repulsive case shows performance boosts of almost $0.50$ points (cross symbol), which is intriguing. This suggests a hidden, but non-absolute advantage of repulsive information injection (see \sectref{\ref{sec:discussion}}). 

The mechanistic outcome is that a repulsive driver immersed in a droplet leads to the self-assembly of an exclusion zone, or, an active interface that synchronizes around the moving driver, even for off-optimal dynamical regimes (see \refref{\cite{Gaimann2025}}, \figref[b]{2}, cross symbol; or the corresponding \href{https://darus.uni-stuttgart.de/file.xhtml?fileId=351510}{Video} \cite{DARUS-4619_2025}). This could sufficiently encode the driver's position, leading to a predictive performance of about $0.60$. For a small parameter region between the circle and nabla symbol, given by a fixed ratio $r$, the attractive driver delivers significantly higher predictive performances than in the repulsive case. This could be due to different onsets of the arrested regime in each case.\\

\subsection{Relating Predictive Performance to Physical Observables}\label{sec:speed-controller-attraction-quantifiers}

\subsubsection{Driver-Agent Time-Series and Correlations}

\begin{figure*}[htbp]
    \centering
    \includegraphics{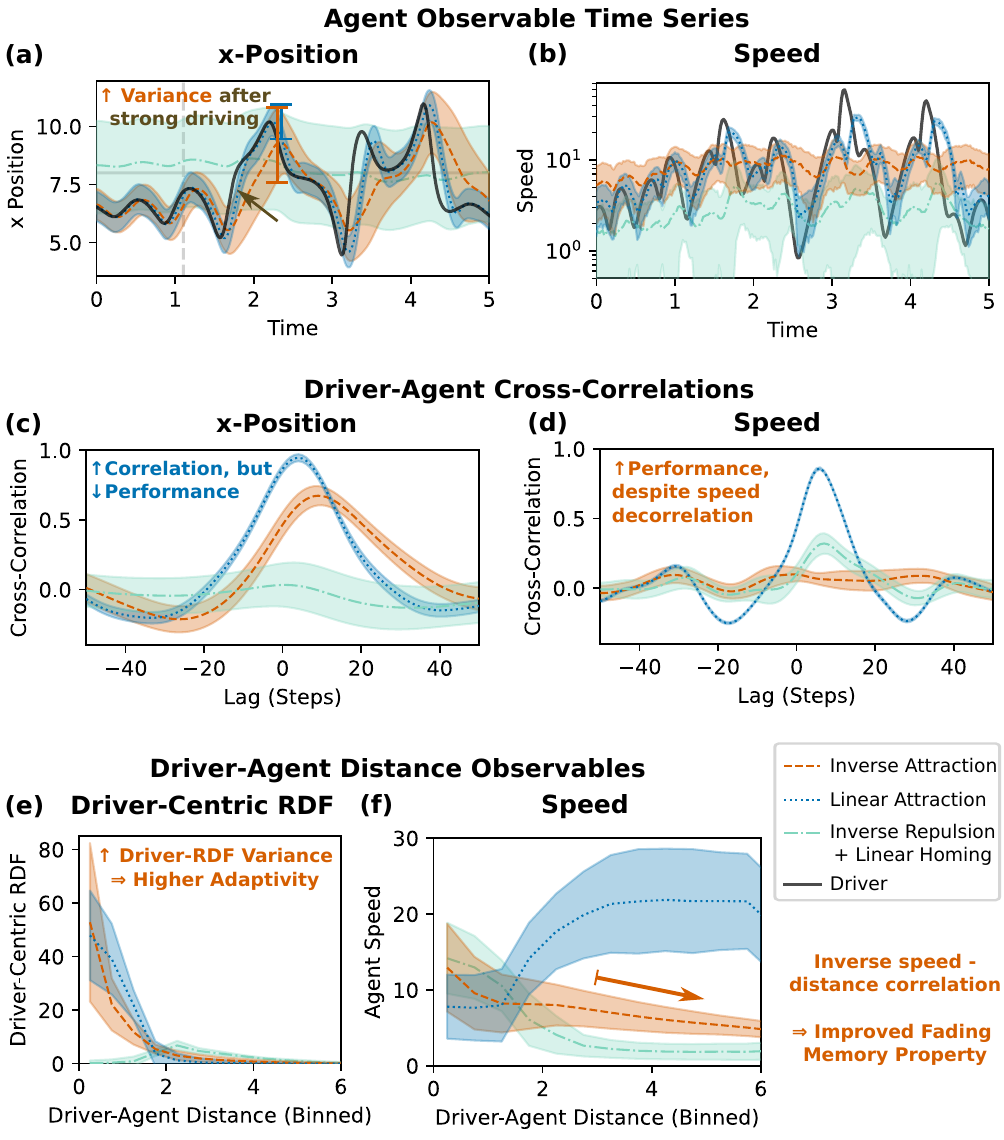}
    \caption{
        \textbf{Characterizing effects of different information injection mechanisms:} inverse attraction ($\sim 1/r_{i,d}$; orange dashed line), linear attraction ($\sim r_{i,d}$; blue dotted line), and inverse repulsion plus linear homing ($\sim 1/r_{i,d} + \sim r_{i,h}$; green dash-dotted line) in the near-critically damped regime (pyramid symbol in \figref{\ref{fig:driver_attraction_speed_controller}} and \figref{\ref{fig:comparison_attractive_to_repulsive_driver}}). (a,b) Mean and standard deviations of agent trajectories for (a) position x coordinate and (b) speed. The driver time series is plotted as a solid, black line. (c,d) Mean and standard deviations of cross-correlations between agent and driver time series. (d) Radial distribution function of agents with respect to the driver position. (e) Agent speed in dependence on agents' distance from the driver. All plots were generated using 1,000 time steps after 1,000 burn-in steps that were not recorded; the bin size for sub-figures (d) and (e) is $\Delta r = 0.5$. The repulsion reference corresponds to the simulations performed in \refref{\cite{Gaimann2025}}.
    }
    \label{fig:driver_attraction_observables}
\end{figure*}

Above observations are quantified in \figref{\ref{fig:driver_attraction_observables}} for a speed-controlling force set in the near-critically damped regime (pyramid symbol) for the two driver-agent attraction interactions (linear ($\sim r_{i,d}$) and inverse ($\sim 1/r_{i,d}$)) as well as the previously studied driver-agent repulsion interaction ($\sim 1/r_{i,d}$) combined with the linear attractioirn force to the center of the simulation box (homing force, $\sim r_{i,h}$) (see \refref{\cite{Gaimann2025}}). 

\figref[a,b]{\ref{fig:driver_attraction_observables}} shows the evolution of the mean and standard deviation of the $N_a = 200$ agent trajectories.
Agents linearly attracted to the driver follow the driver trajectory more closely than in the inverse case, both in terms of x position and speed. Though there are moments of nearly identical mean and variance in positions, this difference becomes especially clear when comparing the swarm response after a phase of strong driver excursions, when the driver's trajectory becomes more volatile (e.g., around $t\approx 1.6$ or $t\approx 3.1$, or $t\approx 4.2$): Here, the variance of agent x positions and speeds increases, and remains increased, for the inversely attractive driver case, which can also be identified as positional spread (fanning out) and color gradient in the screenshots in \figref[c]{\ref{fig:driver_attraction_speed_controller}} (upper panels, pyramid). \figref[b]{\ref{fig:driver_attraction_observables}} shows an interesting feature: The mean speeds are very well regulated during the evolution, having additionally a regulated variance, though wider variance than in the case of linear attraction.  
Notably, the comparison case of inverse driver repulsions renders highly regulated mean agent positions \emph{and} variances; the speed fluctuations are broader than in the performance-comparable inverse attraction case.   

Such responsive increases in the variance of positions (i.e., heterogeneous sub-structures), together with highly regulated mean speeds and speed fluctuations, represent a noteworthy adaptive mechanism. It gets activated whenever the environment becomes more unpredictable; otherwise, the two cases (linear/inverse attraction) look similar in their response. The case of repulsions shows the system behaving very well-regulated in agent position statistics. This matches the observation of more robust morphology we see in the videos of the droplet: There is less pronounced growth and shrinkage of sub-structures (morphogenesis), except for the encapsulating exclusion zone ``bubble'' around the driver. The dynamical susceptibility is much larger and becomes a relevant proxy variable for performance \cite{Gaimann2025}: It is a key descriptor of nonequilibrium condensed matter systems in a bulk, where, notably, repulsions play the dominant role in generating long-ranged correlated many-body dynamics. This points to a complementarity between morphological change abilities and dynamical susceptibility, presenting a basic range of how a system can express its adaptation capabilities. 

Basic driver-agent correlations in structure and dynamics are analyzed in terms of cross-correlations in \figref[c,d]{\ref{fig:driver_attraction_observables}}. 
The swarm attracted via the linear interaction follows the driver more directly: the maximum driver-agent x position cross-correlation of about $0.95$ is observed after a lag of $4.0$ steps. For the linear interaction, the maximum correlation is lower at $0.68$ after a lag of about $9.3$ steps. 
For the case of inverse attraction, the decreased maximum and broader delay-time first peak in the positional cross-correlation (\figref[c]{\ref{fig:driver_attraction_observables}}) attest to structures that emerge later near where the driver arrived (in, e.g., one of the attractive basins of the Lorenz-63 trajectory), and need longer and more diverse timescales. They accumulate and grow in a more complex process (this phenomenology can be observed in the supplementary videos), which generates more diverse spatiotemporal patterns, hypothesized to be advantageous for RC.

For the driver-agent speed cross-correlation in \figref[d]{\ref{fig:driver_attraction_observables}}, the difference between the two attraction mechanisms is particularly stark. While the driver-agent speed cross-correlation for the linear interaction reaches a maximum of about $0.86$ at a lag of $6.0$ steps, the maximum correlation for the inverse attraction case is very weak with a score of about $0.16$ at $0.1$ steps. 
Thus, apart from strength, the nonlinearity of the driver's injection force improves computational quality by decoupling the agent dynamics from those of the driver more strongly: The cross-correlation on speeds is nearly zero for the inverse attraction case (\figref[d]{\ref{fig:driver_attraction_observables}}). There must be shifting relevance towards a richer set of transient mesoscopic, collective entities that carry and process information in ways not linearly correlated with the driver's dynamics (speeds). Notably, the case of the repulsive driver also shares this basic feature on dynamics, though not quite as strongly decoupled, as repulsions fundamentally couple complex systems via momentum exchanges -- the dynamics thus remain minimally coupled at a short lag time. Notably, the position of the peak value in the cross-correlation aligns with that of the linear attraction case, suggesting a characteristic dynamic timescale that may be related to the mechanical properties of the condensed-matter swarm. 

The standard deviation for cross-correlation on both x-position and speed is also significantly smaller for the linear attraction, another indication that the diversity of responses across agents in both dynamics and structure is more limited. For the comparison case of inverse repulsion, the standard deviation (fluctuations) on speed cross-correlations is comparable to that of the inverse attraction case. In contrast, position cross-correlations for the repulsion case have a maximally broad standard deviation, like in sub-plot (a).

As a final remark, it is striking how the x-position cross-correlation peak for the inverse attractive case is more asymmetric than for the linear case with respect to the zero-delay-time line. Possibly, more strongly irreversible dynamics are generated (it is more strongly driven out of equilibrium), but this would need to be verified in other ways.  

The radial distribution function (RDF) of agents with respect to the driver's position is displayed in \figref[e]{\ref{fig:driver_attraction_observables}}. We observe an increased variance for very low driver-agent distances in the case of the inverse attraction, indicating a higher diversity of agent densities directly around the driver. This could imply a potential for a broader variety of responses right around the driver. Indeed, this is supported by 
 measuring the agents' speeds with respect to their radial distance from the driver in \figref[f]{\ref{fig:driver_attraction_observables}}: Very close to the driver, there is a larger variance in agent speed for this case of an inversely attractive driving force.

We uncover another sign of enhanced regulatory mechanisms activated by nonlinear (and not linear) attractive driving forces: For linearly attracted agents, speeds increase with increasing distance to the driver. Conversely, for inversely attractive driving forces, the farther the distance from the driver, the lower the speed measured.
Inertia carries memory of the past of the driver's position across space.
We posit that the nonlinear driving force induces more memory in the reservoir. \new{This hypothesis is supported by the finding that the short-term memory capacity is higher for inverse driver attraction compared to linear driver attraction (see \figref{\ref{fig:speed-controller-memory-capacity-lin-vs-inv}} and \axref{\ref{sec:short-term-mc}} for details). This holds for all parameter combinations broadly around the near-critically damped regime (diamond, square, pyramid symbols) for different input time series.} Overall, \new{higher short-term memory} could \new{contribute to} the increased performance in \new{the inverse driver attraction} case.

\begin{figure*}[htbp]
    \centering
    \includegraphics{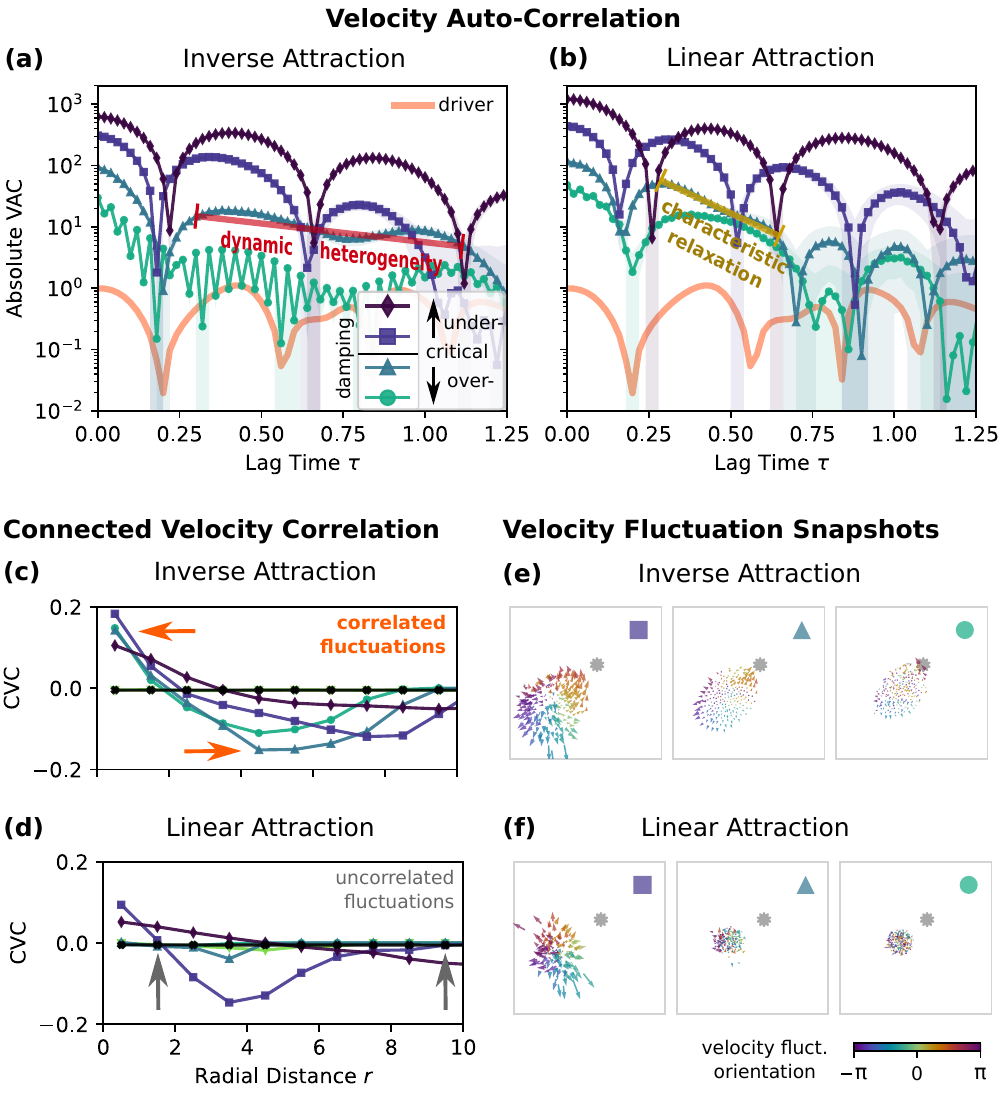}
    \caption{
        \textbf{Agent-agent correlations for driven swarms with two different driver attraction interactions, for different speed-controller settings.} (a,b) Absolute agent-averaged velocity auto-correlations for (a) inversely and (b) linearly attractive drivers. Error bands indicate the standard error of the mean over all evaluated time windows (see \sectref{\ref{sec:mm-simulation-details}} for details). Symbols correspond to parameter combinations changing speed-controller settings $(K_{sc}, s)$ (and resulting damping dynamics) as shown in \figref{\ref{fig:driver_attraction_speed_controller}}. (c,d) Connected velocity correlations for (c) inversely and (d) linearly attractive drivers. (e) Visualizations of agent velocity fluctuations (agent velocities reduced by the center of mass velocity of the swarm). All plots were generated using 1,000 time steps after 1,000 burn-in steps that were not recorded. Refer to \tabref{\ref{tab:supplementary_videos_driver_attraction_inverse_fluctuations}} and \tabref{\ref{tab:supplementary_videos_driver_attraction_linear_flucts}} for corresponding videos of the velocity fluctuations. \new{The Lyapunov timescale of the Lorenz-63 driver is $t^{\text{L63}}_{\text{lyap}} \approx 1.1$ \cite{Viswanath1998}.}
    }
    \label{fig:driver_attraction_correlations}
\end{figure*}

\subsubsection{Correlated Velocities and Velocity Fluctuations}

The nonequilibrium dynamics of the active matter system under continual chaotic driving can be further quantified using time autocorrelation functions. The ``relaxation'' dynamics of the system are captured by the velocity autocorrelation (VAC) (see also the discussion in the prequel in \refref{\cite{Gaimann2025}}). \figref[a,b]{\ref{fig:driver_attraction_correlations}} shows the agent-averaged and time-averaged absolute VACs for different speed-controller parameter combinations (symbols) for an inverse and a linearly attractive driver. Minima in these absolute VAC plots indicate a change from positive to negative correlations. We observe that for an underdamped speed-controller setting (diamond symbol), similar VAC curves are obtained, with minima being located roughly at the same lag times $\tau$, independently of the driver attraction interaction. The system thus does not adapt its nonequilibrium dynamical response as much to the input signal -- compared to more optimally-computing cases described hereafter. This is an indication that these oscillations cannot fully contribute to new information or signal needed to form the computation on the readout layer. The peaks are roughly symmetric (harmonic-like) and wide, which indicates long-range and more persistent oscillations arising due to the high inertia in the system.

In the near-critically damped regime (pyramid symbol), we observe for the inverse driver attraction a stretched absolute VAC after the first minimum, covering a lag time from roughly $\tau \approx 0.3$ to $1.1$. This stretch is characteristic of dynamic heterogeneity: different agents relax at different rates. Visually, this can be related to the emergence of coherent speed gradients present in the active matter system, for example, in \figref[c]{\ref{fig:driver_attraction_speed_controller}} (upper panel, pyramid symbol). This seems to be a unique feature caused by the inverse ($1/r_{i,d}$-decaying) driver attraction in this dynamical regime, as it is not observed for the linear driver attraction or (inverse) driver repulsion (see \refref{\cite{Gaimann2025}}). It is an effect going hand-in-hand with the smooth growth and smooth shrinkage of interfaces, verified via visual inspection, which could be a feature indicating optimal computational abilities. 
The speed gradients thus might express the heightened dynamical ``complexity' (diversity) generating coherent structures that morph, grow, and shrink. 
Coherence and consistency are intuitively helpful for generating semi-repeatable spatiotemporal patterns, thus, signal rather than noise. 

For the linearly attractive driver, the absolute VAC curve in the near-critically damped regime (pyramid symbol) roughly follows an exponential decay from $\tau \approx 0.3$ to $0.7$ after the first minimum. This is a signature of reduced inertia through uniform energy dissipation, compared to a more underdamped speed-controller setting (square symbol). This can be visually confirmed: there is less overshooting and higher damping for swarms at the pyramid symbol parameter combination (\href{https://darus.uni-stuttgart.de/file.xhtml?fileId=401433}{Video} \vidref{\ref{tab:supplementary_videos_driver_attraction_linear}}{4}) compared to the square symbol (\href{https://darus.uni-stuttgart.de/file.xhtml?fileId=401432}{Video} \vidref{\ref{tab:supplementary_videos_driver_attraction_linear}}{3}). These dissipative (incoherent) dynamics could be a signature of improving the \textit{fading memory property} compared to the other explored dynamical regimes; the context of a linear driver is instructive in this regard.
The dynamic heterogeneity indicated by the stretched absolute VAC curve can still be observed in the overdamped regime (circle symbol) for the inversely attractive driver as the envelope curve of an oscillatory curve, which comes from strong speed-control in combination with the chosen integration time step (see \refref{\cite{Gaimann2025}} for a detailed discussion).

Complementary to the VAC are spatial correlation functions of dynamical fluctuations. In \figref[c,d]{\ref{fig:driver_attraction_correlations}} we measure the connected correlation function (CVC) of the active matter system. It captures how the change in the behavior of an agent $i$ correlates with the change in the behavior of another agent $j$ in a certain radial distance $r + dr$. The behavior is characterized by its velocity fluctuation from the center-of-mass velocity of the swarm. We observe the steepest and deepest decay from positive to negative correlation for the parameter combination located at the pyramid symbol for inverse driver attraction (\figref[c]{\ref{fig:driver_attraction_correlations}}). Together with the circle symbol, it features the shortest
correlation length $r_0$ (the first zero crossing of the CVC), indicating a more strongly localized anti-correlative dynamic. The strongest next-nearest neighbor interaction is obtained for the less strongly damped square parameter combination, however, indicated by the higher CVC value close to a radial distance of zero. The weaker CVC here for the diamond symbol could be a signature of the higher inertia and weaker agent-agent interactions. The magnitude of next-nearest neighbor correlations ($\approx 0.2$) is much weaker for the inversely attracting driver compared to the one for inverse driver repulsion ($\approx 0.5$). Driver repulsion induces correlations in dynamics via mechanical compression -- both locally (nearest-neighbor level), but also more long-ranged, as witnessed in a reemergence of the CVC above the zero line in the case of repulsion (see \refref{\cite{Gaimann2025}}), which does not happen here. This compression is lacking in the case of an attractive driving force that, naturally, does not carry a basic kind of correlating effect on dynamics. 

The very long spatial stretching of the negative CVC for diamond and square (more underdamped) cases could be due to more pronounced morphological changes in the system, involving a higher degree of translational motion of transient substructures spanning different regions of space.
The pyramid (optimally computing) case has a wider minimum of the CVC compared to the repulsive driver case (\refref{\cite{Gaimann2025}}), which presents condensed droplets that remain relatively stationary with respect to their center of mass.

For the linear driver attraction (\figref[d]{\ref{fig:driver_attraction_correlations}}), we note that there is only a steep and deep decay of the CVC for the square parameter combination, and the first zero crossing or correlation length is comparable to the optimal computing case of pyramid for the inverse attraction case of the driver; the minimum of the CVC is also comparable with it, but the breadth of the minimum is shorter. For the square and diamond parameter combinations, there is still a significant amount of strong, locally correlated velocity fluctuations in the swarm (also seen in the corresponding snapshot in \figref[e]{\ref{fig:driver_attraction_correlations}}, right panel, square symbol). For higher speed-control (pyramid, square symbols), the active matter system behaves more coherently (uniformly), which is expressed by velocity fluctuations in close particle neighborhoods being barely correlated (see also \figref[e]{\ref{fig:driver_attraction_correlations}}, right panel, pyramid and circle symbols). This also means that the dynamical susceptibility of the swarm is close to zero in this regime (see Appendix \figref[b]{\ref{fig:attraction_observables_2}}). This contrasts to the repulsive-driver case studied in this work's prequel \cite{Gaimann2025}, where the dynamical susceptibility served as a key indicator of high performance.

Together, the optimal performance for the linearly attractive driver is associated with a nearly-zero CVC function, while for the inversely attractive driver (and also for the repulsive driver studied in \refref{\cite{Gaimann2025}}) it is associated with the strongest and steepest decay. Thus, various driver information injection mechanisms lead to significantly different correlated agent dynamics, specifically visible in spatial and temporal correlations of velocity fluctuations. This is another expression of the openness and adaptivity of the system to different types of changing environments -- that it has adapting means of transferring information internally (expressed via these dynamical correlations). 

\textit{Synthesis:}
For a repulsive driver, the swarm must be dynamically susceptible to a driver that is immersed in the swarm. The case of attractive drivers is phenomenologically different. The condition of \emph{immersion} disappears, as seen in the video supplements, as the mode changes in which the swarm generates diversity in structure and dynamics. This is somewhat hindered in the linearly attracted case, where the swarm behaves more uniformly (the agent dynamics follow center-of-mass dynamics closely and have uniformly dissipative decay periods in the VAC); it induces time-delay response and spatial offset compared to the state of the driver.  

Especially the linear attractive case reveals that using the dynamical susceptibility alone, as a singular physical proxy for high performance, is insufficient. As mentioned, this quantity is suitable for expressing a highly correlated condensed state of matter undergoing nonequilibrium dynamics in a bulk. This is complementary to emergent morphological change abilities that are expressed more pronouncedly for the case of the attractive driver, which is chased by the swarm -- outsourcing diverse kinematical modes of locomotion and center-of-mass translations of the full structure and substructures (i.e., little stationary ``bulk'' is apparent).\\

\subsection{Optimal Inverse Driver Attraction Yields Active Droplets with Speed Gradients}

\begin{figure*}[htbp]
    \centering
    \includegraphics{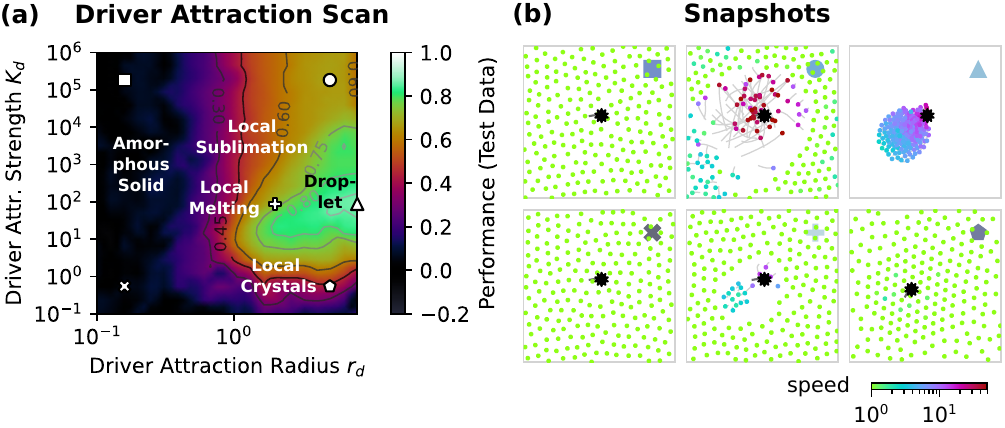}
    \caption{\textbf{Varying the parameters governing the inverse driver attraction interaction}. (a) Predictive performances for varying force strength $K_d$ and interaction radius $r_d$. (b) Snapshots of the active matter system corresponding to symbols in the heatmap in sub-figure (a). Refer to \tabref{\ref{tab:supplementary_videos_driver_attraction_nearcritdamped}} for corresponding videos.}
    \label{fig:driver_attraction}
\end{figure*}

\figref{\ref{fig:driver_attraction}} shows the predictive performance in dependence on varying the parameters of the inverse driver attraction force (see also \eqqref{\ref{eq:driver_attraction_inverse}}). For low driver attraction radii $r_d$ (cross symbol (\href{https://darus.uni-stuttgart.de/file.xhtml?fileId=410070}{Video} \vidref{\ref{tab:supplementary_videos_driver_attraction_nearcritdamped}}{1}) and square symbol (\href{https://darus.uni-stuttgart.de/file.xhtml?fileId=410071}{Video} \vidref{\ref{tab:supplementary_videos_driver_attraction_nearcritdamped}}{3})), only agents located very close to the driver position experience an attractive force. This occurs infrequently and lasts only for a couple of integration time steps, until the attracted agents leave the driver's area of influence. Because of this, the agents loosely arrange across the simulation box, and their agent-agent repulsion interactions give rise to an ``amorphous solid''. The lack of interaction with the driver results in a predictive performance of around $0.0$. High $r_d$ combined with low $K_d$ (pentagon symbol (\href{https://darus.uni-stuttgart.de/file.xhtml?fileId=410074}{Video} \vidref{\ref{tab:supplementary_videos_driver_attraction_nearcritdamped}}{2})) causes the formation of a crystalline solid, covering roughly the area (plus adjacent space) that is touched by the Lorenz-63 driver trajectory. The stronger order around the driver position, as well as small induced velocity fluctuations, could potentially enable a fuzzy driver localization, yielding a performance just under $0.45$. Overall, agent-agent repulsion remains dominant and coins the structure and dynamics of the swarm, leaving little responsiveness to the external environment. Increasing the attraction strength $K_d$ breaks this dominance and causes local (plus, circle symbols) and global (pyramid symbol) attraction to the driver in the swarm. 

In the case of a local attraction, agents constituting the amorphous solid get accelerated towards the driver once they are close enough to it  -- a manifestation of nonlinear thresholding behavior leading to sublimation -- forming a sharp density gap, or a  ``bubble'' in the solid with high surface tension, i.e., a smooth and sharp interface. If the driver is quasi-stationary or slowly-moving, the agents overshoot and are eventually accelerated in the reverse direction towards the driver again, leading to an expansion of the bubble via collisions with particles in the interface. If the driver rapidly switches the basin of attraction, the attracted agents can not follow the driver fast enough and locally disperse again, shrinking and closing the gap and restoring the amorphous solid. A new gap or bubble nucleates and grows around the new driver position once the driver slows down. This phenomenology resembles the self-healing effect previously described in a near-critically damped droplet in \refref{\cite{Gaimann2025}}.

The resulting clear distinction between the background solid and the locally attracted active agents in the case of the bubble (plus symbol (\href{https://darus.uni-stuttgart.de/file.xhtml?fileId=410075}{Video} \vidref{\ref{tab:supplementary_videos_driver_attraction_nearcritdamped}}{6})) could be the main feature explaining the elevated predictive performance of around $0.75$. A similar, albeit stronger and more widely-ranging phenomenon is observed when increasing $K_d$ and $r_d$ (circle symbol (\href{https://darus.uni-stuttgart.de/file.xhtml?fileId=410072}{Video} \vidref{\ref{tab:supplementary_videos_driver_attraction_nearcritdamped}}{4})). Agents here orbit the driver with high speeds, causing orbits spanning up to half of the simulation box, while an amorphous solid remains intact at the box boundaries.

A marked switch in phenomenology/mechanisms is observed once we reach the optimal computing maximum in this cut of the performance landscape (pyramid symbol (\href{https://darus.uni-stuttgart.de/file.xhtml?fileId=410073}{Video} \vidref{\ref{tab:supplementary_videos_driver_attraction_nearcritdamped}}{5})):
For an (almost) global interaction radius of $r_d = 8.0$ and an attraction strength of around $K_d = 100.0$, all agents are attracted to the driver, a droplet forms, and the background solid vanishes: There is an inversion of the roles of the actively responding entity/``body'' near the driver -- which goes from a swarm-vacancy bubble to a swarm droplet. Here, the phenomenology is similar to the one reported in \figref[a]{\ref{fig:driver_attraction_speed_controller}} (pyramid symbol, speed gradients) and a similar predictive performance of $P=0.87$ is reached. The convex shape of the performance landscape in this planar cut is not changed, and there is only a slight increase in performance.

These general findings on the growth, shrinkage, and deformation of coherent structures -- metamorphosis and morphogenesis -- appear significant to computation. However, the optimal droplet case of the pyramid symbol has an additional feature: The smooth and coherent gradients in velocities, which are fundamentally lacking inside a sublimation bubble. Hence, speed gradients are an additional, unique feature generated only when a condensed system is embodied, carrying dynamical information with it that is otherwise absent. 

\subsection{\label{subsec:driver-repulsion} Large and Spatially Extended Repulsive Information Injection Delivers Higher Performance}

\begin{figure*}[htbp]
    
    \centering
    \includegraphics{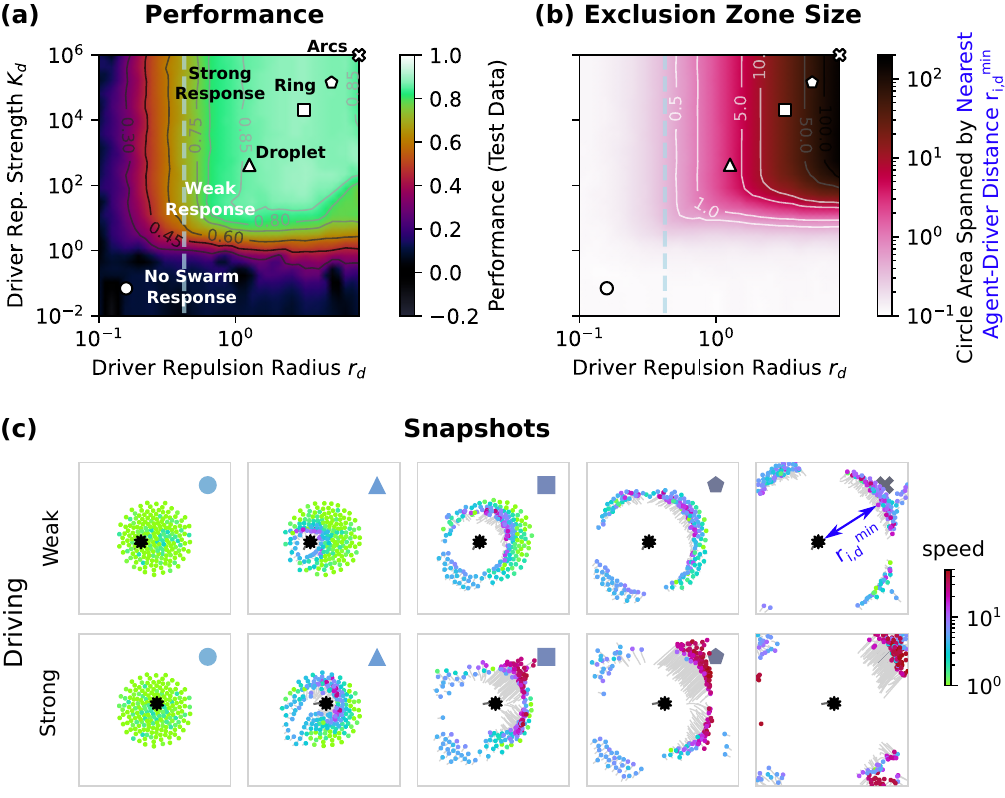}
    \caption{\textbf{Driver repulsion parameter scan using near-critically damped speed-controller settings} ($K_{sc} = 0.02069$, $s = 0.04833$). (a) Predictive performance for varying driver repulsion radius $r_d$ and driver repulsion strength $K_d$ (see \eqqref{\ref{eq:driver_repulsion}}).
    (b) Circle areas spanned by the nearest agent-driver distance $r_{i,d}^{min}$ for different driver repulsion force parameters, averaged over 1,000 time steps. The dashed light-blue vertical line marks the next-nearest neighbor distance $\langle\min(r_{i,j}^{NN})\rangle_{N,T} \approx 0.422$ for each agent $i$ to its next-nearest neighbor $j$, averaged over all $N$ agents and $T$ time steps. This value is calculated using a corresponding undriven steady state simulation (see also \href{https://darus.uni-stuttgart.de/file.xhtml?fileId=415771}{Video} \vidref{\ref{tab:supplementary_videos_undriven}}{1}). (c) Active matter snapshots corresponding to the symbols highlighted in sub-figures (a) and (b). High driver repulsion radii and strengths cause agent exclusion zones around the driver. Low densities correlate with higher predictive performances. Refer to  \figref{\ref{fig:Lymburn_critical_driver_repulsion_scan_heatmap}} for a corresponding parameter scan with speed-controller settings presented in Lymburn \etal \cite{Lymburn2021} ($K_{sc} = 2.0$, $s = 10.0$). Refer to \tabref{\ref{tab:supplementary_videos_driver_repulsion_nearcritdamped}} for the corresponding videos.
    }
\label{fig:nearcritdamped_driver_repulsion_scan_heatmap}
\end{figure*}

The latter section investigated a new variant of active matter reservoir computing using new driver-agent attraction interactions.
In this section, we revisit the information injection mechanism using the previously employed \textit{repulsive} driver-agent interaction \cite{Lymburn2021, Gaimann2025}. The driver repulsion interaction acts within a cut-off radius $r_d$ around the driver on the particles; its strength is scaled with the parameter $K_d$ (see \sectref{\ref{sec:mm-interactions}} for details). Given optimal intrinsic agent dynamics, what are optimal driver repulsion parameters to obtain the best active matter representations and hence the best reservoir computing performances?\\

In \figref{\ref{fig:nearcritdamped_driver_repulsion_scan_heatmap}} we vary $r_d$ and $K_d$ for a swarm with near-critically damped speed-controller settings. For low $r_d$ and $K_d$, the active matter system shows no visible interaction with the driver and enters a configuration resembling an undriven steady state (\figref[c]{\ref{fig:nearcritdamped_driver_repulsion_scan_heatmap}}, circle symbol (\href{https://darus.uni-stuttgart.de/file.xhtml?fileId=401408}{Video} \vidref{\ref{tab:supplementary_videos_driver_repulsion_nearcritdamped}}{1})). Because of this, no input time-series information is injected into the swarm, yielding a reservoir computing performance of around $0.0$ (\figref[a]{\ref{fig:nearcritdamped_driver_repulsion_scan_heatmap}}, circle symbol). 

Increasing $r_d$ and $K_d$ leads to the formation of an exclusion zone around the driver via the repulsion interaction (\figref[c]{\ref{fig:nearcritdamped_driver_repulsion_scan_heatmap}}, pyramid symbol (\href{https://darus.uni-stuttgart.de/file.xhtml?fileId=401409}{Video} \vidref{\ref{tab:supplementary_videos_driver_repulsion_nearcritdamped}}{2})). 
The active matter droplet maintains its shape upon quasi-stationary driver motion (upper panel), while agents become ejected -- followed by a subsequent return -- upon abrupt driver motion (lower panel). This optimized ``localization'' ability of driver information strongly increases performance to values of $P > 0.85$.
When further increasing $r_d$ and $K_d$, the exclusion zone diameter increases and agents assemble in a ring around the driver, which comes with a further increase in performance (\figref[a,c]{\ref{fig:nearcritdamped_driver_repulsion_scan_heatmap}}, square symbol (\href{https://darus.uni-stuttgart.de/file.xhtml?fileId=401410}{Video} \vidref{\ref{tab:supplementary_videos_driver_repulsion_nearcritdamped}}{3}) and pentagon symbol (\href{https://darus.uni-stuttgart.de/file.xhtml?fileId=401411}{Video} \vidref{\ref{tab:supplementary_videos_driver_repulsion_nearcritdamped}}{4})). Speeds of agents ejected by the driver further increase for strong driving (lower panel), which leads to higher speed gradients in the swarm. This could further improve localizing the driver signal in the swarm. For extreme values of $r_d$ and $K_d$ (\figref[a,c]{\ref{fig:nearcritdamped_driver_repulsion_scan_heatmap}}, cross symbol (\href{https://darus.uni-stuttgart.de/file.xhtml?fileId=401412}{Video} \vidref{\ref{tab:supplementary_videos_driver_repulsion_nearcritdamped}}{5})), there are not enough agents to close the active matter ring. Arcs form around the driver in the corners of the simulation box because the driver-agent distance becomes maximal within the periodic boundaries, and the repulsive potential is lowered. The arcs extend across the periodic boundaries, and the performance decreases to $P \approx 0.85$, possibly due to these border effects picked up by the observation kernels.

A lower bound for the exclusion area around the driver can be estimated by calculating the circular area described by the driver-agent distance $r_{i,d}^{min}$ of the closest agent to the driver. \figref[b]{\ref{fig:nearcritdamped_driver_repulsion_scan_heatmap}} shows how this exclusion zone area is associated with the performance in (a); there seems to be a threshold-like increase leading to high performance (above $0.75$) around a minimal exclusion area of $0.5$. In an active matter simulation without the external driver, the average next-nearest neighbor distance in the swarm amounts to around $0.422$, yielding an exclusion area of around $0.554$. This could mean that increasing the driver exclusion area beyond the average exclusion area around each agent significantly improves predictive performance. In other words, the driver signal must become structurally distinguishable from the innate liquid (or solid) structure generated by agent-agent interactions.

We obtain similar results for an active matter system prepared using speed-controller settings presented in Lymburn \etal (2021) (see \figref{\ref{fig:Lymburn_critical_driver_repulsion_scan_heatmap}}). Parameters causing a strong and long-range driver repulsion interaction could be optimal across various other swarm settings.

Effects of periodic boundary conditions (PBCs) were explored for a swarm using speed-controller settings of Lymburn \etal (2021) in \figref{\ref{fig:driver_repulsion_larger_boxes}}. When increasing the simulation box size to $l^{\text{sim}}_{\text{box}} = 32.0$ or $l^{\text{sim}}_{\text{box}} = 64.0$, agents assemble in four independent arcs resembling a cross (see also \figref[f]{\ref{fig:driver_repulsion_larger_boxes}}). Allowing for larger simulation and observer kernel placement box sizes (where PBCs play a lesser or no role anymore) yields qualitatively similar performance heatmaps. Quantitatively, we find that spreading the Gaussian observation kernels more thinly across a larger box strongly reduces the predictive performance (see \figref[c,e]{\ref{fig:driver_repulsion_larger_boxes}}).

\new{Another degree of freedom of how to inject information into the active matter system is the time scale governing the external driving particle dynamics with respect to the time scale of active matter dynamics. In \figref{\ref{fig:speed-controller-different-speeds-same-horizon}} we briefly explored different mean driver speeds $\bar{v}_d \in \{1, 2, 5, 10, 15\}$ for a repulsive driver while ensuring a constant mean driver distance predicted ahead $\bar{s}_{pred} = \bar{v}_d \Delta t_{pred} = 5.0$ to be able to compare the predictive performance across these systems. We find that the driver dynamics time scale plays an important role for predictive performance: the optimal performance varies between $P \approx 0.80$ for a very slow driver ($\bar{v}_d =1$) compared to over $P \approx 0.90$ for faster drivers. This is expected, since the dynamical response of the active matter system (more quasi-stationary for a slow driver with respect to mean agent speed, versus more non-linear and dynamically heterogeneous for a faster driver). Interestingly, the default mean driver speed of $\bar{v}_d \approx 10.0$ chosen throughout this manuscript delivers slightly inferior optimal performance $0.85 < P < 0.90$ than slower or faster drivers ($P > 0.90$). We note that the dynamical regime of optimal performance shifts only slightly for different mean driver speeds and remains in the vicinity of near-critical damping. This underlines once again the previously found robust optimality of this regime \cite{Gaimann2025}.}

The above results on optimal information injection via repulsion are valid for a large number of agents. However, a single-particle reservoir, poised in the near-critically-damped regime, shifts the optimal regime qualitatively (see Appendix \figref[b]{\ref{fig:few_particles}}). This evidences the importance of localizing the source of driving information via a well-defined exclusion zone that is enclosed by a ring comprising multiple agents. An optimal injection mode via driver repulsion must hence be understood through these sorts of collective response mechanisms triggered.

\begin{figure}
    \centering
    \includegraphics{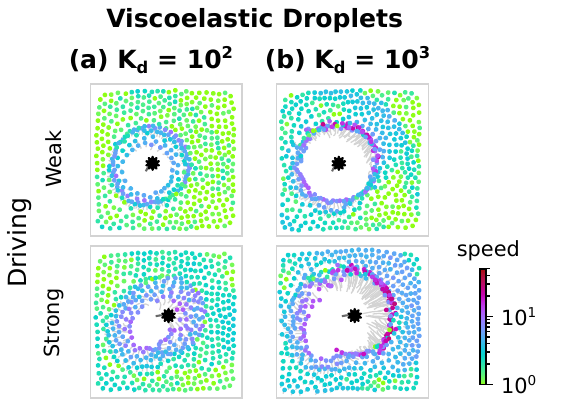}
    \caption{
        \textbf{Liquid droplets with a repulsive driver comprising $\bm{N_a = 500}$ agents.} Shown are two driver repulsion strengths: (a) weak ($K_d = 10^2$) and (b) strong ($K_d = 10^3$). For weak driver repulsion strength, the pressure from outer agents forces agents to invade the agent exclusion zone, which is observed for higher driver repulsion strengths. The liquid droplets are obtained through a near-critically damped speed-controller ($K_{sc} = 0.02069$, $s = 0.04833$), a long-ranged agent-agent repulsion interaction ($r_r = 4.0$, $K_r = 2.0$), and a homing force ($K_{h} = 2.0$). Refer to \tabref{\ref{tab:supplementary_videos_bulk}} for corresponding videos.
    }
    \label{fig:viscoelastic_fluids}
\end{figure}
\subsection{\label{sec:repulsion} Basic \new{Propagation} of the Input Signal via Agent-agent Repulsion}

\begin{figure*}
    \centering
    \includegraphics{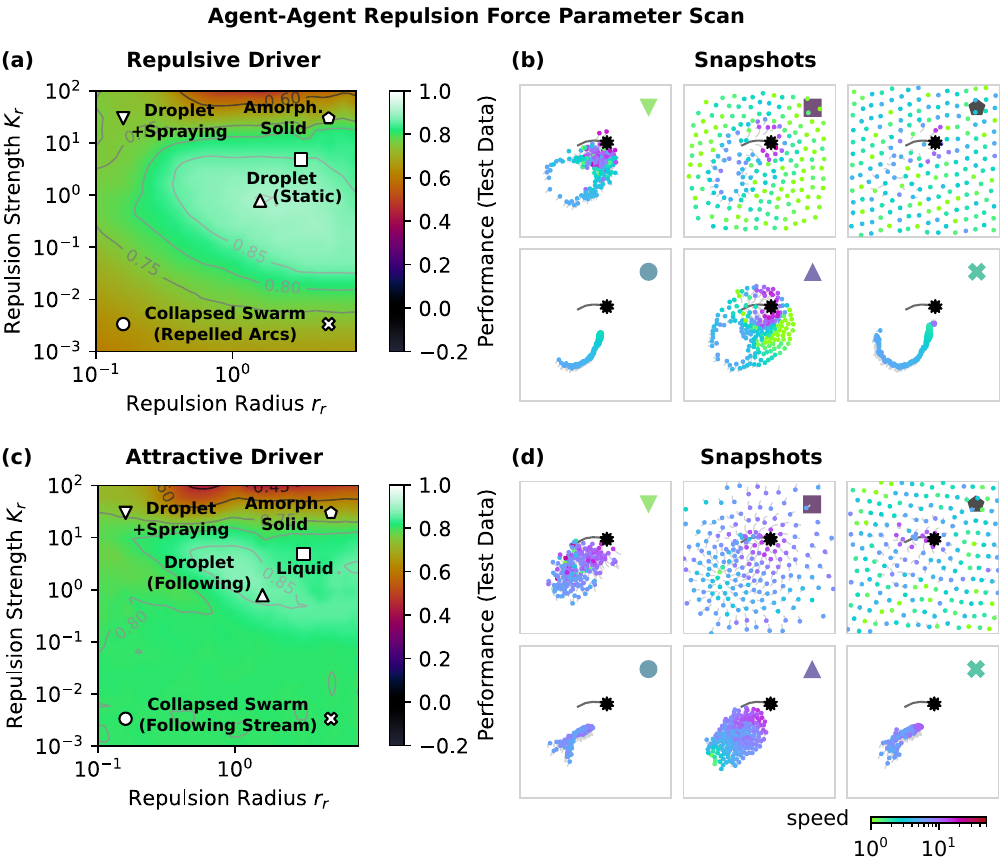}
    \caption{
        \textbf{Critically-damped viscoelastic droplets barely enclosing the driver provide optimal performance.}
        (a,c) Scans over the repulsion strength $K_r$ and the repulsion radius $r_r$ of the agent-agent repulsion force in the critically damped regime ($K_{sc} = 0.02069$, $s = 0.04833$) for (a) a repulsive driver ($r_d = 2.0$) and (c) an attractive driver ($r_d = \infty$), each with $K_d = 100.0$. 
        (b,d) Snapshots of swarms in different dynamical regimes that are marked as symbols in sub-figures (a,c); color indicates agent speed. Refer to \tabref{\ref{tab:supplementary_videos_repulsion_driver_rep}} and \tabref{\ref{tab:supplementary_videos_repulsion_driver_attr}} for corresponding videos. An analogous parameter scan to sub-figure (c,d) but with a restricted driver attraction radius of $r_d = 2.0$ is shown in \figref{\ref{fig:repulsion_attractive_driver_range_2.0}}, recovering effects described in \figref{\ref{fig:driver_attraction}}.}
    \label{fig:repulsion}
\end{figure*}

The significance of spatially ``localizing'' the driver through the swarm is further investigated in this section. The spatial coverage of the swarm is controlled mainly by the agent-agent repulsive force $\mathbf{F}_r$. Tuning its repulsion (cut-off) radius $r_r$ and strength $K_r$ effectively controls the mean swarm density. 

In the first case, we study the previously-employed default repulsive driver ($K_d = 100.0$; $r_d = 2.0$). For low $K_r$, the agent-agent repulsion is very low (\figref[a,b]{\ref{fig:repulsion}}, circle symbol (\href{https://darus.uni-stuttgart.de/file.xhtml?fileId=410021}{Video} \vidref{\ref{tab:supplementary_videos_repulsion_driver_rep}}{4}) and cross symbol (\href{https://darus.uni-stuttgart.de/file.xhtml?fileId=410027}{Video} \vidref{\ref{tab:supplementary_videos_repulsion_driver_rep}}{5})). Agents collapse to a condensed droplet when they experience only weak (quasi-stationary) driving. For stronger driving, the agents become dispersed and form arc-like patterns. 

When increasing $K_r$ and $r_r$, the distance between agents and consequently the area covered by the swarm grows, leading to the formation of a droplet (pyramid symbol (\href{https://darus.uni-stuttgart.de/file.xhtml?fileId=410024}{Video} \vidref{\ref{tab:supplementary_videos_repulsion_driver_rep}}{3}) and square symbol (\href{https://darus.uni-stuttgart.de/file.xhtml?fileId=410018}{Video} \vidref{\ref{tab:supplementary_videos_repulsion_driver_rep}}{2})). Here, the driver remains embedded in the swarm at all times, and an interface forms around the driver. The phenomenology of such near-critically damped droplets has been studied in the prequel to this work \cite{Gaimann2025}. Even larger repulsive forces lead to the formation of a solid state (pentagon symbol (\href{https://darus.uni-stuttgart.de/file.xhtml?fileId=410025}{Video} \vidref{\ref{tab:supplementary_videos_repulsion_driver_rep}}{1})), taking periodic boundary conditions into account. The agents arrange themselves in an ``amorphous solid'' to minimize the total energy of the system. An exclusion zone due to the driver repulsion forms around the driver and resembles a local lattice defect. The driver locally melts the solid state, and the pressure from surrounding agents induces a self-healing process once the driver moves on. 
In the case of low repulsion radii $r_r$ but high strengths $K_r$, agents experience a strong, short-ranged, and short-lived repulsion force (\figref[a,b]{\ref{fig:repulsion}}, nabla symbol (\href{https://darus.uni-stuttgart.de/file.xhtml?fileId=410020}{Video} \vidref{\ref{tab:supplementary_videos_repulsion_driver_rep}}{6})). This causes the assembly of a denser droplet where agents ``spray'' from the center of the simulation box towards the outside of the box. The local speed variance is higher compared to the droplet cases (square, pyramid symbols). \\

In the second case, we study the globally attractive driver introduced in \sectref{\ref{sec:attractive-driver}} ($K_d = 100.0$; $r_d = \infty$). For low $K_r$ (\figref[c,d]{\ref{fig:repulsion}}, circle symbol (\href{https://darus.uni-stuttgart.de/file.xhtml?fileId=412399}{Video} \vidref{\ref{tab:supplementary_videos_repulsion_driver_attr}}{4}), and cross symbol (\href{https://darus.uni-stuttgart.de/file.xhtml?fileId=412401}{Video} \vidref{\ref{tab:supplementary_videos_repulsion_driver_attr}}{3})), the agents oscillate around the driver under weak driver motion, mostly along a horizontal line. The low agent-agent repulsion allows the high concentration of agents on such lines (or streams). They are also formed when following the driver to its next quasi-stationary position after an excursion. 

Increasing the repulsion strength $K_r$ leads to the droplet showing speed gradients in following the driver (pyramid symbol (\href{https://darus.uni-stuttgart.de/file.xhtml?fileId=412400}{Video} \vidref{\ref{tab:supplementary_videos_repulsion_driver_attr}}{3})) (see \sectref{\ref{sec:attractive-driver}}). Very high $K_d$ with low $r_r$ (nabla symbol (\href{https://darus.uni-stuttgart.de/file.xhtml?fileId=412397}{Video} \vidref{\ref{tab:supplementary_videos_repulsion_driver_attr}}{6})) form a following droplet with a spraying effect similar to the repulsive driver case. Increasing $K_d$ and $r_r$ beyond the droplet case leads to an extended low-density droplet (square symbol (\href{https://darus.uni-stuttgart.de/file.xhtml?fileId=412402}{Video} \vidref{\ref{tab:supplementary_videos_repulsion_driver_attr}}{2})). Here, the driver is always immersed in the swarm, but no agent exclusion zone forms, in contrast to the previously studied repulsive driver case. Instead, driver information is transmitted across the swarm via a stronger agent-agent repulsion, which gives rise to currents towards the driver upon rapid motion. Agent speeds adapt to driver speeds and are higher when closely attracted to the driver. Eventually, for very high $K_d$ and $r_r$ an amorphous solid forms (pentagon symbol (\href{https://darus.uni-stuttgart.de/file.xhtml?fileId=412398}{Video} \vidref{\ref{tab:supplementary_videos_repulsion_driver_attr}}{1})), with a small attractive defect forming around the driver position.

\figref[a,c]{\ref{fig:repulsion}} show the predictive performances for these regimes. 
For the collapsed swarms, the driver cannot constantly inject information into the swarm -- especially during its large excursions. The driver's position is also not well resolved spatially, since agent-driver interaction is often restricted to a small spatial angle span. This results in performances around $0.65$ (circle symbol) and $0.72$ (cross symbol) for the repulsive driver, and just above $0.80$ for the attractive driver (circle, cross symbols). The case where a dense stream of agents {continuously follows} the driver seems to render a far stronger reservoir compared to the repelled arcs. The slow return of agents to the center likely hinders continuous information injection by the driver, which may explain the lower performance.

In the droplet regime (pyramid symbols), the driver's position information becomes optimally ``resolved'' by the agents. In the case of a repulsive driving force, the driver is fully immersed in the swarm, and an exclusion zone forms. Interestingly, the performance of the more condensed droplets (pyramid symbol; attractive driver: $P = 0.86$; repulsive driver: $P= 0.904$) is higher than that of the less dense droplets (square symbol; attractive driver: $P = 0.82$; repulsive driver: $P= 0.85$). 
When the swarm forms a condensed droplet, its response is naturally highly collective. All agents are affected by the driver and markedly change their positions and speeds during the simulation. For the repulsive driver, the outermost agents in the less dense droplet feel the driver excitation only weakly, and their speeds and positions remain quasi-static during the simulation. These outermost agents also block a larger spatial response of those agents repelled by the driver, by exerting an (inwards-facing) repulsive force. This could reduce the RC expressivity and cause the observed performance decrease. For the case of the attractive driver, the liquid-like swarm has a comparable but slightly lower performance, which could be due to the lack of a clear interface or the global motion of the total swarm induced by the attractive driver. 

In the regimes where a compressed amorphous solid emerges (pentagon symbols), the number of agents that contribute to information processing is low. The driver only acts locally, while the agents are spread across the whole simulation box, which leads to further performance decrease. 
These solid-state active matter systems have not been previously described or studied for reservoir computing in the literature, to our knowledge. 

We also modeled different agent-agent repulsion interactions and assessed the impact on predictive performance in \figref{\ref{fig:repulsion_variations}}. Overall, the chosen $\sim 1/r^m$ long-range agent-agent repulsion interaction model (see \eqqref{\ref{eq:long_range_repulsion}}) with $m=1$ delivers close to optimal performance in the near-critically damped regime and with a repulsive driver (optimal performance for $r_r = 4.0$ and an exponent of $m\approx 0.55$: $P \approx 0.89$). For a short-range interaction (see \eqqref{\ref{eq:short_range_repulsion}}), the optimal performance was similar ($r_r = 4.0$ and a decay constant of $\lambda \approx 0.55$: $P \approx 0.90$). This suggests that the exact type of microscopic interaction may be less relevant for optimal reservoir computing performance, as long as they generate favorable macroscopic swarm patterns. 

\subsection{\label{sec:n_agents}The Significance of Number of Agents and Agent-agent Repulsion for Emergent Collectivity}

\begin{figure*}
    \centering
    \includegraphics{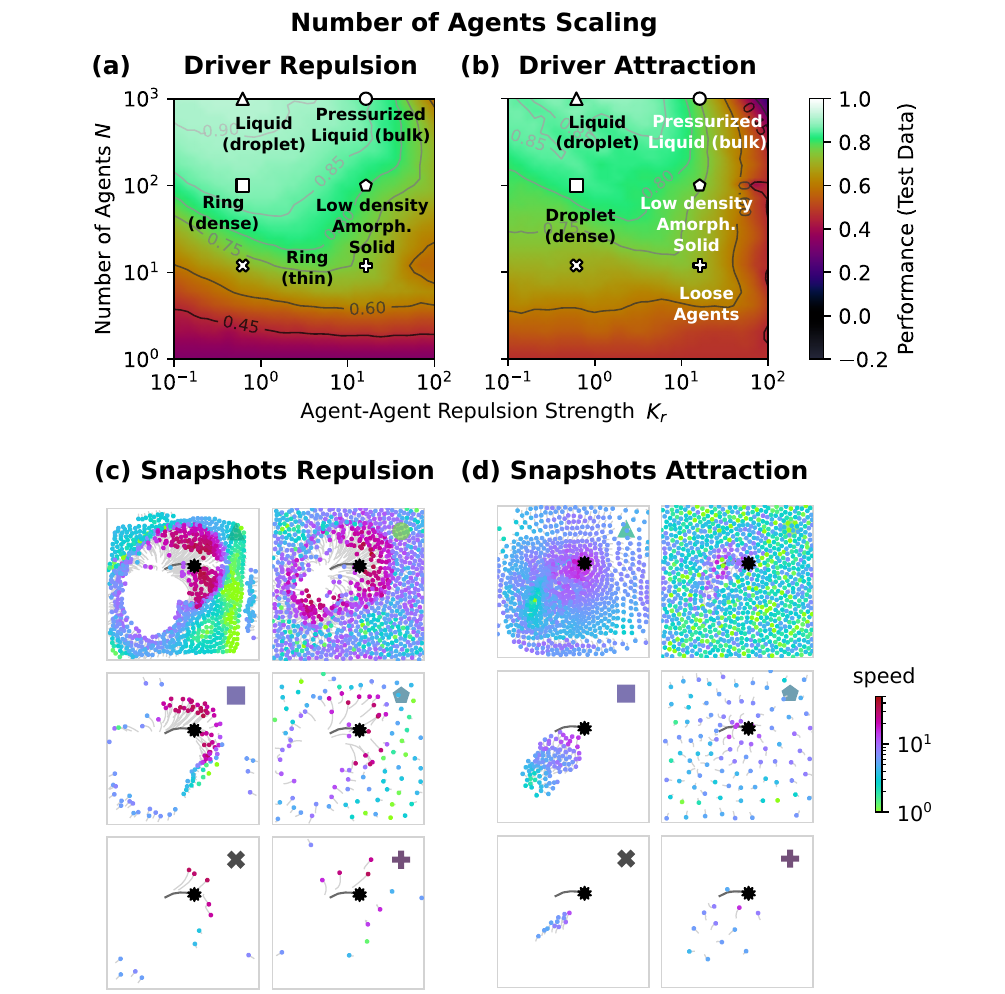}
    \caption{
        \textbf{Active matter reservoir computing performance scaling in dependence of the swarm size and the inter-agent repulsion strengths} in the near-critically damped regime and using a long-ranged agent-agent repulsion interaction ($r_r = 4.0$). (a,b) Predictive performances versus employed number of agents $N_a$ and agent-agent repulsion strengths $K_r$ for (a) a repulsive ($K_d = 10^3$; $r_d = 4.0$) and (b) an inversely attractive driver ($K_d = 10^2; r_d = \infty$). (c,d) Snapshots corresponding to parameter combinations (symbols) marked in sub-figures (a,b). Refer to \tabref{\ref{tab:supplementary_videos_repulsion-nagents-driver-repulsion}} and \tabref{\ref{tab:supplementary_videos_repulsion-nagents-driver-attraction}} for corresponding videos.
    }
    \label{fig:n_agents_repulsion_strength}
\end{figure*}

The last section revealed the importance of ensuring optimal information transmission in the swarm through generally strong and long-ranged agent-agent repulsion interactions. This raises the question of how the number of responding agents impacts the predictive performance, which needs to be understood in relation to the phases that appear in each case. Very different nonequilibrium responses appear and, hence, information processing mechanisms, for either type of driving force (repulsive/attractive). 

\figref{\ref{fig:n_agents_repulsion_strength}} varies the number of agents $N_a$ and the agent-agent repulsion strength $K_r$.
We choose a driver repulsion strength of $K_d = 10^3$ and a driver attraction strength of $K_d = 10^2$ based on findings in previous sections (\figref[a]{\ref{fig:nearcritdamped_driver_repulsion_scan_heatmap}} and \figref[a]{\ref{fig:driver_attraction}}).

We first analyze the repulsive driver case in \figref[a,c]{\ref{fig:n_agents_repulsion_strength}}.
For a low number of $N_a \approx 10$ agents (cross symbol, plus symbol), the predictive performance remains just under $0.75$. While agents assemble in small arcs consisting of a few particles for low $K_r$ (cross symbol (\href{https://darus.uni-stuttgart.de/file.xhtml?fileId=409928}{Video} \vidref{\ref{tab:supplementary_videos_repulsion-nagents-driver-repulsion}}{5})), a larger angular space is covered for higher $K_r$ and the angular agent distribution around the driver is more uniform (plus symbol (\href{https://darus.uni-stuttgart.de/file.xhtml?fileId=409929}{Video} \vidref{\ref{tab:supplementary_videos_repulsion-nagents-driver-repulsion}}{6})). Despite this difference, the contrast in performance is only marginal. This indicates that in this few-agent case, uniform ring-like structures (and hence collectivity) do not yield a performance benefit yet. 

Increasing the number of agents towards $N_a \approx 100$ leads to the formation of an agent ring around the driver for low $K_r$ (square symbol (\href{https://darus.uni-stuttgart.de/file.xhtml?fileId=409921}{Video} \vidref{\ref{tab:supplementary_videos_repulsion-nagents-driver-repulsion}}{3})) and for higher $K_r$ to a large agent exclusion zone (defect) in a low-density amorphous solid (pentagon symbol (\href{https://darus.uni-stuttgart.de/file.xhtml?fileId=409926}{Video} \vidref{\ref{tab:supplementary_videos_repulsion-nagents-driver-repulsion}}{4})). We note that the ring features a bulge, originating from an excess number of agents compared to an (almost) perfect uniform ring structure. This bulge switches sides when the driver switches to its next quasi-stationary resting position (the basin of attraction of the Lorenz-63 attractor). The performance difference between these two parameter combinations (about $0.86$ versus $0.80$) signifies that a high-density ring around the driver that can eject agents towards the borders of the simulation box under driving is superior to the ``solid'', which restricts ejections.

Further increasing the agent number to $N_a = 1,000$ leads to the formation of a dense liquid droplet for low $K_r$ that almost covers the complete simulation box (pyramid symbol (\href{https://darus.uni-stuttgart.de/file.xhtml?fileId=409930}{Video} \vidref{\ref{tab:supplementary_videos_repulsion-nagents-driver-repulsion}}{1})). It is important to note that the system here does not form a full bulk, but rather a small gap exists at the periodic boundaries, which allows the system to relax. It is highly susceptible to the driver and shows the propagation of changes in the driver's position via stationary-like waves synchronized with it, in a quasi-stationary regime. There is a clear interface between the driver and the agents, which quickly re-establishes even after strong, abrupt driving. An interface-enabled ``localization'' of the driver is supported by higher agent speeds in the vicinity of this interface, forming a gradient radially. All of these observations lead to the highest predictive performance to date of $P = 0.9156$, reported for reservoir computing with active matter simulations for the Lorenz-63 prediction task. 

Increasing $K_r$ for the 1,000 agent system leads to the formation of a bulk, pressurized liquid (circle symbol (\href{https://darus.uni-stuttgart.de/file.xhtml?fileId=409924}{Video} \vidref{\ref{tab:supplementary_videos_repulsion-nagents-driver-repulsion}}{2})). The high agent-agent repulsion force enforces a high pressure, which causes fluctuating velocity. It also pushes agents inside the area spanned by the driver repulsion radius $r_d$. For a stationary driver, agents residing around $r_d$ and experiencing the driver repulsion force obtain temporarily higher speeds and form a ring of high speeds around the driver, while agents inside this ring have lower speeds. Higher agent speeds caused by rapid driver motion are rapidly transmitted and affect almost the entire bulk. The significant difference in performance (here $P \approx 0.85$) could originate from the smaller driver exclusion zone in the bulk, which could be less sensitive. Moreover, the high-pressure bulk liquid does not exhibit coherent information transmission via waves, which could be a crucial feature for optimal active matter reservoir computing.

In the attraction case in \figref[b,d]{\ref{fig:n_agents_repulsion_strength}}, we observe the formation of small, high-density (cross symbol (\href{https://darus.uni-stuttgart.de/file.xhtml?fileId=415746}{Video} \vidref{\ref{tab:supplementary_videos_repulsion-nagents-driver-attraction}}{5})) and loose, low-density (plus symbol (\href{https://darus.uni-stuttgart.de/file.xhtml?fileId=415730}{Video} \vidref{\ref{tab:supplementary_videos_repulsion-nagents-driver-attraction}}{6})) droplets for low agent numbers $N_a \approx 10$ with performances of around $0.70$. Increasing the number of agents to $N_a \approx 100$ strongly increases performance to over $0.80$ for low $K_r$, where a droplet with a visible speed gradient is obtained that has been studied in \sectref{\ref{sec:attractive-driver}} (square symbol (\href{https://darus.uni-stuttgart.de/file.xhtml?fileId=415734}{Video} \vidref{\ref{tab:supplementary_videos_repulsion-nagents-driver-attraction}}{3})). A weaker driver attraction in this scenario further increases the performance (see \figref[b,d]{\ref{fig:n_agents_repulsion_strength_supplement}}, square symbol), which could be due to a fine-tuned (slower) time scale of agents returning. Further increasing $K_r$, however, leads to a low-density amorphous solid that fully covers the simulation box (pentagon symbol (\href{https://darus.uni-stuttgart.de/file.xhtml?fileId=415745}{Video} \vidref{\ref{tab:supplementary_videos_repulsion-nagents-driver-attraction}}{4})), similar to \figref[c,d]{\ref{fig:repulsion}}, square symbol). The agent density and agent speeds are higher around the driver, and the swarm responds globally to the driver's motion. 
For an even higher number of $N_a \approx 1,000$ agents and low $K_r$, we observe a liquid droplet that again spans almost the complete simulation box (pyramid symbol (\href{https://darus.uni-stuttgart.de/file.xhtml?fileId=415744}{Video} \vidref{\ref{tab:supplementary_videos_repulsion-nagents-driver-attraction}}{1})). It features again wave-like patterns and is highly susceptible, because they are formed even for minor driver motion. Agent speeds are higher around the driver, and the formation of a circular interface in the liquid can be observed. This interface forms upon small, quasi-stationary motion.
In contrast, a fully immersed driver in an almost static liquid droplet is observed for the same scenario and a weaker driver attraction strength of $K_d \approx 11.2$ (see \figref[b,d]{\ref{fig:n_agents_repulsion_strength_supplement}}, pyramid symbol). Here, agent wave dynamics can again be observed: The driver leaves traces in the liquid in the form of propagating spiral waves. While this coherent form of information transfer seems to be useful for reservoir computing with a performance just under $0.85$, it is not as effective as the repulsive driver. It could be that the performance is lower compared to the case with $K_d = 100.0$ because the Gaussian observation kernels cannot properly process the information transmitted through the spiral wave process due to its alternating nature.  
For high $K_r$ (circle symbol (\href{https://darus.uni-stuttgart.de/file.xhtml?fileId=415736}{Video} \vidref{\ref{tab:supplementary_videos_repulsion-nagents-driver-attraction}}{2})), again a pressurized liquid with high local speed fluctuations is obtained, where the attractive driver only leaves traces of higher speeds in the otherwise homogeneous bulk when moving rapidly, yielding a still surprisingly high performance of around $0.75$.

Overall, the interplay between a repulsive component from the driver that injects information, an attractive resetting component from the homing force, and another repulsive component that distributes the information in space delivers optimal predictive performance.


In the extreme case of using a single agent as a reservoir (see \figref{\ref{fig:few_particles}}), we find qualitative differences in their predictive performance heatmaps compared to their $N_a=200$ counterparts. Comparing the driver repulsion parameter scan with $N_a = 200$ in \figref[a]{\ref{fig:nearcritdamped_driver_repulsion_scan_heatmap}} to the one with $N_a = 1$ in \figref[b]{\ref{fig:few_particles}}, we observe that the qualitative trend (the stronger and the more long-ranged the driver repulsion the higher the predictive performance) observed for $N_a = 200$ does not hold for the single particle case. There, a performance optimum is reached at a relatively low driver repulsion strength of $K_d \approx 1.0$ and driver repulsion radii of $r_d \gtrapprox 1.0$ (\figref[b]{\ref{fig:few_particles}}, plus symbol). This suggests that the optimal information processing mechanism for a single particle is to track the driver more closely. Strong single-agent repulsion causes a temporary gap in information injection because the agent requires more time to return to the driver in this near-critically damped regime. For a larger number of particles, however, the optimal mechanism is to push the agents further away, such that an agent droplet or ring forms around the driver. This mechanism remains receptive to new driver information even in the case of strong driving, which could be why high $K_d$ and $r_d$ are preferred in this case. This means that a near-critically damped speed controller seems to be only optimal under the condition of sufficient collectivity, which is not given for a single agent.
In contrast, the optimal speed-controller parameters could be inferred already from a single particle reservoir, since the speed-controller affects individual, \textit{intrinsic} agent dynamics. 
This is also true for a repulsive driver in the speed-controller setting used in \refref{\cite{Lymburn2021}} (see \figref[a]{\ref{fig:Lymburn_critical_driver_repulsion_scan_heatmap}} for $N_a = 200$ and \figref[a]{\ref{fig:few_particles}} for $N_a = 1$).\\

We note that the existence of an extended driver exclusion zone void of agents is not a prerequisite for a high predictive performance. In \figref{\ref{fig:viscoelastic_fluids}} we present two active matter systems with two different driver repulsion strengths. For a high driver repulsion strength $K_d = 10^3$ (right panel; \href{https://darus.uni-stuttgart.de/file.xhtml?fileId=410252}{Video} \vidref{\ref{tab:supplementary_videos_bulk}}{2}), we recover the driver exclusion zone free of agents together with the wave-like interface formation, previously presented in \figref[c]{\ref{fig:n_agents_repulsion_strength}} (pyramid symbol).
On the other hand, for a lower driver repulsion strength $K_d = 10^2$ (left panel; \href{https://darus.uni-stuttgart.de/file.xhtml?fileId=410253}{Video} \vidref{\ref{tab:supplementary_videos_bulk}}{1}), the agents ``invade'' the previously present agent-free exclusion zone. We report the establishment of three interfaces: the first interface separates a vacuum (circular area without agents / exclusion zone) from the liquid droplet. A second interface consists of a ring of agents with higher speeds, compared to agents closer or further away from the driver. It forms because agents in the vicinity of the circle described by the driver repulsion radius become repelled here in a discontinuous fashion.
Lastly, a third interface similar to a wavefront forms under sufficiently strong driving in the outer layers of the droplet (not in the quasi-stationary case). It results from the propagation of information via waves induced by the driver.
Agents inside the second ring obtain higher speeds under strong driving and then form a sharp interface. Interestingly, the interface moves smoothly through the swarm. This is possible because agents located at the interface become closer to the driver; they are ``squeezed'' between the inner and outer layers of agents. Agents outside of the interface then seamlessly form the new interface. Even under fast driving conditions, this mechanism generates a sharp interface in the swarm. The predictive performances for both the lower and higher driver repulsion strengths are very high, with $0.896$ and $0.902$, respectively. Comparing these two cases, we learn that an interface \textit{embedded} in an active matter background provides sufficient contrast for high reservoir computing performance. 
\endgroup

\begingroup
\let\clearpage\relax
\section{\label{sec:discussion} Discussion and Conclusion}

In this paper, we performed reservoir computing using an out-of-equilibrium driven active matter system as the central computing element (the reservoir). We looked at two important interactions that are essential for its functioning. Firstly, we focused on the interaction between a driving particle and the active matter system, which determines how external information is initially injected into the reservoir -- the main interaction force that governs the direction of the nonequilibrium evolution. Secondly, we addressed the repulsive interactions between neighboring particles, which are key in how information propagates within the swarm.

By switching the type of driving force from repulsive to attractive, the resulting phenomenology and thus, mechanisms of information processing, are markedly different. Yet, the performance landscape cuts in parameter space can be highly similar, at least for the most important control factors. For inverse attraction forces, peak performance is identical -- essentially the same landscape of the optimally computing dynamical regime is recovered in the speed controlling parameters, which appear the most essential for active matter RC and have to do with intrinsic self-stabilization or -damping abilities \cite{Gaimann2025}. For linear attraction, some limitations appear, suggesting that the nonlinearity of information injection is key.

By observing the mean and variance of positions together with speeds, in the time evolution, we indicated noteworthy signs of adaptive mechanisms in each case of driver force. These become activated whenever the environment becomes more unpredictable, and are mostly visible in the fluctuations and statistics on dynamical quantities. For example, speed and speed variances are highly regulated in cases with maximal performance and robust shapes of the performance landscape (in planar cuts of it) -- the cases of inverse repulsion and inverse attraction driving forces, for swarms poised near critical damping, are very similar in this regard.  

Cross-correlations between driver and agent are also useful measures to distinguish the effects of different driving. 
We noted that an inverse/nonlinear driving force appears to drive the system more strongly out of equilibrium; there is a stronger asymmetry in delay-time. This is the subject of a future study.
We further characterized this phenomenon using velocity autocorrelations, which show a characteristic stretch that captures the dynamic heterogeneity of the gradient. Spatial correlations of velocity fluctuations (connected velocity correlations (CVC)) show that the optimal regime is also characterized by the strongest and deepest change from a positive to a negative CVC among different speed-controller settings.

In the broad underdamped (sub-optimal) regime, the repulsive case showed a significant performance boost over the attractive case, which is intriguing. This suggests a hidden, but not absolute, advantage of repulsive interactions. From a basic physical perspective, repulsions are fundamental to detection, as they transfer momenta -- in fact, they usually reverse momenta of agents, generating contrast in the kinematical and dynamical information, in this case of strong non-reciprocal driver-agent interactions, which treats the driver as immutable.  From a biological perspective, the apparent advantage might ultimately be about ``triangulating'' the position of the driving point source, which is a robust, and fundamentally collective mechanism, rather than localizing by chasing it, which is the optimal strategy at the individual (single agent) level. 

The proposed attractive driver interaction is also a candidate for a more minimal active matter reservoir computing model, because the previously employed homing force required to compensate a repulsive driver can be removed.
We confirmed that to obtain optimal performance for an attractive driver, the interaction range should then span the complete simulation box to form a droplet comprising all agents. For smaller ranges, we observe phenomenologically the local melting, sublimation, or crystallization of an otherwise ``amorphous solid'' through the driver.

We revisited the repulsive driver interaction and found phenomenologically -- for increasing driver repulsion strength and range -- a non-responding droplet, the formation of an exclusion zone in the droplet, a ring of agents, and finally the splitting of this ring into arcs. The highest predictive performance can here be obtained with a large ring of agents around the driver. Why it performs almost equally well to the condensed droplet is not intuitively clear. We suggest
there may be a tradeoff between morphological and dynamical diversity at play. This could be verified by quantifying structural versus dynamical information (subject to future work). 

For the role of agent-agent repulsion interactions (basic tuners of information \new{propagation capabilities}), we confirm that optimal performance is obtained with a dense droplet that immerses the driver -- independently of the driver interaction (attractive or repulsive). Notably, a decrease in density decreases the predictive performance, possibly due to a reduction in response modes. This could be further quantified in future research, for example, by measuring the variance of the Gaussian activation kernels for low/high density droplets. We report that sufficiently high agent-agent repulsion forces lead to the formation of a bulk solid-state reservoir, extending through the periodic boundaries. The performance we found for bulk systems was lower than for droplets, possibly because of suppressed response modes and a lack of contrast between a well-defined undriven swarm ``ground state'' and a driven ``excited state''. The impact of going from the liquid droplet to a bulk system while not changing agent-agent repulsion interactions or the number of agents could be further assessed by changing the simulation box size.

The number of agents also tunes the basic information transmission properties by defining the density together with the agent-agent repulsion interactions. We revealed that the highest predictive performances of $P = 0.9156$ reported to date for reservoir computing with active matter simulations is obtained using $N_a = 10^3$ agents, a strong and long-ranged repulsive driver ($r_d = 4.0; K_d = 10^3$) and a weak, long-ranged agent-agent repulsion interaction ($r_r = 4.0; K_r \approx 0.62$). With these settings, phenomenologically, a dense liquid droplet covering almost the complete simulation box is obtained. Small changes in the driver position excite coherent waves in the droplet, which seems to be a favorable information transmission property useful for reservoir computing.\\

With our work, we add a key hypothesis to the broad discourses on collectivity and computation, and how to physically quantify computation quality: Information processing is enhanced through spatially and temporally coherent structure \emph{and} dynamics. But, any mesoscopic structure needs to be transient, as signals need to fade in time for real-time processing (fading memory). There appears to be an optimal tradeoff between these two factors, ultimately enabling a complex system to process (parallel) information in real-time -- effectively via, e.g., amplification, decomposition, and aggregation \cite{Daniels2016,Flack2017}. The diversity of structure and dynamics in space \emph{and} time, and, secondly, their coherence in space in time,  may together be the key to optimality of self-organizing systems for reservoir computing. Indeed, we discussed how measuring dynamical susceptibility alone has shortcomings, as it is reduced for attractive driver cases, but a comprehensive analysis on multiple variables, including the VAC, cross-correlations with the driver, and statistics over time series, was insightful. One question is how to measure the morphological changes and translational motion of mesoscopic structures.
Picking up signals of correlated motion upon different coarse-graining operations might be one fruitful approach, or using morphometric measures \cite{Gring2013}.
Finally, studying the full $N$-body phase space (nonequilibrium) trajectories may be insightful, which is a topic that we are exploring.

\new{The effect of collectivity in RC was measured in the prequel to this manuscript \cite{Gaimann2025} and here in the Appendix, where the performance of a single-agent substrate (no collectivity) was compared.  Hence, a performance of around $P\approx0.55$ for single-particle reservoirs is a rough lower bound on the performance of active-matter RC in our setup. One may refer to the Discussion section of \cite{Gaimann2025}.}
\\

\new{The physical analyses demonstrated in this paper are important steps towards an understanding of higher-level functional properties of the computing system (like nonlinearity, fading memory, emergent recurrence) in terms of basic principles and mechanisms of the dynamics.}

\new{\emph{Nonlinearity and nonlinear response}: 
The evolution of microstates in the many-body system is always nonlinear in our model due to the integration of the particles' equations of motion. The active drive and the moving driving force couple their motion very strongly and induce a highly nonlinear response.
Generally, the response functions of a condensed material (correlations of key observables in time and space) are considered `non-linear' when they deviate from an equilibrium case. This is highly characteristic of soft matter \cite{Komura2012}. Determining the degree of nonlinearity of this response under controlled conditions is accessible, e.g., via rheological experiments, and, theoretically, would require an approximate theory \cite{HansenMcDonald-book,KuboToda-book}. 
In our model, the combination of internal driving (`active' forces in the speed controller), as well as external driving forces and the particle interactions are essential. 
We calculated response functions such as the velocity autocorrelation function, here, from microscopic particle states. We also measured time-cross-correlations between the driver and particles.}

\new{\emph{Memory and emergent recurrence:}
Recurrence can be generally understood as the dependence of the current state on past states; in RC, it is achieved through effective feedback from components. In physical systems, particle interactions generate feedback through their coupled equations of motion in state space. When out of equilibrium, like in the models studied here, the system's feedback and the persistence of past events (memory) become very noticeable, i.e., the system response becomes highly nonlinear (viscoelastic liquids or solids are interesting examples \cite{Christensen1982}), and effective dynamics non-Markovian: Effective equations of motion (the theoretical description) would involve an integration over past states through a so-called memory kernel or involve a memory function \cite{Schilling2022,Mori1965,Levesque1970,Dieterich1979,Cichocki1987}. Non-Markovian dynamics and memory are identifiable from non-exponential decay of the response functions of currents in systems, such as the velocity autocorrelation function (VAC) \cite{Das2004}. As in this work, oscillations around the zero line of response functions are a particular feature of complex fluid or solid-like systems, as are generally non-exponential decays and stretched plateau-like regions, both of which we observed in our model under continual driving. We also studied the intrinsic response or relaxation from an initial condition in Ref.~\cite{Gaimann2025}. }
\\

So far, research into reservoir computing with active matter simulations has focused on agent-agent and driver-agent interactions, because these are essential for forming high-dimensional mappings of the input time series through spatio-temporal patterning. For a direct understanding of how certain (observed) patterns relate to specific predictive performances, the next step is to study the information at the level of the coarse-grained observations layer, and that in the trained readout layer. This could include measuring quantities such as correlations of activations \cite{Lukosevicius2009}, their eigenvalue spread \cite{Lukosevicius2009}, or the entropy of their distribution \cite{Jaeger2005}. Other useful measures would be relating reservoir states to different input time series, to capture the short-term memory capacity \cite{Jaeger2001-short_term_memory} \new{beyond what we investigated}, the information processing capacity \cite{Dambre2012}, \new{and a different measure of the} separation property \cite{Maass2002-real-time_computing}, or kernel quality \cite{Maass2004}. By connecting these quantities to physical properties of different active matter systems presented in \refrefs{\cite{Gaimann2025}, \cite{Lymburn2021}}, and in this paper, a more fundamental understanding of the computational properties of matter (\textit{in materio} computing) could be obtained.

Our work adds to the growing diversity of reservoir computing examples using physical systems, each coming with a unique combination of inherent nonlinearity and memory capabilities, transient response times, signal transmission speed, computational complexity, and energy efficiency \cite{Tanaka2019}. Specifically, the computational power of excitable continuous medium reservoirs such as the active matter droplets that we have presented remains poorly understood \cite{Tanaka2019}. 

Methodically, instead of a $2D$ time series input signal, one could extend our method to $2D$ spatiotemporally modulated input fields covering the simulation canvas, as proposed in \refref{\cite{Jeggle2025}}. 
Furthermore, the different driven active matter states that we have presented here could be inspiring for other bio-inspired machine learning paradigms, such as (multi-agent) reinforcement learning (RL) \cite{Marl2024-book,Bloembergen2015}, or useful for physical reinforcement learning with active matter \cite{Tovey2024_chemotaxis, Tovey2025, Falk2021, Cai2025,Tovey2023}. 

We showed that emergence arising from simple, self-organizing agents is a powerful bio-inspired design principle for reservoir computing and beyond. 
Our work thus contributes to broader interdisciplinary fields of both \textit{intelligent matter} \cite{Jeggle2025, Kaspar2021-intelligent_matter}, and morphological computation \cite{Hauser2011,Hauser2012,Hauser2021}, (in essence, ``physical intelligence'' \cite{Sitti2021}), to inspire applications of autonomous collective systems \cite{Casadei2023}, soft robotics \cite{Mertan2024}, and bio-inspired unconventional computing.

\section*{Author Contributions}

Miriam Klopotek: Conceptualization, Funding acquisition, Methodology, Investigation, Project administration, Supervision, Writing – review \& editing.

Mario U. Gaimann: Conceptualization, Data curation, Formal analysis, Investigation, Methodology, Software, Visualization, Writing – original draft, Writing – review \& editing.
\begin{acknowledgments}

Funded by Deutsche Forschungsgemeinschaft (DFG, German Research Foundation) under Germany's Excellence Strategy -- EXC 2075 -- 390740016. We acknowledge the support of the Stuttgart Center for Simulation Science (SimTech), the Interchange Forum for Reflecting on Intelligent Systems (IRIS) at the University of Stuttgart, and the International Max Planck Research School for Intelligent Systems (IMPRS-IS).
We thank Christian Holm, Mathias Niepert, Anna Levina, Francesco Ginelli, Arantzazu Saratxaga, Patrick Egenlauf, and Max Weinmann for fruitful discussions.

\end{acknowledgments}
\endgroup

\bibliography{export, miriamsrefs}

\clearpage

\appendix
\begin{widetext}
\raggedbottom
\section{Additional Performance Measures}
\label{sec:additional-performance-measures}

\begin{figure}[H]
    \centering
    \subfloat[Pearson Correlation (y)]{\includegraphics[scale=1.0]{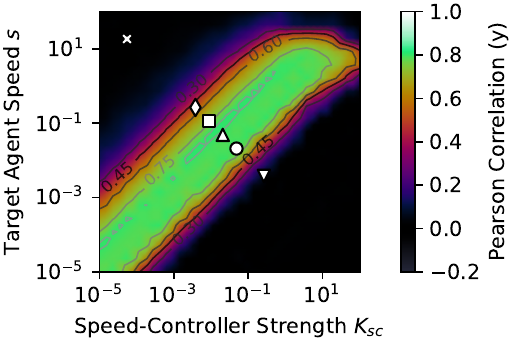}}\hfill
    \subfloat[NRMSE]{\includegraphics[scale=1.0]{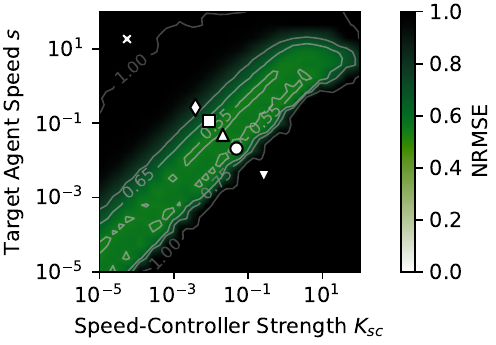}}\\
    \vspace{1em}
    \subfloat[NMSE]{\includegraphics[scale=1.0]{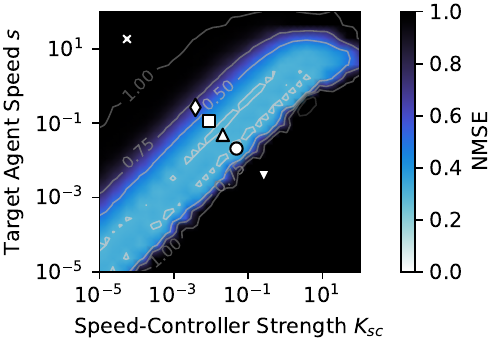}}\hfill
    \subfloat[sMAPE]{\includegraphics[scale=1.0]{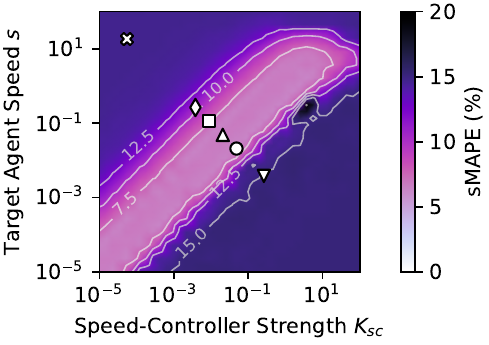}}

    \caption{\new{\textbf{Different performance measures for the speed-controller parameter scan with an inversely attractive driver} shown in \figref[a]{\ref{fig:driver_attraction_speed_controller}} in the main text: (a) Pearson correlation coefficient for the second dimension ($y$), (b) Normalized Root Mean Squared Error by standard deviation (NRMSE), (c) Normalized Mean Squared Error (NMSE) by variance, and (d) Symmetric Mean Absolute Percentage Error (sMAPE).}}
\label{fig:driver_attraction_inverse_speed_controller-more-metrics}
\end{figure}

\begin{figure}[H]
    \centering
    \subfloat[Pearson Correlation (y)]{\includegraphics[scale=1.0]{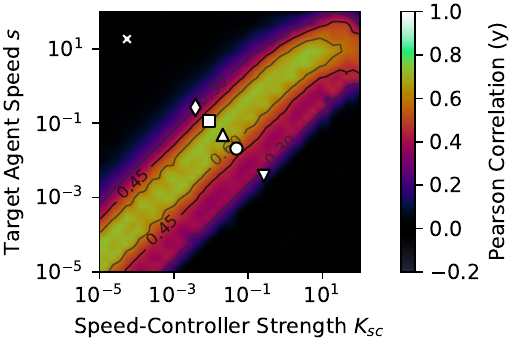}}\hfill
    \subfloat[NRMSE]{\includegraphics[scale=1.0]{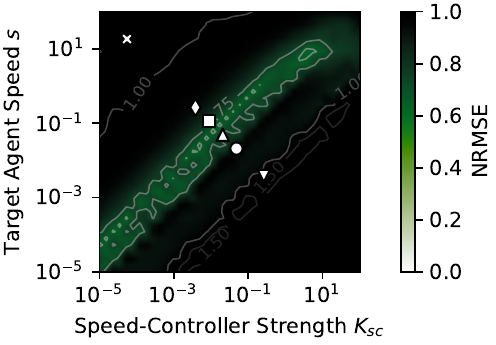}}\\
    \vspace{1em}
    \subfloat[NMSE]{\includegraphics[scale=1.0]{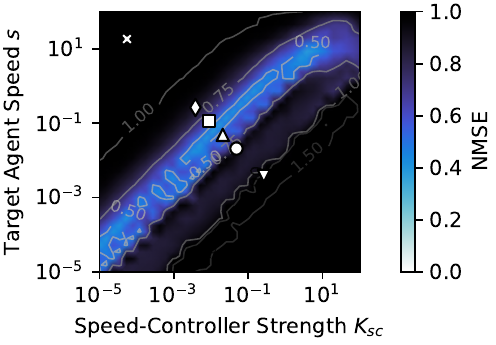}}\hfill
    \subfloat[sMAPE]{\includegraphics[scale=1.0]{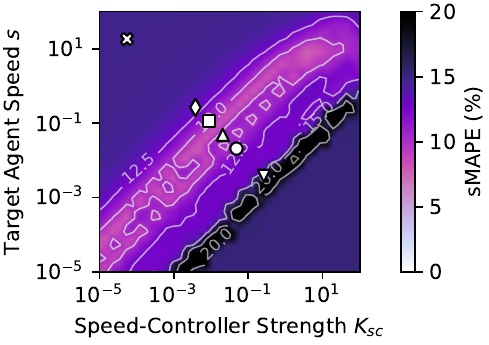}}

    \caption{\new{\textbf{Different performance measures for the speed-controller parameter scan with a linearly attractive driver} shown in \figref[b]{\ref{fig:driver_attraction_speed_controller}}.}}
    \label{fig:driver_attraction_linear_speed_controller-more-metrics}
\end{figure}

\begin{figure}[H]
    \centering
    \subfloat[Pearson Correlation (y)]{\includegraphics[scale=1.0]{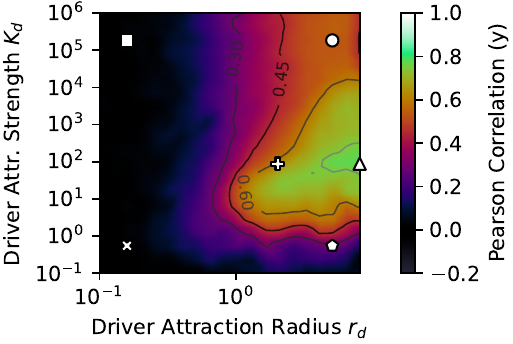}}\hfill
    \subfloat[NRMSE]{\includegraphics[scale=1.0]{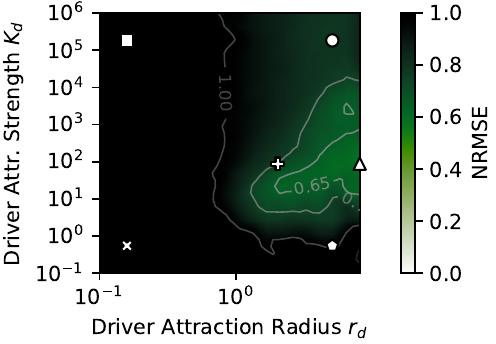}}\\
    \vspace{1em}
    \subfloat[NMSE]{\includegraphics[scale=1.0]{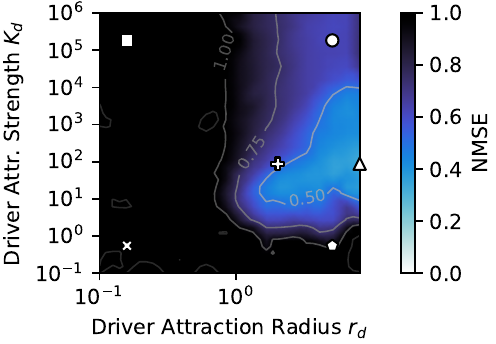}}\hfill
    \subfloat[sMAPE]{\includegraphics[scale=1.0]{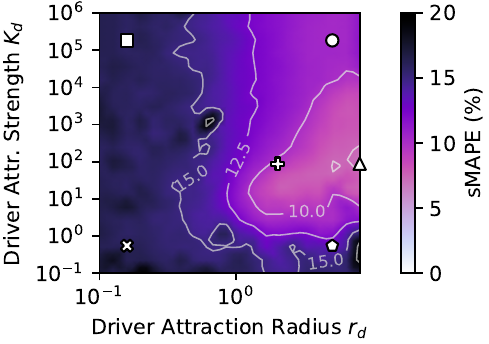}}

    \caption{\new{\textbf{Different performance measures for the driver attraction parameter scan, for an inversely attractive driver} shown in \figref[a]{\ref{fig:driver_attraction}}.}}
    \label{fig:driver_attraction-more-metrics}
\end{figure}

\begin{figure}[H]
    \centering
    \subfloat[Pearson Correlation (y)]{\includegraphics[scale=1.0]{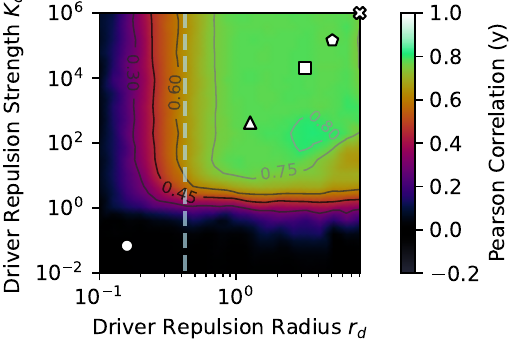}}\hfill
    \subfloat[NRMSE]{\includegraphics[scale=1.0]{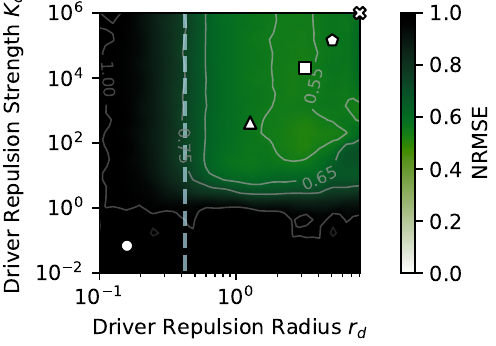}}\\
    \vspace{1em}
    \subfloat[NMSE]{\includegraphics[scale=1.0]{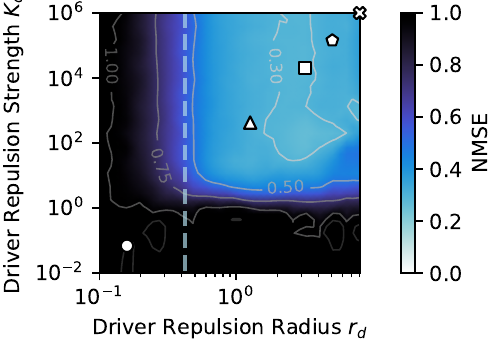}}\hfill
    \subfloat[sMAPE]{\includegraphics[scale=1.0]{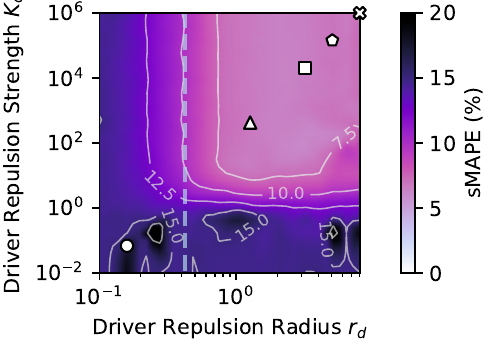}}

    \caption{\new{\textbf{Different performance measures for the driver attraction parameter scan, for an inversely repulsive driver} shown in \figref[a]{\ref{fig:nearcritdamped_driver_repulsion_scan_heatmap}}.}}
    \label{fig:driver_repulsion-more-metrics}
\end{figure}

\begin{figure}[H]
    \centering
    \subfloat[Pearson Correlation (y)]{\includegraphics[scale=1.0]{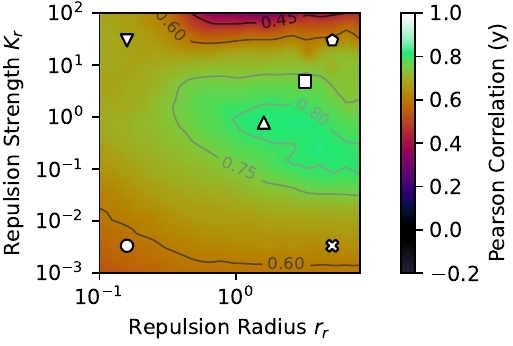}}\hfill
    \subfloat[NRMSE]{\includegraphics[scale=1.0]{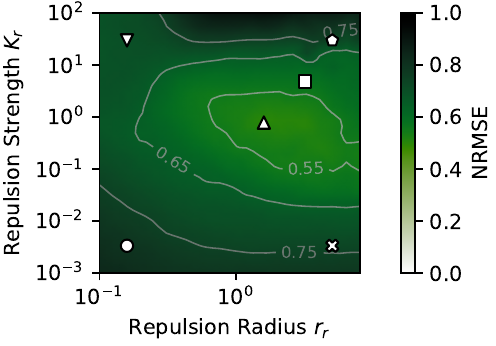}}\\
    \vspace{1em}
    \subfloat[NMSE]{\includegraphics[scale=1.0]{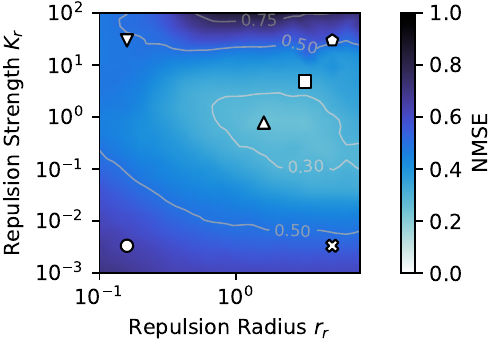}}\hfill
    \subfloat[sMAPE]{\includegraphics[scale=1.0]{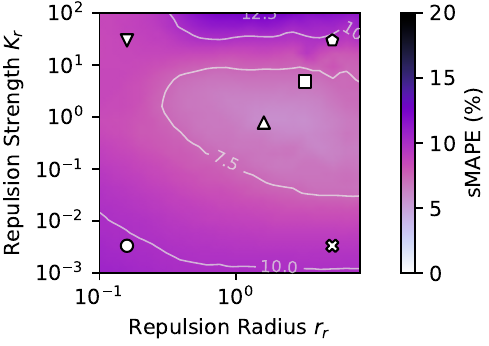}}

    \caption{\new{\textbf{Different performance measures for the agent-agent repulsion parameter scan, for a repulsive driver} shown in \figref[a]{\ref{fig:repulsion}}.}}
    \label{fig:repulsion-more-metrics-lin-driver}
\end{figure}

\begin{figure}[H]
    \centering
     \subfloat[Pearson Correlation (y)]{\includegraphics[scale=1.0]{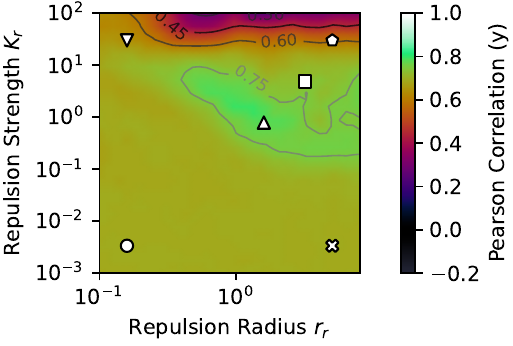}}\hfill
    \subfloat[NRMSE]{\includegraphics[scale=1.0]{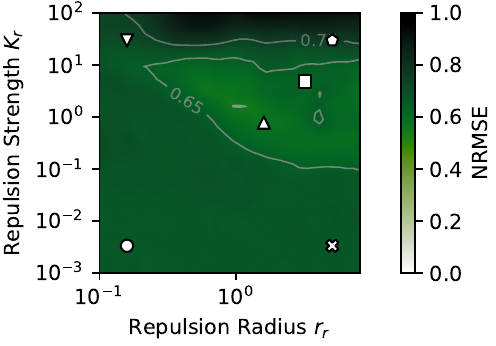}}\\
    \vspace{1em}
    \subfloat[NMSE]{\includegraphics[scale=1.0]{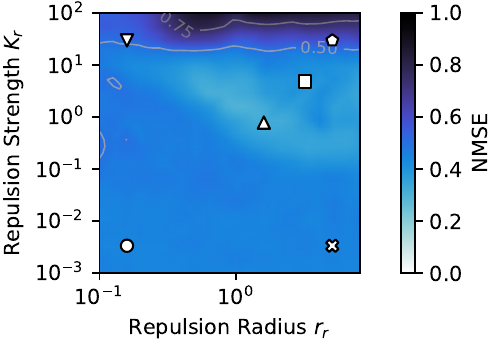}}\hfill
    \subfloat[sMAPE]{\includegraphics[scale=1.0]{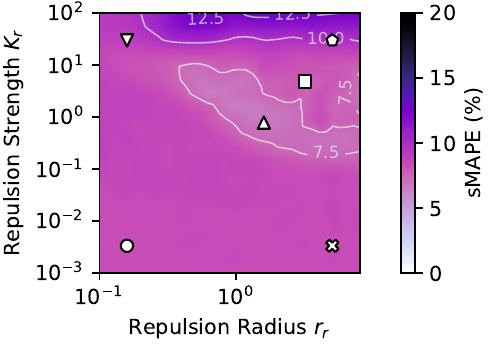}}

    \caption{\new{\textbf{Different performance measures for the agent-agent repulsion parameter scan, for an inversely attractive driver} shown in \figref[b]{\ref{fig:repulsion}}.}}
    \label{fig:repulsion-more-metrics-inv-driver}
\end{figure}

\begin{figure}[H]
    \centering
    \subfloat[Pearson Correlation (y)]{\includegraphics[scale=1.0]{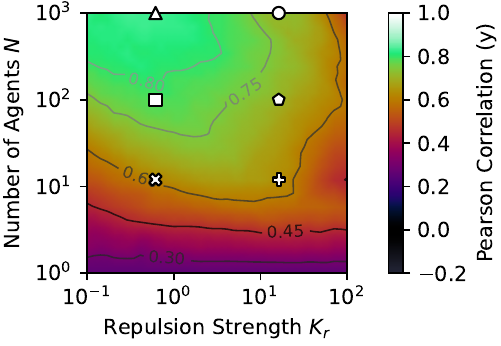}}\hfill
    \subfloat[NRMSE]{\includegraphics[scale=1.0]{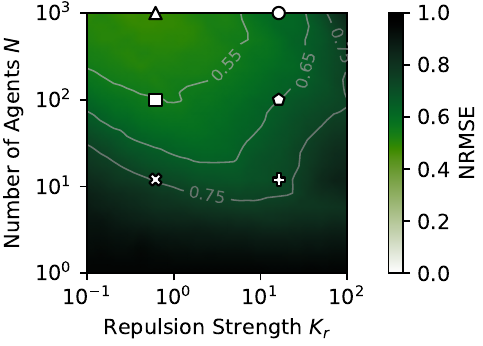}}\\
    \vspace{1em}
    \subfloat[NMSE]{\includegraphics[scale=1.0]{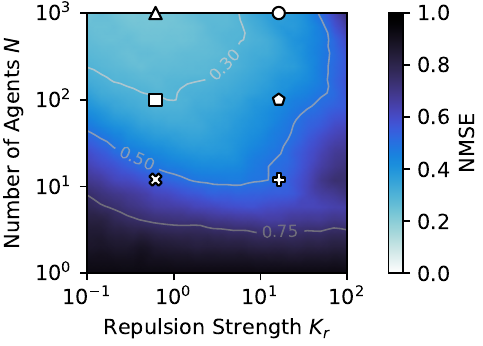}}\hfill
    \subfloat[sMAPE]{\includegraphics[scale=1.0]{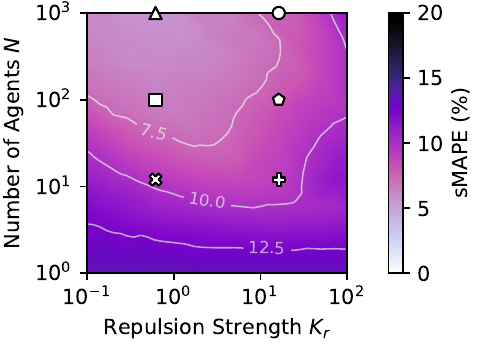}}

    \caption{\new{\textbf{Different performance measures for the agent-agent repulsion versus number of agents parameter scan, with a repulsive driver} shown in \figref[a]{\ref{fig:n_agents_repulsion_strength}}.}}
    \label{fig:n_agents_repulsion_strength_repulsive-more-metrics}
\end{figure}

\begin{figure}[H]
    \centering
    \subfloat[Pearson Correlation (y)]{\includegraphics[scale=1.0]{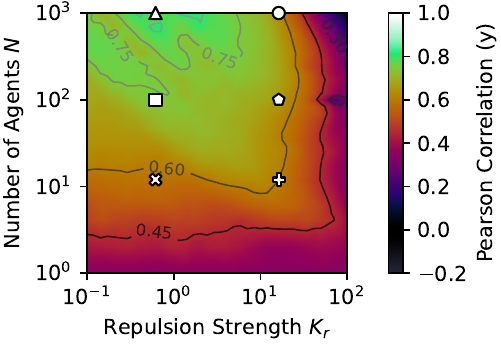}}\hfill
    \subfloat[NRMSE]{\includegraphics[scale=1.0]{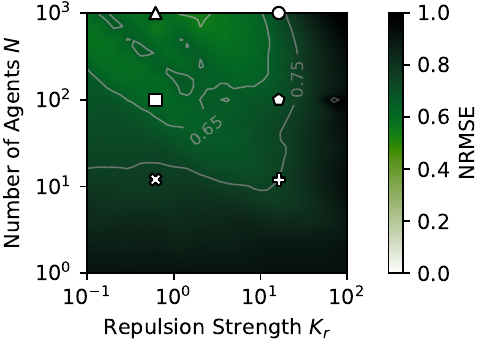}}\\
    \vspace{1em}
    \subfloat[NMSE]{\includegraphics[scale=1.0]{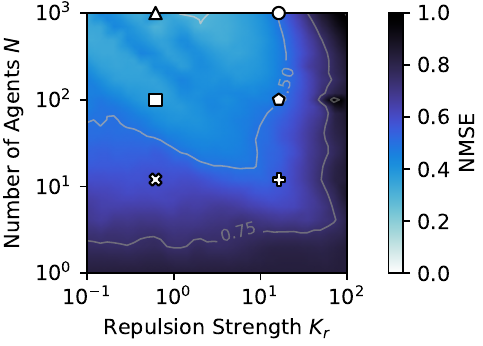}}\hfill
    \subfloat[sMAPE]{\includegraphics[scale=1.0]{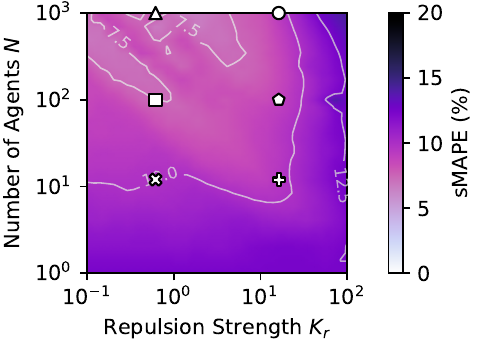}}

    \caption{\new{\textbf{Different performance measures for the agent-agent repulsion versus number of agents parameter scan, with an inversely attractive driver} shown in \figref[b]{\ref{fig:n_agents_repulsion_strength}}.}}
    \label{fig:n_agents_repulsion_strength_attractive-more-metrics}
\end{figure}

\section{\label{ax:supplementary_statistics}Supplementary Statistics}

\new{In this section, we report statistics of predictive performances obtained for active matter reservoir computing with 100 different random seeds. Each seed yields different initial particle positions (placed uniformly at random in the simulation box), different initial particle velocities (determined as uniformly at random direction with a fixed magnitude (speed) of 1.0), as well as different initial driver positions (sampled uniformly at random within an empirical bounding box that fully contains the Lorenz-63 attractor ($x_0,y_0 \in [-20,20], z_0\in [0,50]$) and then rescaled to fit a box with length $l_{\text{box}}^{\text{driver}}$ centered inside the simulation box as described in \sectref{\ref{sec:mm-simulation-details}}).
In \figref{\ref{fig:driver_attraction_inv_speed_controller_ensemble_details}} we show the predictive performance statistics for 100 random seeds (different driver and agent initial positions for train and test runs) for the speed-controller parameter scan with an inversely attractive driver shown in \figref[a]{\ref{fig:driver_attraction_speed_controller}}. The box plot in sub-figure (a) confirms that parameter combination ID 209 (circle symbol, near-critically damped regime) delivers the highest mean predictive performance. We note that there is a significant number of outliers for all relevant parameter combinations with a mean performance of $P > 0.6$. These outliers are more severe for the parameter combinations with the best mean predictive performances (square, pyramid, circle symbols), with some performance drops reaching down to almost zero performance. In the transition regime (diamond symbol), these drops are not that severe (down to about $P \approx 0.40$), albeit significant.
To understand whether these outliers appear for the same driver and agents' initial positions, we plot the predictive performances for each of the 100 seeds per parameter combination in sub-figure (b). This plot reveals that there seem to be specific seeds that yield a deterioration of performance independent of the parameter combination at hand.
When comparing the predicted train and test time series for the strongest outlier in the near-critically damped regime (pyramid symbol) in sub-figure (c), it becomes clear that the performance on train data is much higher ($P \approx 0.90$) than on test data ($P \approx 0.058$). This suggests that the readout layer could have been overfit on the training data.
\figref{\ref{fig:driver_attraction_inv_speed_controller_ensemble_details_randagentsonly}} shows the statistics for 100 seeds where only the agent initial positions were randomized and the driver initial positions were kept constant. The inter-quartile ranges are much smaller here, and severe outliers are absent, indicating that the cause of them is indeed the choice of different driver initial conditions. 
Overall, this analysis suggests that the initial conditions used throughout the main text are representative of the ensemble-averaged findings.}\\

\new{In the following, we show means and standard deviations of predictive performance across 100 random seeds, measured using the default method (Pearson correlation coefficient of the $x$ dimension determined as described in \sectref{\ref{sec:rc_setup}}) and using the NRMSE, for all parameter scans in the main text.}

\begin{figure}[H]
    \centering
    \subfloat[Statistics]{\includegraphics[scale=1.0]{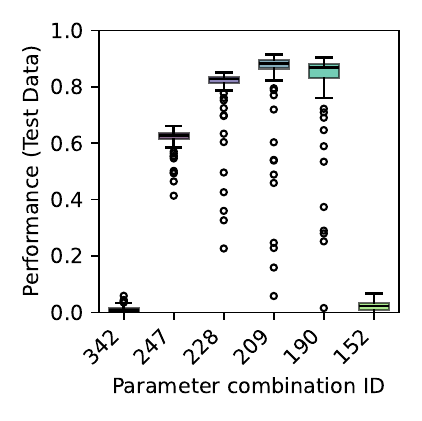}}\hfill
    \subfloat[Performance per Seed]{\includegraphics[scale=1.0]{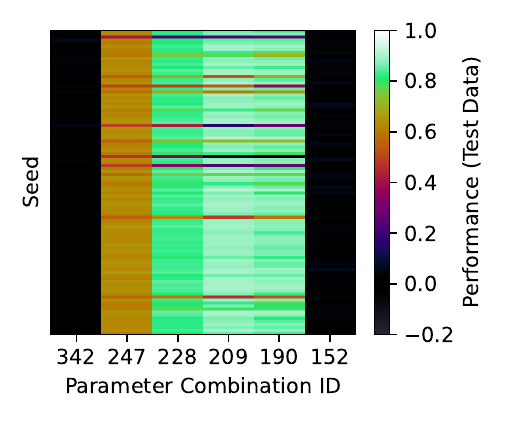}}\\
    \vspace{1em}
    \subfloat[Strongest Outlier Time Series Comparison]{\includegraphics[scale=1.0]{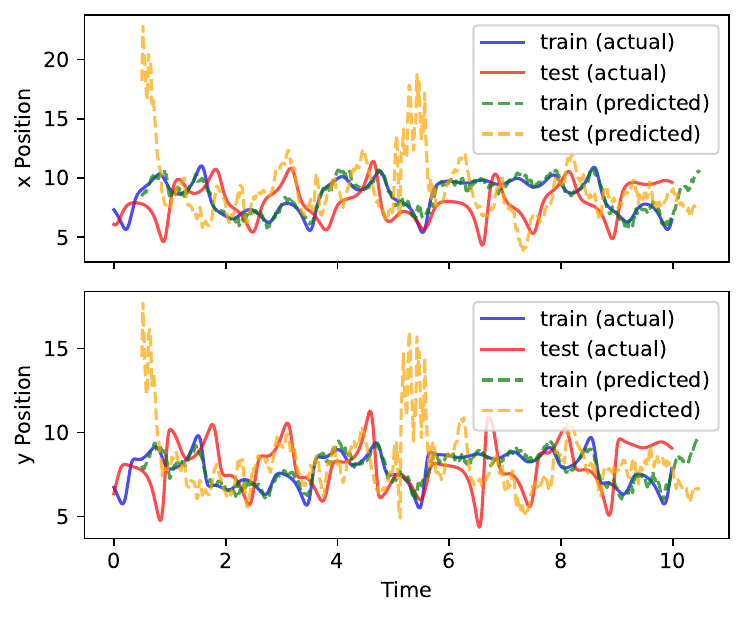}}\\
        
    \caption{\new{\textbf{Detailed analysis for the reservoir computing performance across 100 random seeds (random initial driver and agent positions)}, corresponding to the speed-controller parameter scan with an inversely attractive driver shown in \figref[a]{\ref{fig:driver_attraction_speed_controller}}. (a) Box plot indicating the median (black line), the 25th to 75th percentile (colored boxes; colors correspond to parameter combination symbol colors), the 1.5 IQR (inter-quartile range; whiskers), and outliers (black circles) of predictive performances. (b) Heatmap showing for each random seed the predictive performance for each parameter combination. (c) Time series for the strongest outlier for parameter combination 209 (performances: $P^{\text{train}} \approx 0.90$; $P^{\text{test}} \approx 0.06$). Parameter combination IDs are displayed in order of increased damping (increasing speed-controller strength, decreasing target agent speed) corresponding to the symbols shown in the main text (342: cross, 247: diamond, 228: square, 209: pyramid, 190: circle, 152: nabla).}}
    \label{fig:driver_attraction_inv_speed_controller_ensemble_details}
\end{figure}

\begin{figure}[H]
    \centering
    \subfloat[Statistics]{\includegraphics[scale=1.0]{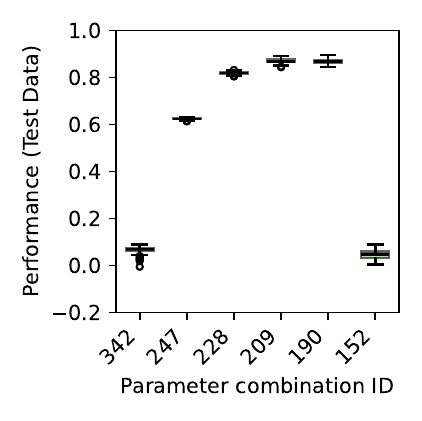}}\hfill
    \subfloat[Performance per Seed]{\includegraphics[scale=1.0]{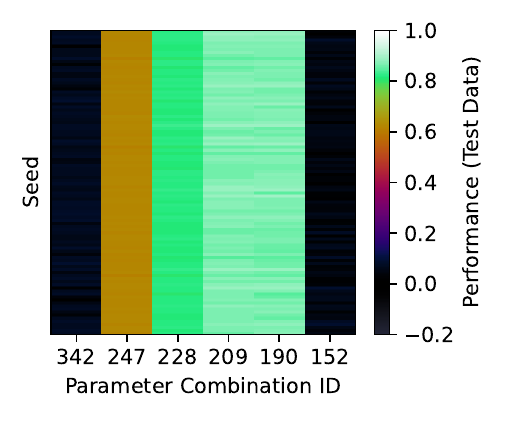}}
    
    \caption{\new{\textbf{Detailed analysis for the reservoir computing performance across 100 random seeds (only random initial agent positions)}, corresponding to the speed-controller parameter scan with an inversely attractive driver shown in \figref[a]{\ref{fig:driver_attraction_speed_controller}}. Driver initial states are $(x,y,y) = (0.0, 1.0, 1.05)$ (training) and $(0.0, 1.1, -1.5)$ (testing). (a) Box plot indicating the median (black line), the 25th to 75th percentile (colored boxes; colors correspond to parameter combination symbol colors), the 1.5 IQR (inter-quartile range; whiskers), and outliers (black circles) of predictive performances. (b) Heatmap showing for each random seed the predictive performance for each parameter combination. Parameter combination IDs are displayed in order of increased damping (increasing speed-controller strength, decreasing target agent speed) corresponding to the symbols shown in the main text (342: cross, 247: diamond, 228: square, 209: pyramid, 190: circle, 152: nabla).}}
    \label{fig:driver_attraction_inv_speed_controller_ensemble_details_randagentsonly}
\end{figure}

\begin{figure}[H]
    \centering
    \subfloat[Mean Pearson Correlation (x)]{\includegraphics[scale=1.0]{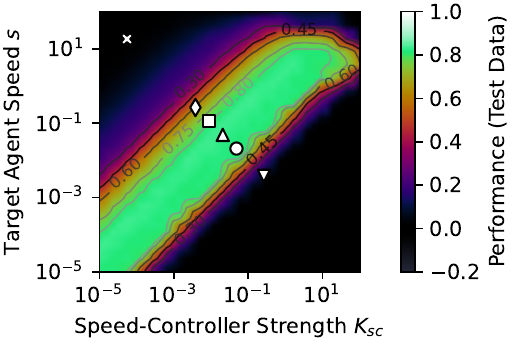}}\hfill
    \subfloat[Std.dev.\ Pearson Correlation (x)]{\includegraphics[scale=1.0]{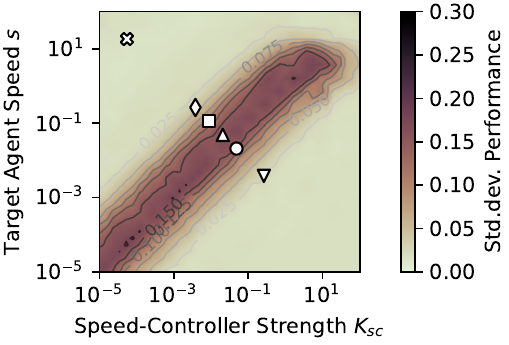}}\\
    \vspace{1em}
    \subfloat[Mean NRMSE]{\includegraphics[scale=1.0]{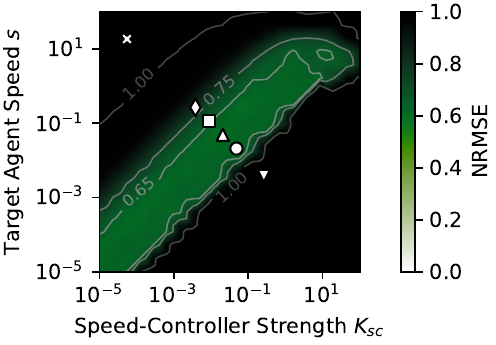}}\hfill
    \subfloat[Std.dev.\ NRMSE]{\includegraphics[scale=1.0]{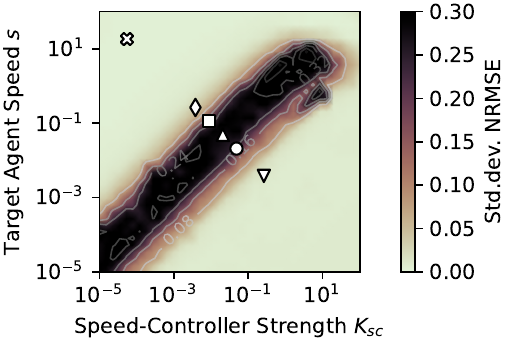}}

    \caption{\new{\textbf{Mean and standard deviation of performance metrics for the speed-controller parameter scan with an inversely attractive driver} shown in \figref[a]{\ref{fig:driver_attraction_speed_controller}}, ensemble averaged over 100 different driver and agent initial conditions.}}
    \label{fig:driver_attraction_inverse_speed_controller-mean-std}
\end{figure}

\begin{figure}[H]
    \centering
    \subfloat[Mean Pearson Correlation (x)]{\includegraphics[scale=1.0]{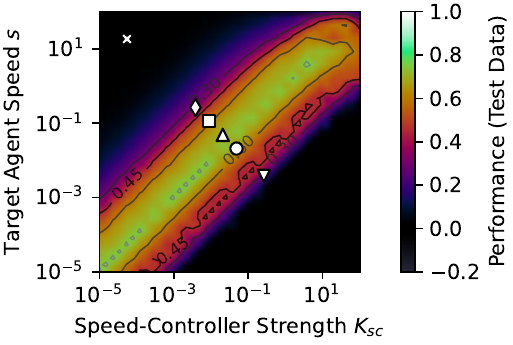}}\hfill
    \subfloat[Std.dev.\ Pearson Correlation (x)]{\includegraphics[scale=1.0]{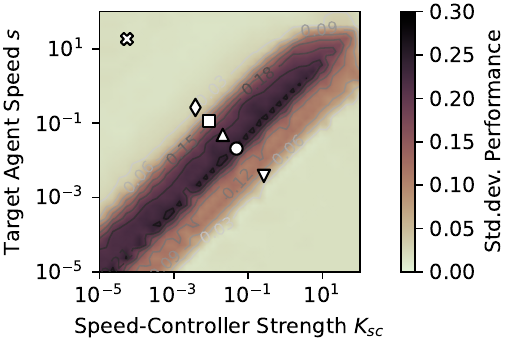}}\\
    \vspace{1em}
    \subfloat[Mean NRMSE]{\includegraphics[scale=1.0]{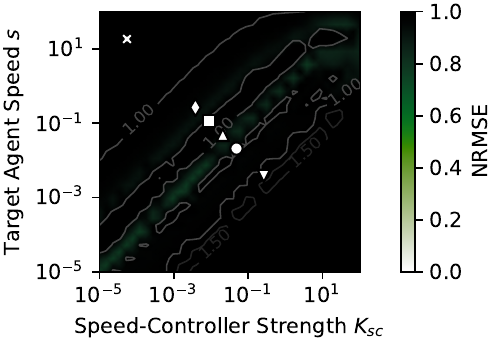}}\hfill
    \subfloat[Std.dev.\ NRMSE]{\includegraphics[scale=1.0]{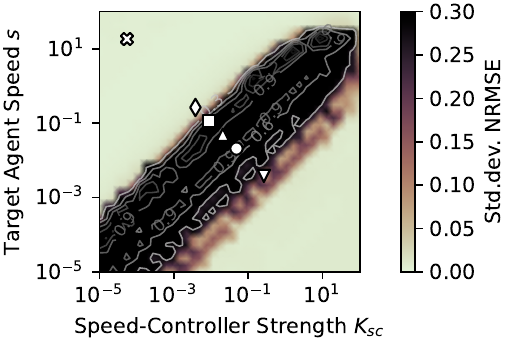}}

    \caption{\new{\textbf{Mean and standard deviation of performance metrics for the speed-controller parameter scan with a linearly attractive driver} shown in \figref[b]{\ref{fig:driver_attraction_speed_controller}}, ensemble averaged over 100 different driver and agent initial conditions.}}
    \label{fig:driver_attraction_linear_speed_controller-mean-std}
\end{figure}

\begin{figure}[H]
    \centering
    \subfloat[Mean Pearson Correlation (x)]{\includegraphics[scale=1.0]{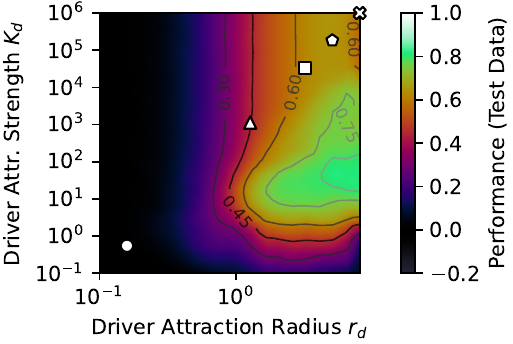}}\hfill
    \subfloat[Std.dev.\ Pearson Correlation (x)]{\includegraphics[scale=1.0]{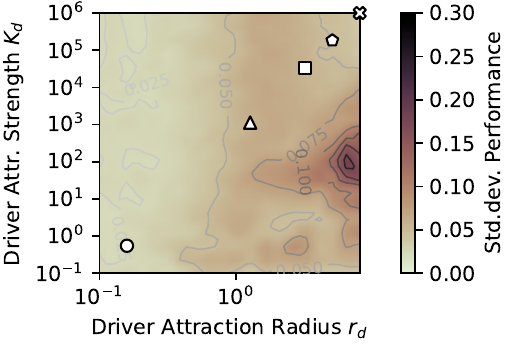}}\\
    \vspace{1em}
    \subfloat[Mean NRMSE]{\includegraphics[scale=1.0]{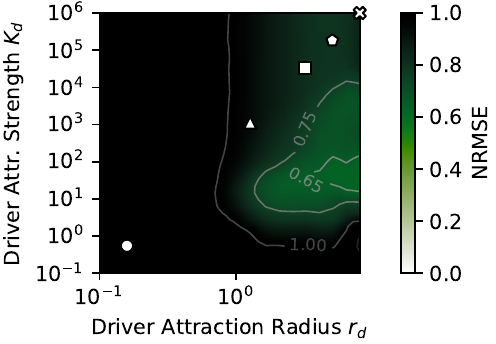}}\hfill
    \subfloat[Std.dev.\ NRMSE]{\includegraphics[scale=1.0]{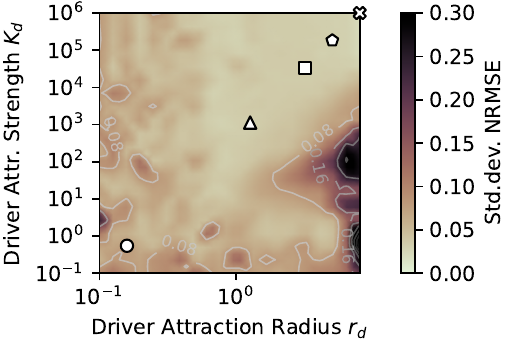}}

    \caption{\new{\textbf{Mean and standard deviation of performance metrics for the driver attraction parameter scan, for an inversely attractive driver} shown in \figref[a]{\ref{fig:driver_attraction}}, ensemble averaged over 100 different driver and agent initial conditions.}}
    \label{fig:driver_attraction-mean-std}
\end{figure}

\begin{figure}[H]
    \centering
    \subfloat[Mean Pearson Correlation (x)]{\includegraphics[scale=1.0]{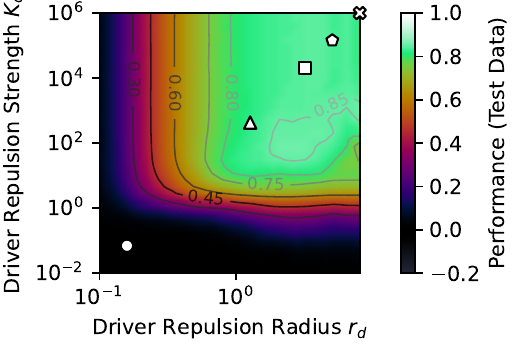}}\hfill
    \subfloat[Std.dev.\ Pearson Correlation (x)]{\includegraphics[scale=1.0]{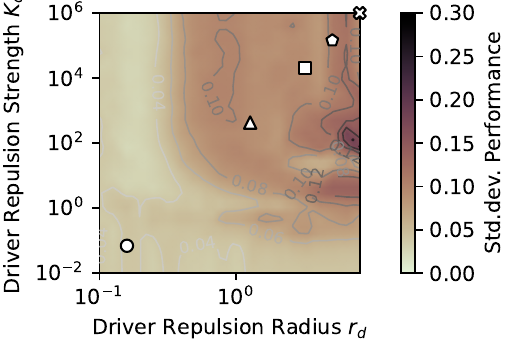}}\\
    \vspace{1em}
    \subfloat[Mean NRMSE]{\includegraphics[scale=1.0]{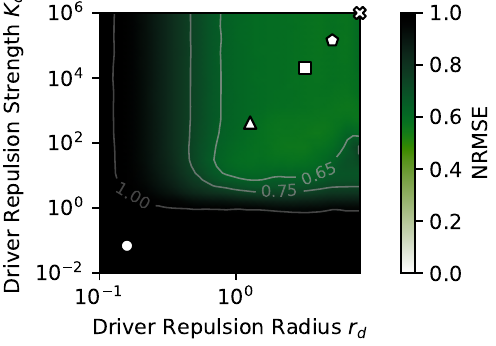}}\hfill
    \subfloat[Std.dev.\ NRMSE]{\includegraphics[scale=1.0]{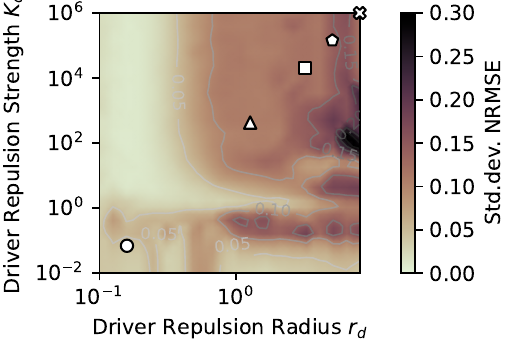}}

    \caption{\new{\textbf{Mean and standard deviation of performance metrics for the driver attraction parameter scan, for an inversely repulsive driver} shown in \figref[a]{\ref{fig:nearcritdamped_driver_repulsion_scan_heatmap}}, ensemble averaged over 100 different driver and agent initial conditions.}}
    \label{fig:driver_repulsion-mean-std}
\end{figure}

\begin{figure}[H]
    \centering
    \subfloat[Mean Pearson Correlation (x)]{\includegraphics[scale=1.0]{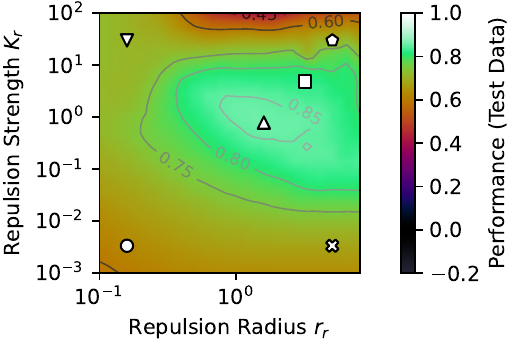}}\hfill
    \subfloat[Std.dev.\ Pearson Correlation (x)]{\includegraphics[scale=1.0]{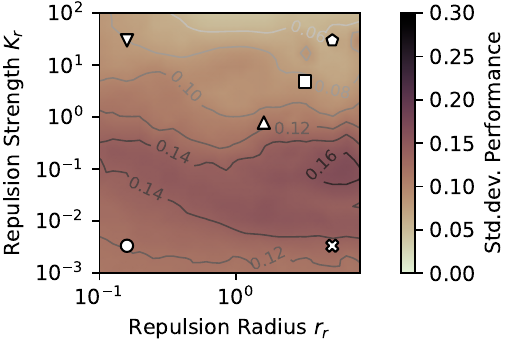}}\\
    \vspace{1em}
    \subfloat[Mean NRMSE]{\includegraphics[scale=1.0]{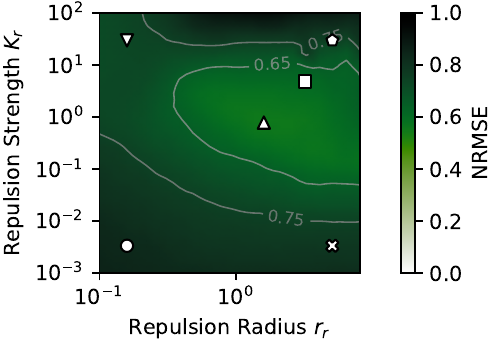}}\hfill
    \subfloat[Std.dev.\ NRMSE]{\includegraphics[scale=1.0]{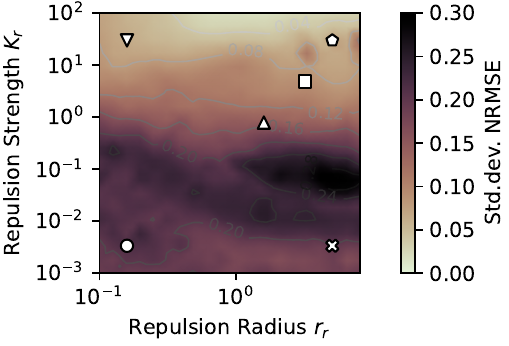}}

    \caption{\new{\textbf{Mean and standard deviation of performance metrics for the agent-agent repulsion parameter scan, for a repulsive driver} shown in \figref[a]{\ref{fig:repulsion}}, ensemble averaged over 100 different driver and agent initial conditions.}}
    \label{fig:repulsion-mean-std}
\end{figure}

\begin{figure}[H]
    \centering
    \subfloat[Mean Pearson Correlation (x)]{\includegraphics[scale=1.0]{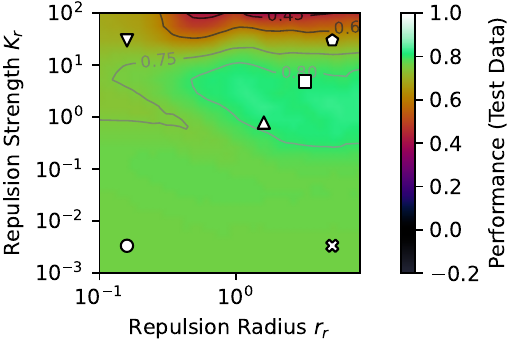}}\hfill
    \subfloat[Std.dev.\ Pearson Correlation (x)]{\includegraphics[scale=1.0]{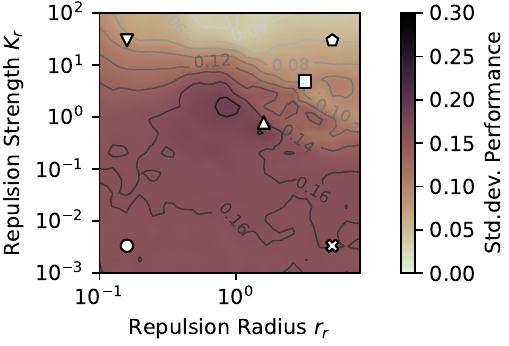}}\\
    \vspace{1em}
    \subfloat[Mean NRMSE]{\includegraphics[scale=1.0]{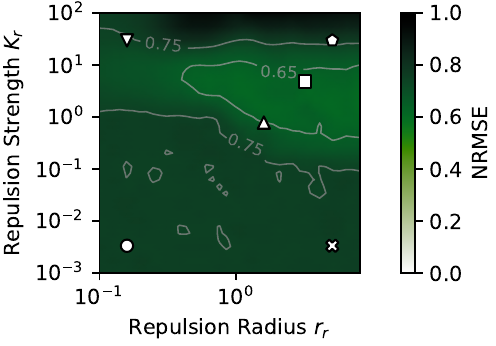}}\hfill
    \subfloat[Std.dev.\ NRMSE]{\includegraphics[scale=1.0]{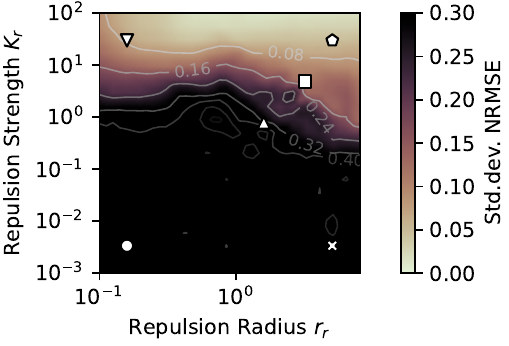}}

    \caption{\new{\textbf{Mean and standard deviation of performance metrics for the agent-agent repulsion parameter scan, for an inversely attractive driver} shown in \figref[b]{\ref{fig:repulsion}}, ensemble averaged over 100 different driver and agent initial conditions.}}
    \label{fig:repulsion-attractive-mean-std}
\end{figure}

\begin{figure}[H]
    \centering
    \subfloat[Mean Pearson Correlation (x)]{\includegraphics[scale=1.0]{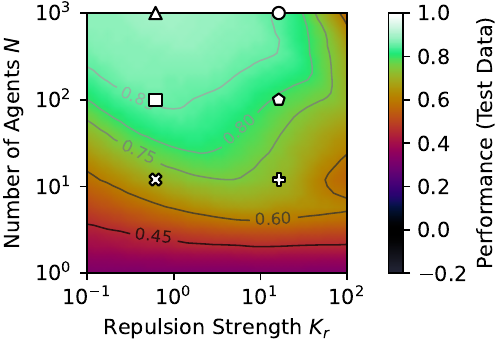}}\hfill
    \subfloat[Std.dev.\ Pearson Correlation (x)]{\includegraphics[scale=1.0]{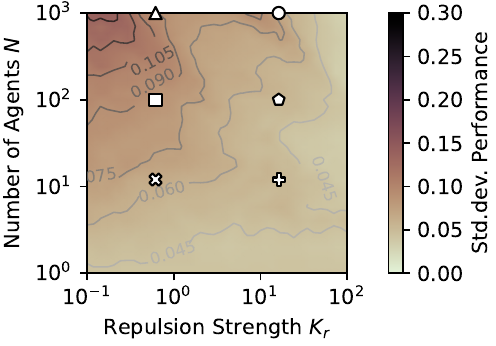}}\\
    \vspace{1em}
    \subfloat[Mean NRMSE]{\includegraphics[scale=1.0]{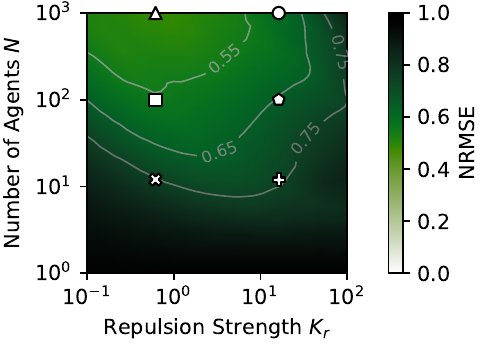}}\hfill
    \subfloat[Std.dev.\ NRMSE]{\includegraphics[scale=1.0]{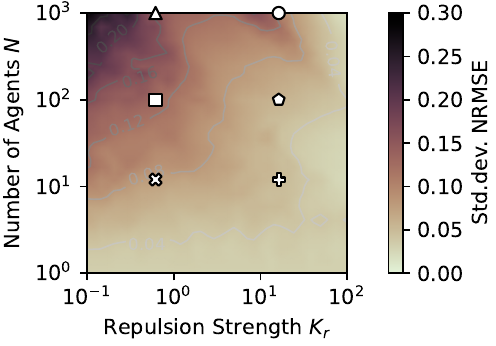}}

    \caption{\new{\textbf{Mean and standard deviation of performance metrics for the agent-agent repulsion versus number of agents parameter scan, with a repulsive driver} shown in \figref[a]{\ref{fig:n_agents_repulsion_strength}}, ensemble averaged over 100 different driver and agent initial conditions.}}
    \label{fig:n_agents_repulsion_strength_repulsive-mean-std}
\end{figure}

\begin{figure}[H]
    \centering
    \subfloat[Mean Pearson Correlation (x)]{\includegraphics[scale=1.0]{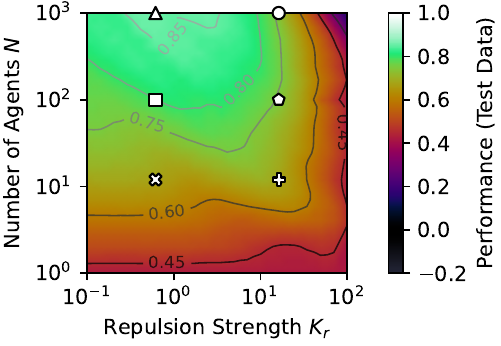}}\hfill
    \subfloat[Std.dev.\ Pearson Correlation (x)]{\includegraphics[scale=1.0]{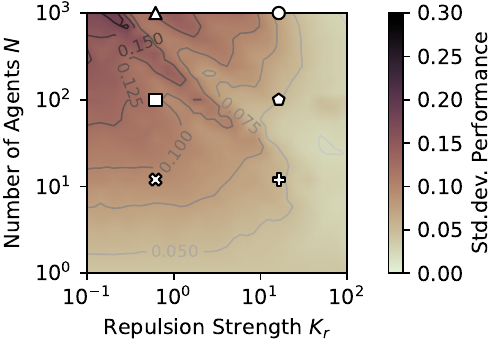}}\\
    \vspace{1em}
    \subfloat[Mean NRMSE]{\includegraphics[scale=1.0]{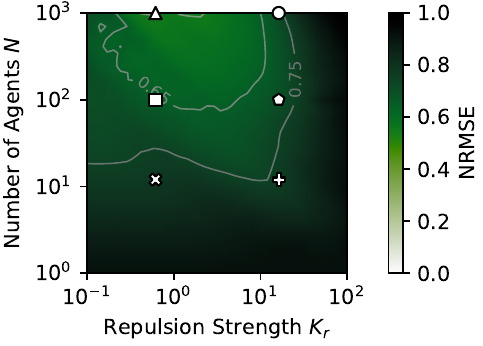}}\hfill
    \subfloat[Std.dev.\ NRMSE]{\includegraphics[scale=1.0]{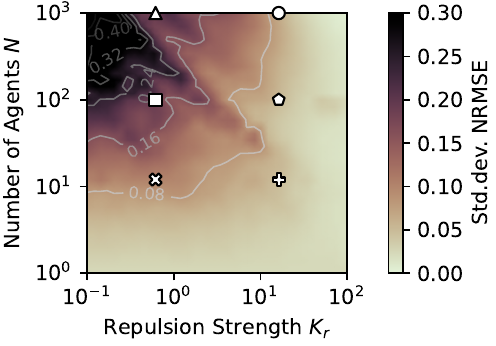}}

    \caption{\new{\textbf{Mean and standard deviation of performance metrics for the agent-agent repulsion versus number of agents parameter scan, with an inversely attractive driver} shown in \figref[b]{\ref{fig:n_agents_repulsion_strength}}, ensemble averaged over 100 different driver and agent initial conditions.}}
    \label{fig:n_agents_repulsion_strength_attractive-mean-std}
\end{figure}

\newpage
\section{Short-term memory capacity}
\label{sec:short-term-mc}

 \new{We quantify the memory capacity, a central performance metric for reservoir computers, following the definition of Jaeger (2001) \cite{Jaeger2001-short_term_memory}.  
    The \textit{k-delay Short Term Memory Capacity} (STMC) is given as
    \begin{equation}
        MC_{k,d} = \frac{\operatorname{cov}^2\!\bigl(y_{\text{target,d}}(t-k\Delta t),\, y^{k}_{\text{pred,d}}(t)\bigr)}
    {\sigma^2\!\bigl(y_{\text{target,d}}(t-k\Delta t)\bigr)\,\sigma^2\!\bigl(y^k_{\text{pred,d}}(t)\bigr)}\,,
    \end{equation}
    where $y^{k}_{\text{pred,d}}(t)$ denotes the time series predicted by the trained readout for delay $k$ in dimension $d$, $y_{\text{target,d}}(t)$ is the corresponding target time series, and $\Delta t$ is the integration time step. Perfect reconstruction of the signal at delay $k$ yields $MC_{k,d} = 1.0$, while the absence of usable memory leads to $MC_{k,d} = 0.0$.  
    The total STMC is obtained by summing over all delays and both spatial dimensions,
    \begin{equation}
        MC = \sum_{d=1}^2 \sum_{k=1}^{k_{\max}} MC_{k,d},
    \end{equation}
    with $k_{\max} = 100$, implying a maximal possible memory capacity of 200.0 in our setting. This quantity measures how much information about past inputs the reservoir can retain.\\
    In standard reservoir computing, the memory capacity is typically evaluated using a time series sampled uniformly at random. This approach may lead to specific issues in our setup, however: A trajectory sampled uniformly over the entire simulation box can place the driver in corners, where interactions with the swarm are rare, if the driver-agent interaction is repulsive. To avoid this, we draw the driver trajectory from a disk of radius $R$ according to
    \begin{equation}
        U \sim \mathrm{Uniform}(0,1), \quad r = R \sqrt{U}, \quad \theta \sim \mathrm{Uniform}(0, 2\pi),
    \end{equation}
    }\new{convert to Cartesian coordinates via
    \begin{equation}
        x_d = r \cos(\theta), \quad y_d = r \sin(\theta),
    \end{equation}
    and then shift the $(x,y)$ coordinates such that the origin $(0.0,0.0)$ is mapped to the center of the simulation box of side length $\ell_{\text{box}} = 16.0$, i.e., $(8.0, 8.0)$ (see \refref{\cite{Gaimann2025}}, \figref{45} for a visualization). We set $R = 4.0$, which closely matches the maximal radial extent $R_{\max} \approx 3.94$ of an undriven swarm in the near-critically damped regime.  
    A second complication arises when the driver position is resampled at every integration step, since the swarm then has too little time to respond and build up memory. Therefore, we vary the update interval of the driver position and consider changes every $\Delta s \in \{1, 2, 5, 10\}$ integration steps, with example trajectories shown in \refref{\cite{Gaimann2025}}, \figref{45}. In addition, we also compute the STMC using our continuous default Lorenz-63 driving protocol.}\\

    \new{In both \figref[a]{\ref{fig:speed-controller-memcap-total-inv-attr}} (inverse attraction) and \figref[a]{\ref{fig:speed-controller-memcap-total-lin-attr}} (linear attraction) we observe that for the driver change interval with the highest frequency ($\Delta s = 1$) we observe low maximum STMCs of $< 5.0$, most likely because the active matter system cannot build up memory due to the low residence times (high jump frequency) of the driver. For larger residence times $\Delta s > 1$, the active matter system has more time to build memory, and the STMC increases towards up to around $36$ for inverse driver attraction and around $20$ for linear driver attraction, respectively (sub-figures (d)). The highest STMCs for the highlighted parameter combinations are found here around the square symbol (slightly above the near-critically damped regime towards underdamping) for the inverse driver attraction case, and around the pyramid symbol and (interestingly) between the square and the nabla symbol. The high STMC in the latter regime could stem from the buildup of a heavy tail in the $k$-delay MC in \figref[d]{\ref{fig:speed-controller-attr-lin-memcap-k-delay}} for stronger overdamping (but not too strong to cause arrestedness) (circle symbol to nabla symbol). The local minimum for the pyramid symbol could arise from a fast erasure of memory due to a fast relaxation; the small increase for the diamond symbol could stem from an overshooting of the agents over the driver position.
    For the modified STMC measurement using the continuous Lorenz-63 trajectory as input we obtain a similar qualitative picture in \figref[a,b]{\ref{fig:speed-controller-memory-capacity-L63}} as for $\Delta s = 10$, but with much higher absolute values of up to $120$ for inverse driver attraction and up to $105$ for linear driver attraction. This is likely due to the high correlation of subsequent driver positions in the continuous trajectory, compared to uncorrelated random jumps investigated before.
    Comparing the STMC differences for all scenarios in \figref{\ref{fig:speed-controller-memory-capacity-lin-vs-inv}} we find that for a broad parameter range (cross symbol to pyramid symbol) the STMC is higher in the inverse driver attraction case compared to the linear driver attraction case. The opposite is true for the overdamped region between the circle and nabla symbols, which could be due to the fact that the active matter system in the inverse attraction case becomes arrested using this integration time step of $\Delta t = 0.02$ (see also \figref[c]{\ref{fig:driver_attraction_speed_controller}}, nabla symbol).
    To conclude, given enough time for the active matter system to adapt in the random jump case, both discrete and continuous analyses show the same distinct regions in the speed-controller parameter scan that indicate maximum STMC. In particular, for the linear driver attraction, these regions do not coincide with high predictive performance; in fact, there is a local minimum (pyramid symbol, \figref[c]{\ref{fig:speed-controller-memory-capacity-L63}}). This suggests that the active matter system in this configuration has other favourable properties that deliver high predictive performance.
    }\\


\begin{figure*}[htbp]
    \centering
    \subfloat[$\Delta s = 1$]{\includegraphics[scale=1.0]{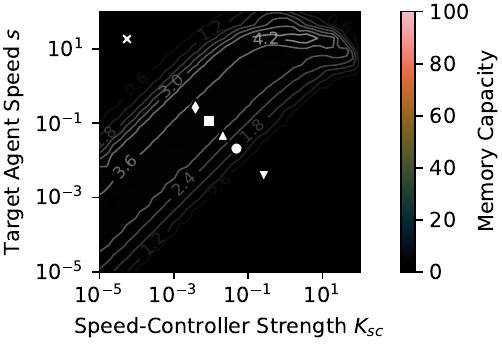}}\hfill
    \subfloat[$\Delta s = 2$]{\includegraphics[scale=1.0]{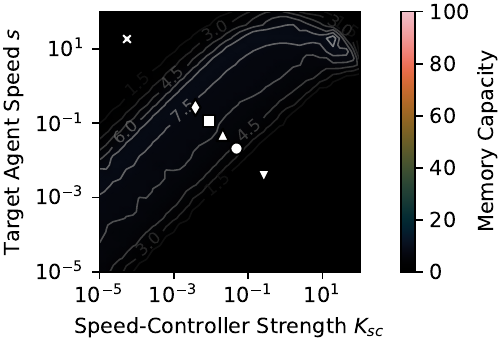}}\\
   \vspace{1em}
    \subfloat[$\Delta s = 5$]{\includegraphics[scale=1.0]{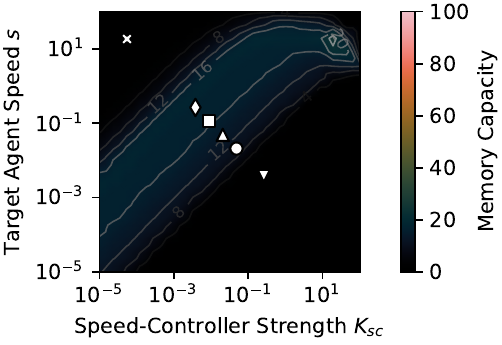}}\hfill
    \subfloat[$\Delta s = 10$]{\includegraphics[scale=1.0]{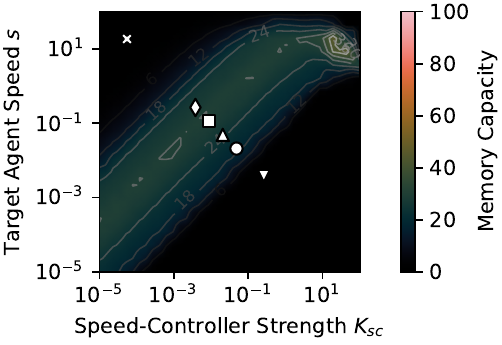}}

  \caption{\new{\textbf{Short-term memory capacity measurements for varying speed controller parameters with an inversely attractive driver and different driver change intervals $\Delta s$ for driver positions sampled uniformly at random from a circle with radius $R = 4$ around the center of the simulation box} (see also \refref{\cite{Gaimann2025}}, \figref{45}). }}
    \label{fig:speed-controller-memcap-total-inv-attr}
\end{figure*}

\begin{figure*}[htbp]
    \centering
    \subfloat[$\Delta s = 1$]{\includegraphics[scale=1.0]{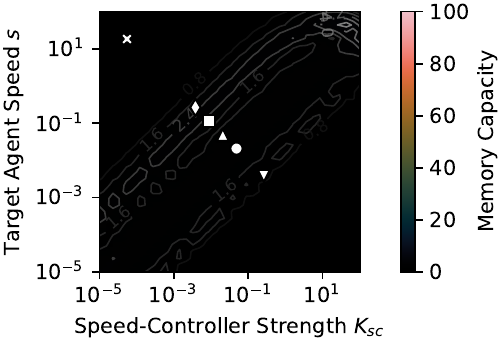}}\hfill
    \subfloat[$\Delta s = 2$]{\includegraphics[scale=1.0]{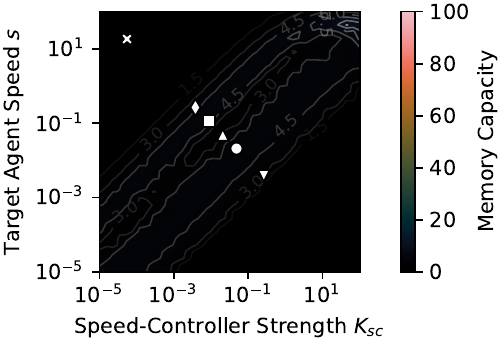}}\\
   \vspace{1em}
    \subfloat[$\Delta s = 5$]{\includegraphics[scale=1.0]{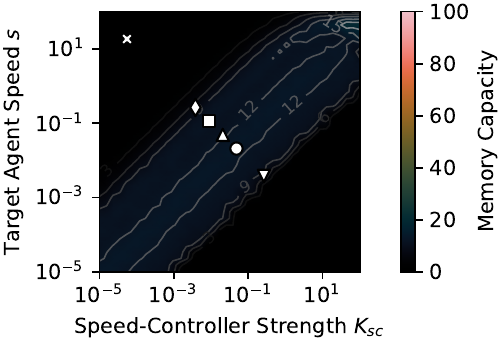}}\hfill
    \subfloat[$\Delta s = 10$]{\includegraphics[scale=1.0]{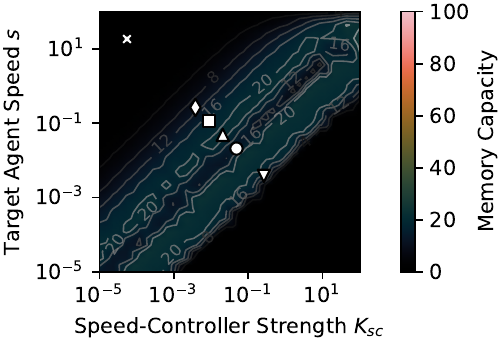}}

  \caption{\new{\textbf{Short-term memory capacity measurements for varying speed controller parameters with a linearly attractive driver and different driver change intervals $\Delta s$ for driver positions sampled uniformly at random from a circle with radius $R = 4$ around the center of the simulation box} (see also \refref{\cite{Gaimann2025}}, \figref{45}). }}
    \label{fig:speed-controller-memcap-total-lin-attr}
\end{figure*}


\begin{figure*}[htbp]
    \centering
    \subfloat[$\Delta s = 1$]{\includegraphics[scale=1.0]{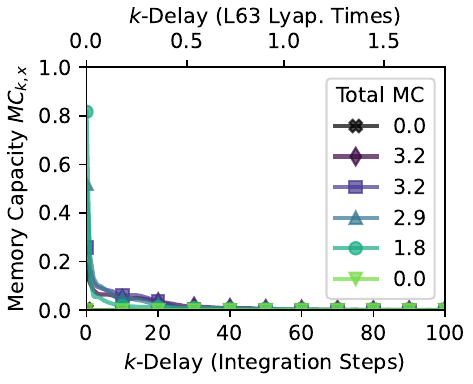}}\hfill
    \subfloat[$\Delta s = 2$]{\includegraphics[scale=1.0]{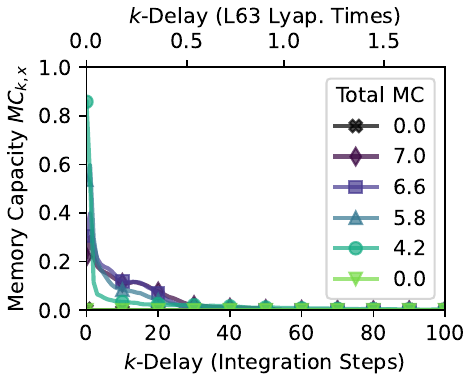}}\\
   \vspace{1em}
    \subfloat[$\Delta s = 5$]{\includegraphics[scale=1.0]{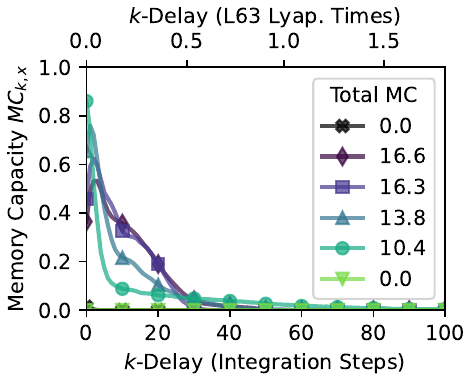}}\hfill
    \subfloat[$\Delta s = 10$]{\includegraphics[scale=1.0]{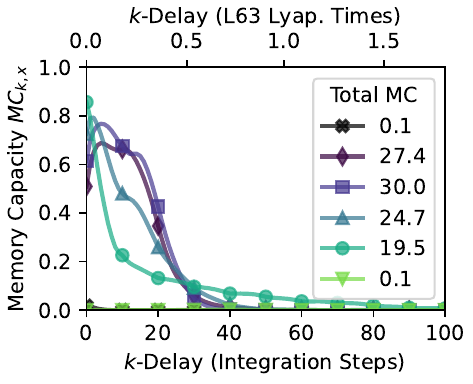}}

  \caption{\new{\textbf{$k$-delay memory capacity measurements for varying speed controller parameters using an inversely attractive driver and different uniformly at random driver change intervals $\Delta s$}, matching the total memory capacities reported in \figref{\ref{fig:speed-controller-memcap-total-inv-attr}}. }}
    \label{fig:speed-controller-attr-inv-memcap-k-delay}
\end{figure*}

\begin{figure*}[htbp]
    \centering
    \subfloat[$\Delta s = 1$]{\includegraphics[scale=1.0]{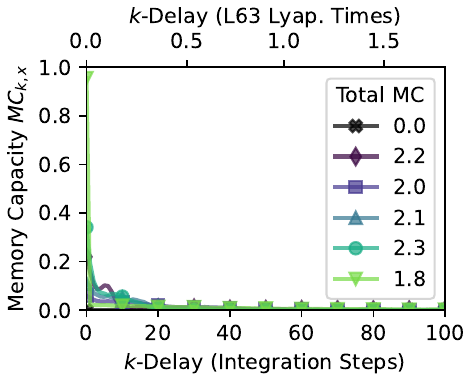}}\hfill
    \subfloat[$\Delta s = 2$]{\includegraphics[scale=1.0]{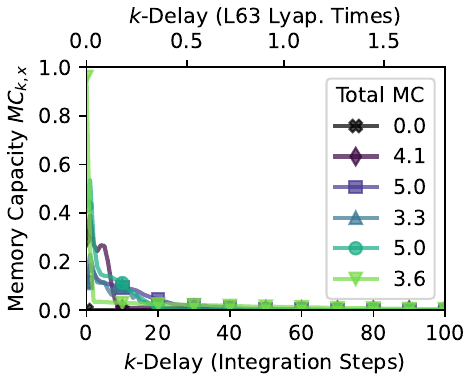}}\\
   \vspace{1em}
    \subfloat[$\Delta s = 5$]{\includegraphics[scale=1.0]{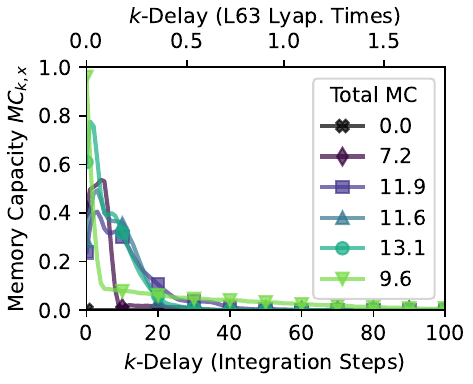}}\hfill
    \subfloat[$\Delta s = 10$]{\includegraphics[scale=1.0]{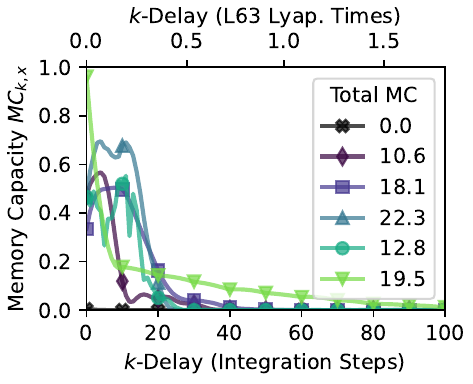}}

  \caption{\new{\textbf{$k$-delay memory capacity measurements for varying speed controller parameters using a linearly attractive driver and different uniformly at random driver change intervals $\Delta s$}, matching the total memory capacities reported in \figref{\ref{fig:speed-controller-memcap-total-lin-attr}}. }}
    \label{fig:speed-controller-attr-lin-memcap-k-delay}
\end{figure*}


\begin{figure*}[htbp]
    \centering
    \subfloat[Total MC, inverse attr.]{\includegraphics[scale=1.0]{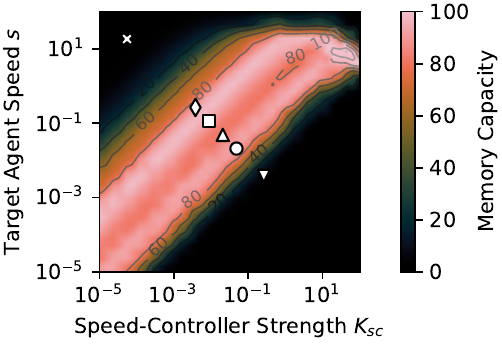}}\hfill
    \subfloat[$k$-delay MC, $x$ dim, inverse attr.]{\includegraphics[scale=1.0]{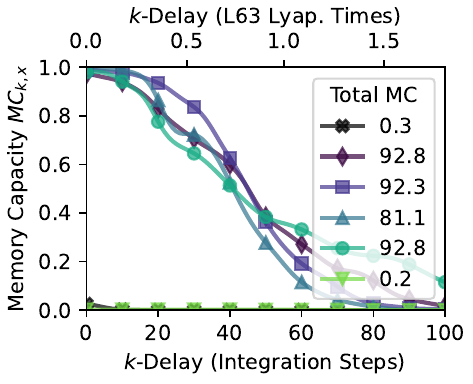}}
       \vspace{1em}
    \subfloat[Total MC, linear attr.]{\includegraphics[scale=1.0]{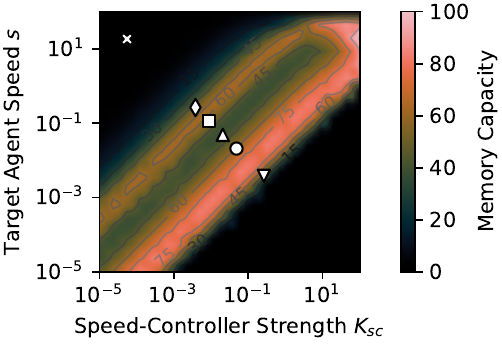}}\hfill
    \subfloat[$k$-delay MC, $x$ dim, linear attr.]{\includegraphics[scale=1.0]{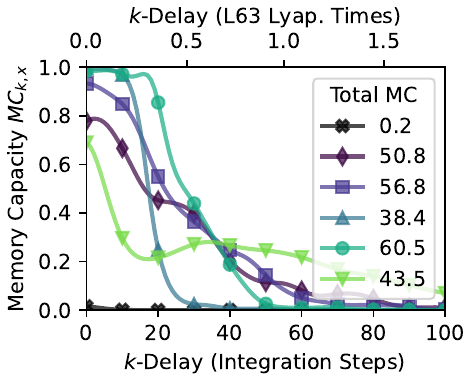}}
  \caption{\new{\textbf{Short-term memory capacity (MC) measured using the default Lorenz-63 trajectory as input signal instead of the random uniform signal} used in \figref{\ref{fig:speed-controller-memcap-total-inv-attr}} and \figref{\ref{fig:speed-controller-memcap-total-lin-attr}}, for varying speed-controller parameters using an (a,b) inversely attractive and an (c,d) linearly attractive driver, corresponding to the predictive performance plots shown in \figref[a,b]{\ref{fig:driver_attraction_speed_controller}}.}}
    \label{fig:speed-controller-memory-capacity-L63}
\end{figure*}


\begin{figure}[H]
    \centering
    \subfloat[$\Delta s = 1$]{\includegraphics[scale=0.9]{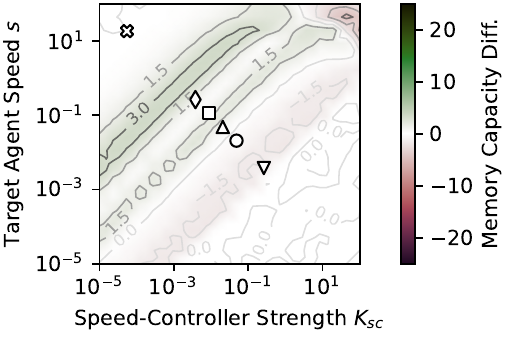}}\hfill
    \subfloat[$\Delta s = 2$]{\includegraphics[scale=0.9]{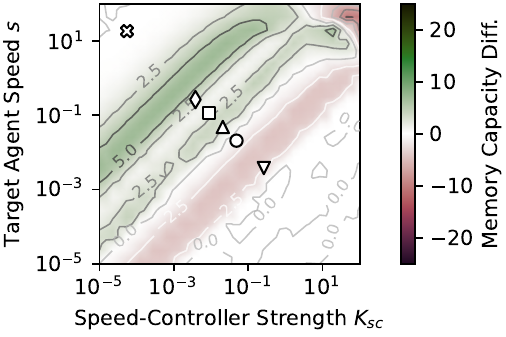}}\\
   \vspace{1em}
    \subfloat[$\Delta s = 5$]{\includegraphics[scale=0.9]{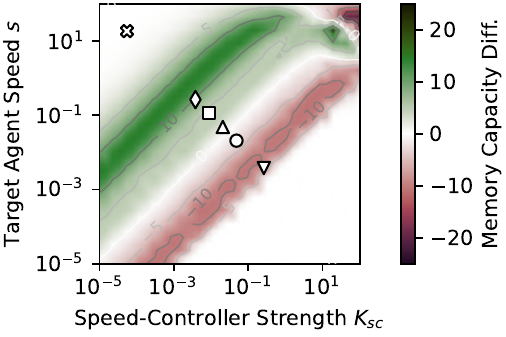}}\hfill
    \subfloat[$\Delta s = 10$]{\includegraphics[scale=0.9]{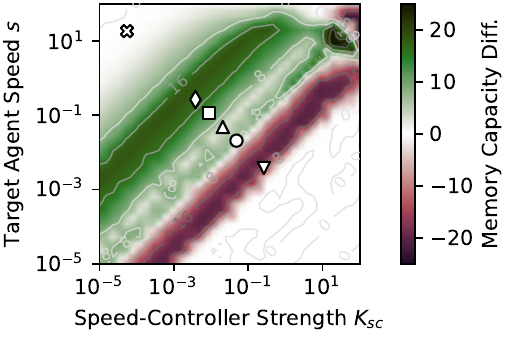}}\\
   \vspace{1em}
    \subfloat[Lorenz-63]{\includegraphics[scale=0.9]{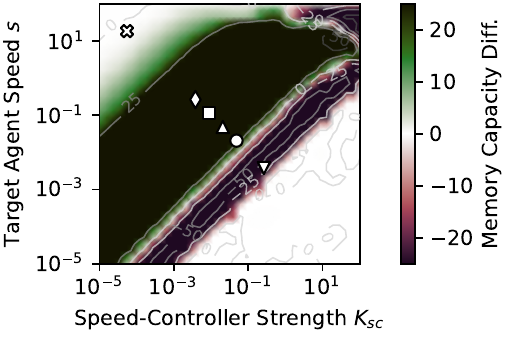}}\hfill
  \caption{\new{\textbf{Short-term memory capacity difference $\text{MC}^{\text{inv}} - \text{MC}^{\text{lin}}$ between inverse and linear driver attraction in a speed-controller parameter scan} using random uniform signals with different change intervals $\Delta s$ and the default Lorenz-63 trajectory as input signal. Refer to \figref{\ref{fig:speed-controller-memcap-total-inv-attr}}, \figref{\ref{fig:speed-controller-memcap-total-lin-attr}} and \figref[a,c]{\ref{fig:speed-controller-memory-capacity-L63}} to view the respective absolute $\text{MC}$ values. }}
    \label{fig:speed-controller-memory-capacity-lin-vs-inv}
\end{figure}

\section{Separability}\label{sec:kernel_rank}

\begin{figure*}[htbp]
    \centering
    \subfloat[Random unif., inverse attr.]{\includegraphics[scale=1.0]{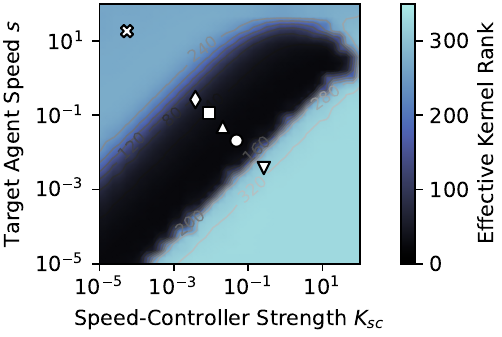}}\hfill
    \subfloat[Lorenz-63, inverse attr.]{\includegraphics[scale=1.0]{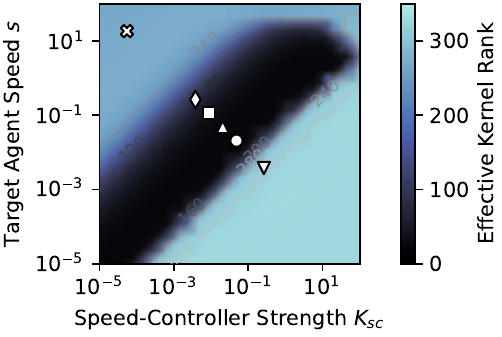}}\\
   \vspace{1em}
    \subfloat[Random unif., linear attr.]{\includegraphics[scale=1.0]{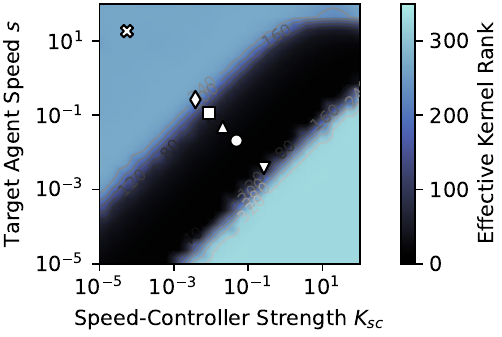}}\hfill
    \subfloat[Lorenz-63, linear attr.]{\includegraphics[scale=1.0]{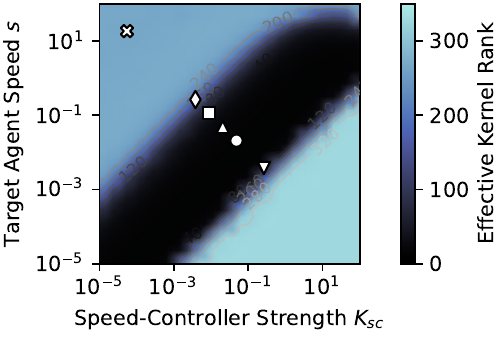}}

  \caption{\new{\textbf{Separability (kernel rank) for varying speed-controller parameters measured using (a,c) a random uniform input signal and (b,d) the default Lorenz-63 trajectory for (a,b) an inversely attractive driver and an (c,d) a linearly attractive driver.} The heatmaps correspond to the predictive performance plots shown in \figref[a,b]{\ref{fig:driver_attraction_speed_controller}}.}} \label{fig:speed-controller-separability}
\end{figure*}

\new{We measured the kernel rank (also known as separation rank) as a measure of separability. High separability means that the inputs are projected into a rich, high-dimensional space, where they can be linearly separated using the trained readout layer. 
To determine the kernel rank, we use a simple method described in \refrefs{\cite{Dale2019_framework, Wringe2024}}. We inject a time series of $T = 50,000$ inputs after a burn-in phase of $1,000$ steps into our reservoir. We use two types of time series: (1) canonically a discrete, randomly uniform time series confined to a radius of $4.0$ and centered in the simulation box, which was also used for the memory capacity computation (see \axref{\ref{sec:short-term-mc}}); and (2) the continuous Lorenz-63 trajectory. The observations are collected as described in \sectref{\ref{sec:rc_setup}} in the observation matrix $\mathbf{X} \in \mathbb{R}^{3M \times T}$ using $M = 200$ Gaussian observation kernels. The effective rank of this matrix is then determined \cite{Roy2007, Wringe2024}.
We performed these measurements for the speed-controller parameter scans with linear and inverse driver attraction interactions.}
\new{The results in \figref[]{\ref{fig:speed-controller-separability}} clearly show that the kernel rank is lowest around the near-critically damped regime, while the underdamped regime and the arrested regime show the highest values. This can be rationalized by the fact that the observation state space becomes reduced for stronger damping. The stronger damping causes agent trajectories with low mean squared displacements and overall a condensed agent distribution, which in turn limits the possible states created through the Gaussian kernels. Despite the low effective kernel rank, the predictive performance has an optimum in this regime. This could be because the higher kernel rank in the underdamped and the arrested regimes comes with the caveat that is similar input signals are mapped to greatly different high dimensional representations. The higher expressivity may therefore come at the cost of reduced consistency.}

\section{\label{ax:narma10}NARMA-10 Benchmarks}

    \new{To assess the long short‑term memory and the nonlinear interaction between past outputs and inputs, we define the 10th order Nonlinear Auto Regressive Moving Average (NARMA-10) time series
    \begin{equation}
      y(t) = 0.3\,y(t-1)
        + 0.05\,y(t-1)\sum_{i=1}^{10} y(t-i)
        + 1.5\,\tilde{u}(t-10)\,\tilde{u}(t-1)
        + 0.1,\quad t \ge 10\,,  
    \end{equation}
    where $\tilde{u}(t)$ is typically the input to the reservoir at a time $t$ and drawn uniformly at random from the interval $[0.0,0.5)$. To fit the input time series to our active matter reservoir computing setup and for better comparability with the prequel to this work, \refref{\cite{Gaimann2025}}, we generate 2D input time series (driver trajectories $(x_d(t), y_d(t))$) as
    \begin{equation}
        U(t) \sim \mathrm{Uniform}(0,1), \quad r = R \sqrt{U}, \quad \theta \sim \mathrm{Uniform}(0, 2\pi),
    \end{equation}
    \begin{equation}
        x_d = r \cos(\theta), \quad y_d = r \sin(\theta)\,.
    \end{equation}
    The time series is confined to a circle with radius $R = 4.0$ and centered in the simulation box (see also \refref{\cite{Gaimann2025}}, Appendix C and \figref[a]{45} for a visualization).
    We obtain for each $x$ and $y$ dimension the input $\tilde{u}(t)$ used to compute NARMA-10 by rescaling $(x_d(t), y_d(t))$ to the interval $[0.0, 0.5)$. Here, we predict one step ahead ($y(t+1)$) for each of the $x$ and $y$ dimensions of the 2D input time series (the driver position).
    Unlike for the computation of the short-term memory capacity in \axref{\ref{sec:short-term-mc}}, we only consider driver change intervals of $\Delta s = 1$ here. For larger change intervals $\Delta s > 1$, we observe that the corresponding NARMA-10 time series diverges ($y(t) \rightarrow \infty$). 
    Because of this potential for divergence and other reasons NARMA-10 has drawn criticism as benchmark \cite{Wringe2024}.}\\
    
    \new{In \figref{\ref{fig:speed-controller-driver-narma10}} we show the NRMSEs for the speed-controller parameter scan with inverse and linear driver attraction, respectively.
    For an inversely attractive driver in sub-figure (a), we confirm that the optimal performance $\text{NRMSE} \approx 0.72$ is obtained in the near-critically damped regime.
    For the linearly attractive driver in sub-figure (c), however, we observe an optimal performance regime of $\text{NRMSE} \approx 0.75$ in the overdamped regime, between the circle and the nabla symbols. In contrast, the performance for the Lorenz-63 prediction task (see \figref[b]{\ref{fig:driver_attraction_speed_controller}} and \figref[b]{\ref{fig:driver_attraction_linear_speed_controller-more-metrics}}) was optimal around the near-critically damped regime (between pyramid and square symbols). This agrees with previous observations \cite{Gaimann2025} that for different tasks, the optimal regime may be slightly shifted from the near-critically damped regime, as is expected from the memory-nonlinearity tradeoff \cite{Schrauwen2007a}. 
    We also found that in the region between the circle and the nabla symbol, the short-term memory capacity has a local optimum (see also \figref[c]{\ref{fig:speed-controller-memory-capacity-L63}}), coinciding with the optimum in NARMA-10 predictive performance. This agrees with the finding that NARMA-10 may serve as a proxy for the short-term memory capacity \cite{Wringe2024}.}

\begin{figure*}[htbp]
    \centering
    \subfloat[Inv. driver attr., NRMSE]{\includegraphics[scale=1.0]{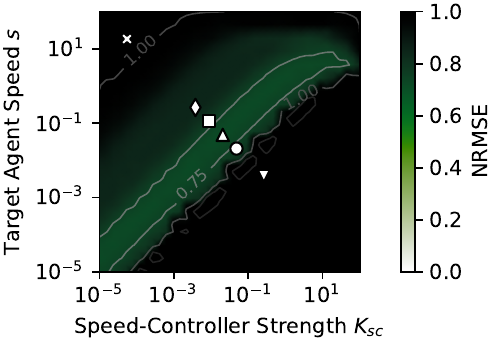}}\hfill
    \subfloat[Inv. driver attr., NMSE]{\includegraphics[scale=1.0]{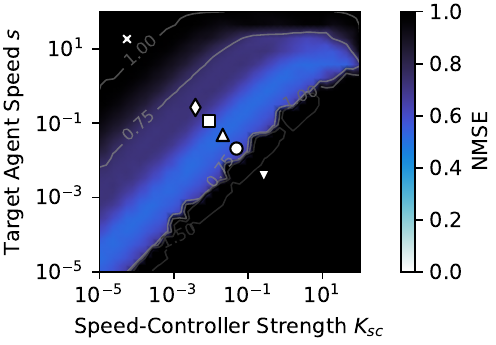}}\\
   \vspace{1em}
    \subfloat[Lin. driver attr., NRMSE]{\includegraphics[scale=1.0]{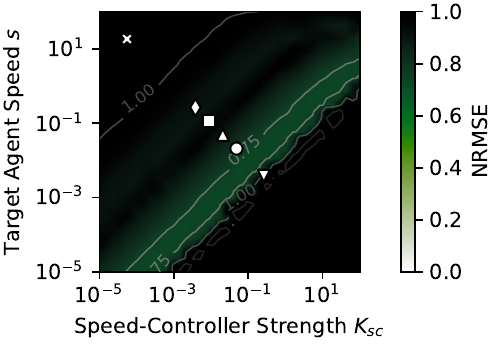}}\hfill
    \subfloat[Lin. driver attr., NMSE]{\includegraphics[scale=1.0]{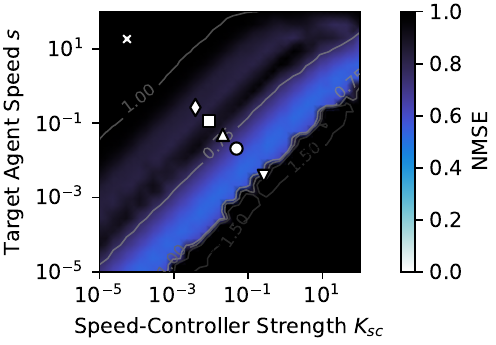}}

  \caption{\new{\textbf{(a,c) NRMSE and (b,d) NMSE for the NARMA-10 benchmark for varying speed controller parameters with (a,b) an inversely attractive driver and (c,d) a linearly attractive driver}, corresponding to \figref[a,b]{\ref{fig:driver_attraction_speed_controller}} in the main text. Refer to \axref{\ref{ax:narma10}} for details. }}
    \label{fig:speed-controller-driver-narma10}
\end{figure*}

\section{Data Availability}

Data generated in this study can be accessed in the Data Repository of the University of Stuttgart (DaRUS), in the dataverse Stuttgart Center for Simulation Science EXC 2075 / Project 6-15:\\
\href{ https://doi.org/10.18419/DARUS-4805}{\texttt{ https://doi.org/10.18419/DARUS-4805}} \cite{DARUS-4805_2025}\\
All data in this manuscript was generated using the ResoBee software for active matter reservoir computing. It is currently being prepared for publication in an open-source software journal.\\

\section{Supplementary Figures}

\subsection{Driver Repulsion}

\begin{figure*}[htbp]
    \centering
    \includegraphics{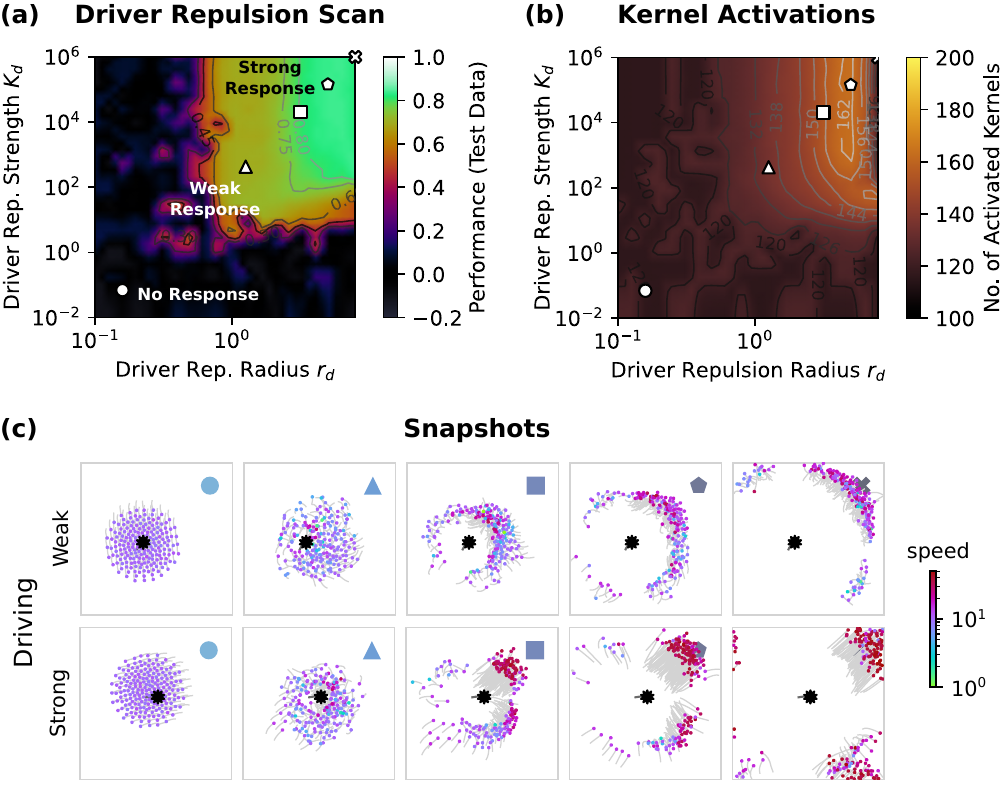}
    \caption{\textbf{(a) Scan over the parameters of the driver repulsion interaction: the interaction (cut-off) radius $r_d$  around the driver and the strength $K_d$ of the interaction.} Stronger driver repulsion and higher interaction radii yield better predictive performances. (b) Stronger repulsions and higher repulsion radii correlate with a higher number of Gaussian kernel activations on particle counts. Kernel activation is computed as a time-average over $T = 50,000$ integration time steps here and measured using thresholding: a kernel is defined as activated when the agent count observed by the Gaussian kernel exceeds a threshold value of $0.001$. We note that different activation thresholds in the range $[10\textsuperscript{-5}; 10\textsuperscript{-1}]$ yield qualitatively similar pictures. (c) Snapshots of driven swarms. Symbols indicate parameter combinations in the heatmaps. For intermediate to large interaction radii and strengths, agent exclusion zones form around the driver. Refer to \tabref{\ref{tab:supplementary_videos_driver_repulsion}} for corresponding videos. The same scan using near-critically damped speed-controller settings is shown in \figref{\ref{fig:nearcritdamped_driver_repulsion_scan_heatmap}}.}
    \label{fig:Lymburn_critical_driver_repulsion_scan_heatmap}
\end{figure*}

\begin{figure}[H]
    \centering
    \subfloat[$l_{\text{box}}^{\text{sim}} = 32.0$, $l_{\text{box}}^{\text{obs}} = 16.0$]{\includegraphics{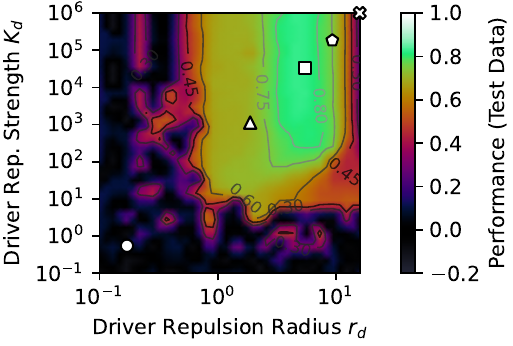}}\hfill
    \subfloat[Snapshot corresponding to (a)]{\includegraphics[width=.25\textwidth]{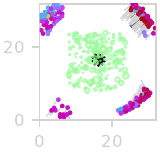}}\\
    \vspace{1em}
    \subfloat[$l_{\text{box}}^{\text{sim}} = 32.0$, $l_{\text{box}}^{\text{obs}} = 32.0$]{\includegraphics{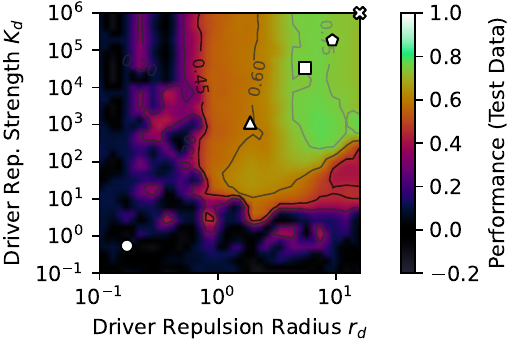}}\hfill
    \subfloat[Snapshot corresponding to (c)]{\includegraphics[width=.25\textwidth]{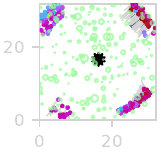}}\\
    \vspace{1em}
    \subfloat[$l_{\text{box}}^{\text{sim}} = 64.0$, $l_{\text{box}}^{\text{obs}} = 64.0$]{\includegraphics{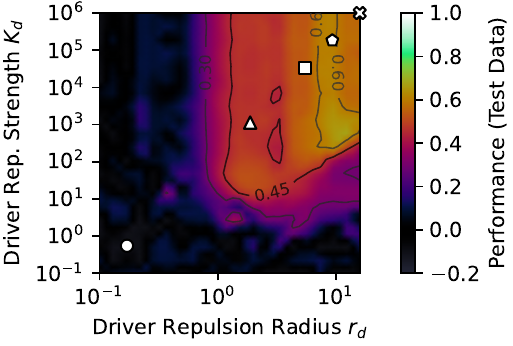}}\hfill
    \subfloat[Snapshot corresponding to (e)]{\includegraphics[width=.25\textwidth]{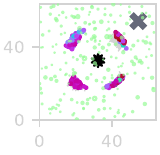}}
    
    \caption{\textbf{Driver repulsion parameter scans with different simulation box sizes and Gaussian kernel observer box sizes.} (a,c,e) Predictive performances with (b,d,f) corresponding system snapshots for the parameter combination indicated by the cross symbol. Simulation parameters are analogous to \figref{\ref{fig:Lymburn_critical_driver_repulsion_scan_heatmap}}, but here different simulation box sizes and Gaussian kernel observer box sizes ($l_{\text{box}}^{\text{sim}} \in \{32.0; 64.0\}$; $l_{\text{box}}^{\text{obs}} \in \{16.0; 32.0; 64.0\}$) are chosen compared to \figref{\ref{fig:Lymburn_critical_driver_repulsion_scan_heatmap}} ($l_{\text{box}}^{\text{sim}} = l_{\text{box}}^{\text{obs}} = 16.0$). Light green circles indicate Gaussian observation kernels, with radius indicating half kernel width. Refer to \tabref{\ref{tab:supplementary_videos_driver_repulsion}} for corresponding videos. }
    \label{fig:driver_repulsion_larger_boxes}
\end{figure}

\subsection{Variations of Agent-Agent Repulsion Interactions}

\begin{figure}[H]
    \centering
    \subfloat[Long Range Repulsion ($r_r = 1.0$)]{\includegraphics{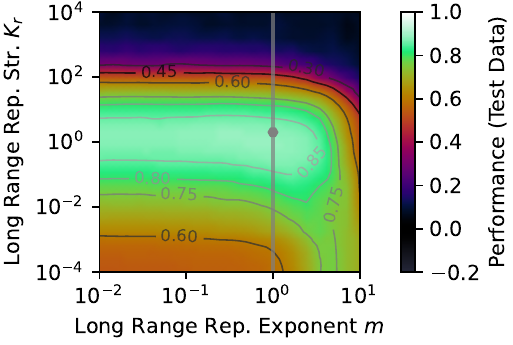}}\hfill
    \subfloat[Long Range Repulsion ($r_r = 4.0$)]{\includegraphics{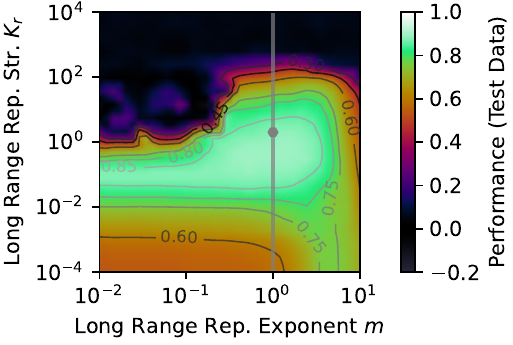}}\\
          \vspace{1em}
    \subfloat[Short Range Repulsion ($r_r = 1.0$)]{\includegraphics{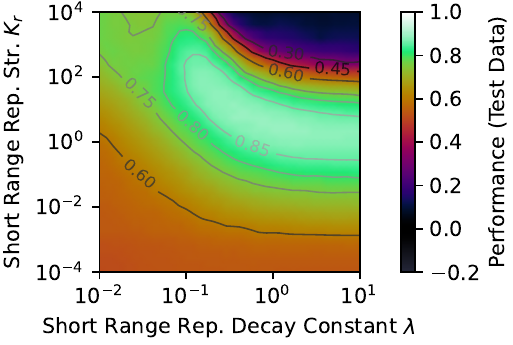}}\hfill
    \subfloat[Short Range Repulsion ($r_r = 4.0$)]{\includegraphics{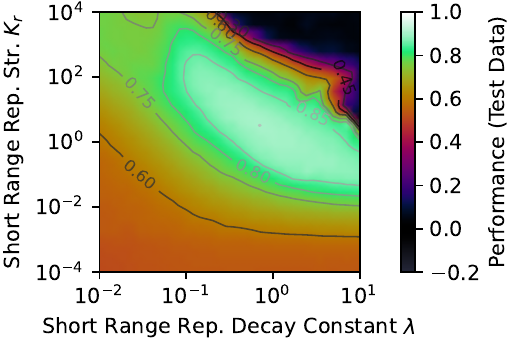}}

    \caption{\textbf{Variations of the basic agent-agent repulsion interaction with a repulsive driver in the near-critically damped regime} ($K_{sc} = 0.02069$, $s = 0.04833$) in \eqqref{\ref{eq:repulsion_force}}. (a,b) Long range repulsion interactions (\eqqref{\ref{eq:long_range_repulsion}}) and (c,d) short range repulsion interactions (\eqqref{\ref{eq:short_range_repulsion}}) for (a,c) default interaction radii ($r_r = 1.0$) and (b,d) higher interaction radii ($r_r = 4.0$). The gray vertical lines indicate the default exponent $m=1$ used in \refref{\cite{Lymburn2021} and \cite{Gaimann2025}}, the hexagon indicates the default strength $K_r = 2.0$. Optimal performances $P$: (a) $P = 0.88817$ @ $(m, K_r) = (0.2636650899, 1.6237767392)$; (b) $P = 0.89232$ @ $(m, K_r) = (0.5455594781, 0.2335721469)$; (c) $P = 0.88829$ @ $(\lambda, K_r) = (1.6237767392, 1.6237767392)$; (d) $P = 0.90186$ @ $(\lambda, K_r) = (0.5455594781, 4.2813323987)$.}
    \label{fig:repulsion_variations}
\end{figure}

\subsection{Single Agent Scans}

\begin{figure}[H]
    \centering
    \subfloat[Driver Repulsion, Lymburn \etal ``critical'' speed-controller]{\includegraphics{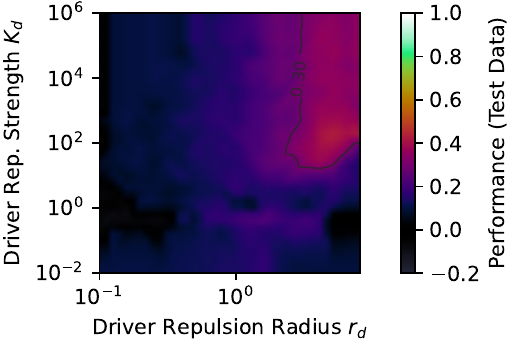}}\hfill
    \subfloat[Driver Repulsion, Near-critically damped speed-controller]{\includegraphics{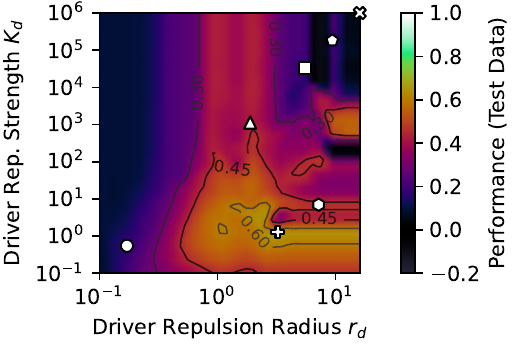}}\\
    \subfloat[Inverse Driver Attraction, Near-critically damped speed-controller]{\includegraphics{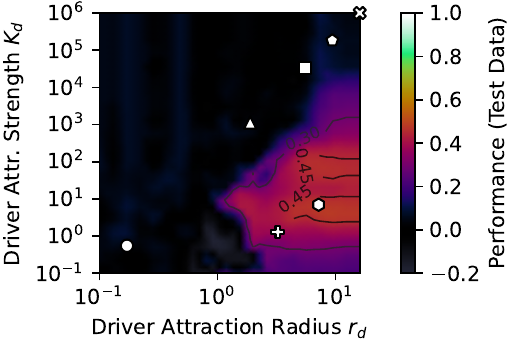}}
    
    \caption{\textbf{Predictive performance for reservoir computing with a single agent for varying (a,b) driver repulsion and (c) inverse driver attraction parameters} analogous to the 200 particle systems in \figref[a]{\ref{fig:Lymburn_critical_driver_repulsion_scan_heatmap}}, \figref[a]{\ref{fig:nearcritdamped_driver_repulsion_scan_heatmap}} and \figref[a]{\ref{fig:driver_attraction}}. Refer to \tabref{\ref{tab:supplementary_videos_driver_repulsion_nearcritdamped_1_agent}} and \tabref{\ref{tab:supplementary_videos_driver_attraction_inverse_nearcritdamped_1_agent}} for corresponding videos.}
    \label{fig:few_particles}
\end{figure}

\subsection{Driver Attraction}

\begin{figure}[H]
    \centering
    \includegraphics{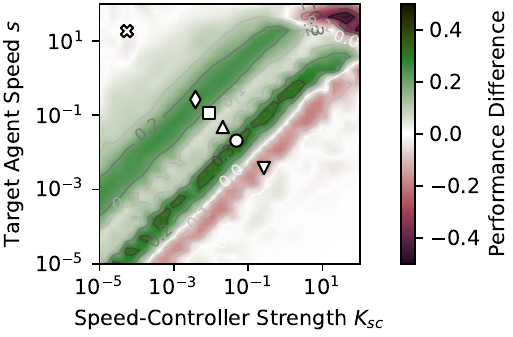}
    \caption{\textbf{Performance difference $P^{\text{attr}}_{\text{inverse}}- P^{\text{attr}}_{\text{linear}}$ between inverse and linear driver attraction} for the speed-controller parameter scans presented in \figref{\ref{fig:driver_attraction}}. Positive values indicate higher performance for the inversely attractive driver compared to the linearly attractive driver.}
    \label{fig:driver_attraction_performance_differences}
\end{figure}

\begin{figure}[H]
    \centering
    \subfloat[Linear]{\includegraphics{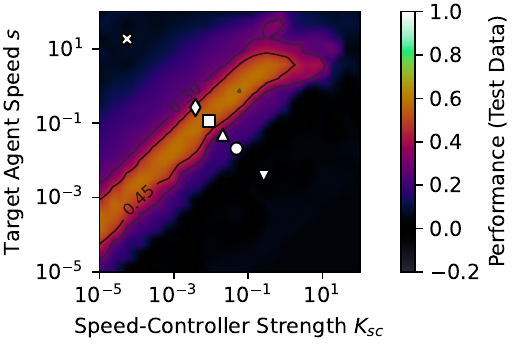}}\hfill
    \subfloat[Inverse]{\includegraphics{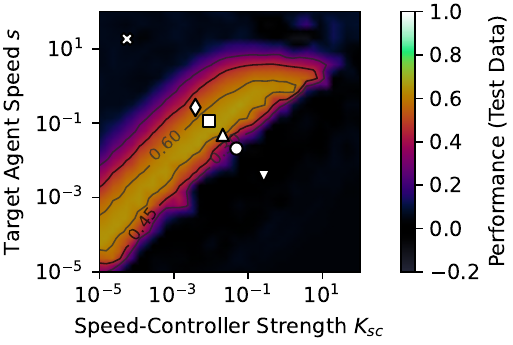}}\hfill
    \caption{
        \textbf{Weaker force strengths $K_{d} = 2.0$ for different driver attraction mechanisms} analogously to \figref{\ref{fig:driver_attraction}}. This force parameter corresponds to the default homing force strength $K_h$ used in this paper and in \refref{\cite{Lymburn2021}, \cite{Gaimann2025}}.
    }
    \label{fig:driver_attraction_weak}
\end{figure}

\begin{figure}[H]
    \centering
    \subfloat[Repulsive driver speed-controller parameter scan]{\includegraphics{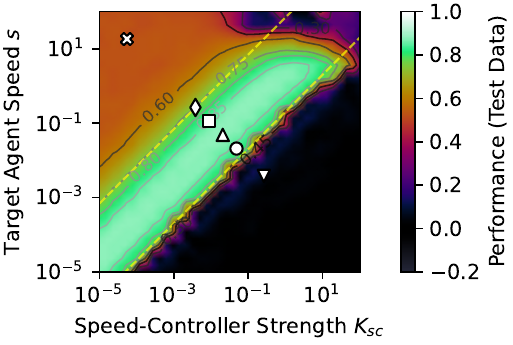}}\hfill
    \subfloat[Performance difference]{\includegraphics{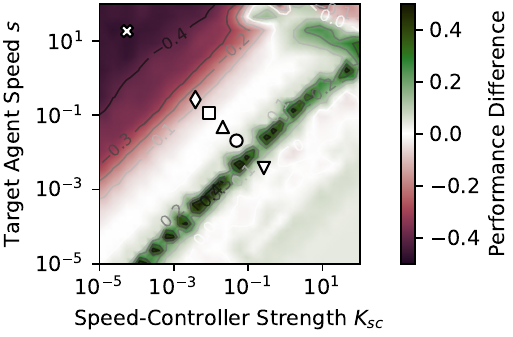}}\\
    \caption{\textbf{Comparison of an inversely repulsive driver with an inversely attractive driver.} (a) Reproduction of the speed-controller scan with a repulsive driver, agent-agent repulsion and alignment interactions, and a homing interaction presented in \figref{2a} in \refref{\cite{Gaimann2025}}. (b) Performance difference $P^{\text{attr}}_{\text{inverse}}- P^{\text{rep}}$ between inversely attractive and (inversely) repulsive drivers for the analogous parameter scan shown in \figref[a]{\ref{fig:driver_attraction}}. Positive values indicate better performance of the attractive driver compared to the repulsive driver for the same parameter combination.}
\label{fig:comparison_attractive_to_repulsive_driver}
\end{figure}

\begin{figure}[H]
    \centering
    \subfloat[$\sim 1/r_{i,d}$, Polarity]{\includegraphics{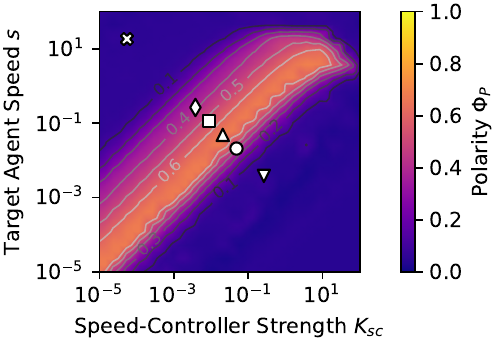}}\hfill
    \subfloat[$\sim r_{i,d}$, Polarity]{\includegraphics{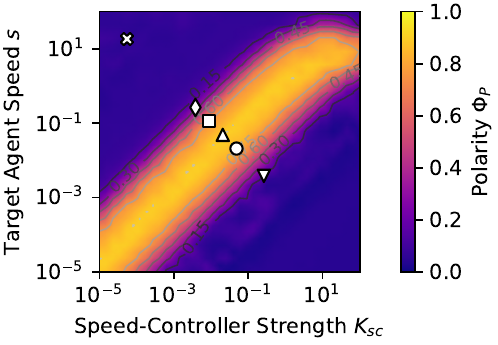}}\\
       \vspace{1em}
    \subfloat[$\sim 1/r_{i,d}$, Rotation]{\includegraphics{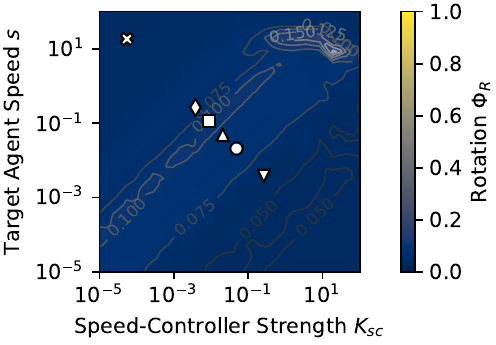}}\hfill
    \subfloat[$\sim r_{i,d}$, Rotation]{\includegraphics{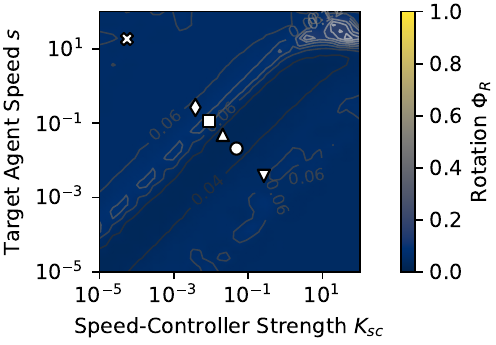}}\\
       \vspace{1em}
    \subfloat[$\sim 1/r_{i,d}$, Mean Squared Displacements]{\includegraphics{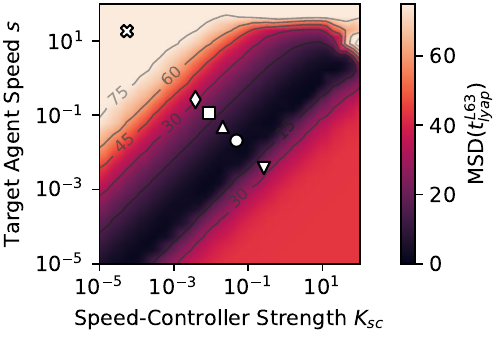}}\hfill
    \subfloat[$\sim r_{i,d}$, Mean Squared Displacements]{\includegraphics{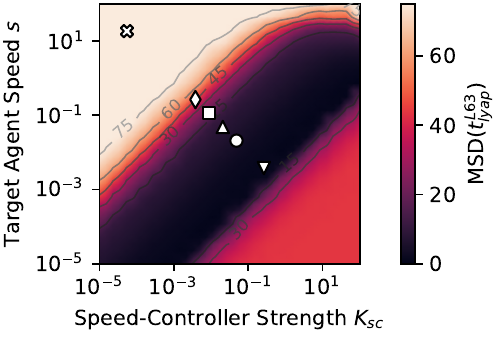}}

    \caption{\textbf{Various observables for a speed-controller parameter scan for inverse ($\sim 1/r_{i,d}$, left column) and linear ($\sim r_{i,d}$, right column) driver attractions}, corresponding to the performance heatmaps shown in \figref[a,b]{\ref{fig:driver_attraction}}. }
    \label{fig:attraction_observables_1}
\end{figure}

\begin{figure}[H]
    \centering
    \subfloat[$\sim 1/r_{i,d}$, Dynamical Susceptibility\newline]{\includegraphics{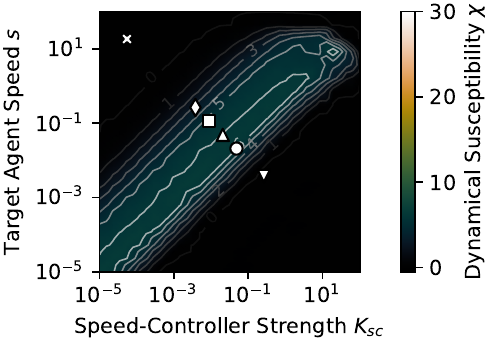}}\hfill
    \subfloat[$\sim r_{i,d}$, Dynamical Susceptibility\newline]{\includegraphics{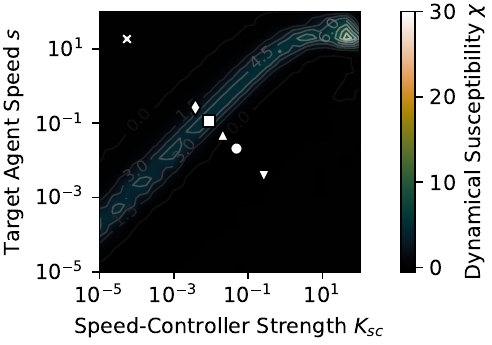}}\\
       \vspace{1em}
    \subfloat[$\sim 1/r_{i,d}$, Connected Velocity Correlation Function\newline at First Local Minimum]{\includegraphics{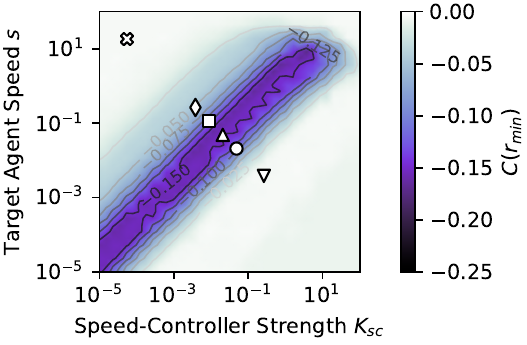}}\hfill
    \subfloat[$\sim r_{i,d}$, Connected Velocity Correlation Function\newline at First Local Minimum]{\includegraphics{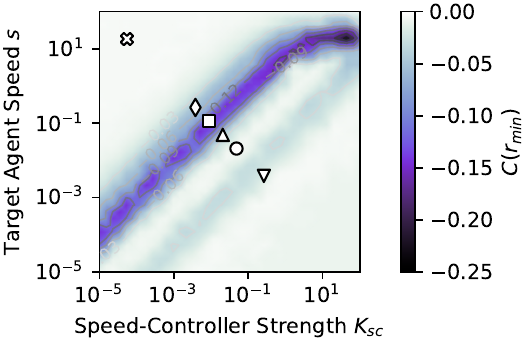}}\\
       \vspace{1em}
    \subfloat[$\sim 1/r_{i,d}$, Circle Area Spanned by Nearest\newline Agent–Driver Distance]{\includegraphics[width=.49\textwidth]{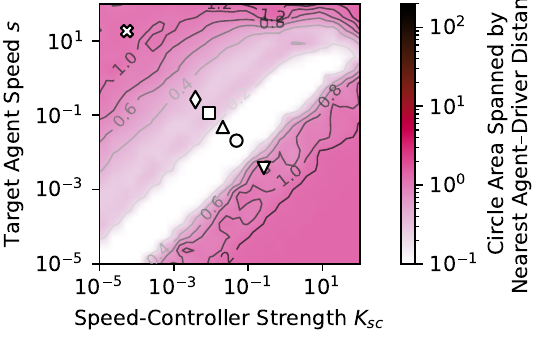}}\hfill
    \subfloat[$\sim r_{i,d}$, Circle Area Spanned by Nearest\newline Agent–Driver Distance]{\includegraphics[width=.49\textwidth]{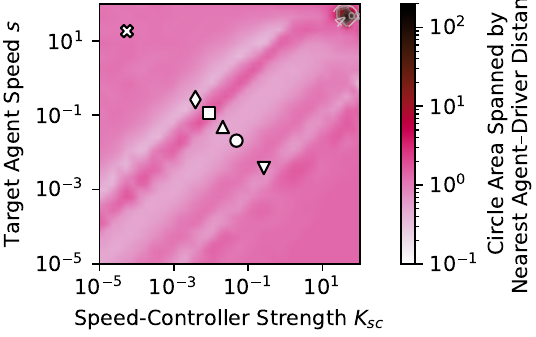}}

    \caption{\textbf{Various observables for a speed-controller parameter scan for inverse ($\sim 1/r_{i,d}$, left column) and linear ($\sim r_{i,d}$, right column) driver attractions (continued)}, corresponding to the performance heatmaps shown in \figref[a,b]{\ref{fig:driver_attraction}}. }
    \label{fig:attraction_observables_2}
\end{figure}

\begin{figure}[H]
    \centering
    \subfloat[$\sim 1/r_{i,d}$ driver attraction]{\includegraphics{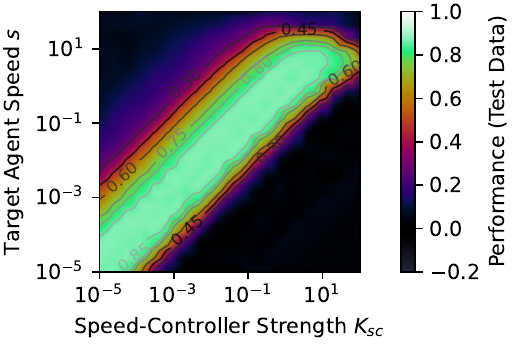}}\hfill
    \subfloat[$\sim r_{i,d}$ driver attraction]{\includegraphics{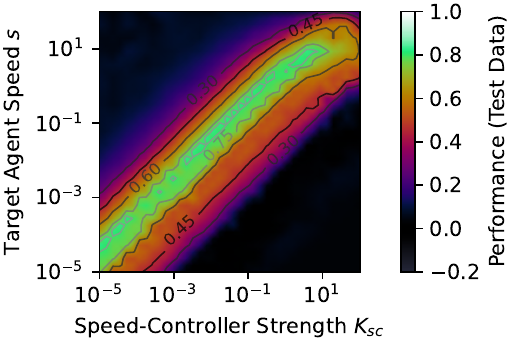}}\hfill
    \caption{
        \textbf{Predictive performances for linearly ($\sim r_{i,d}$) and inversely ($\sim 1/r_{i,d}$) attractive drivers with a different Ridge coefficient $\lambda_{\text{Ridge}} = 200.0$}, analogously to \figref[a,b]{\ref{fig:driver_attraction}} (where $\lambda_{\text{Ridge}} = 1.0$).
    }
    \label{fig:driver_attraction_lambda_ridge_200}
\end{figure}

\begin{figure}[H]
    \centering
    \includegraphics{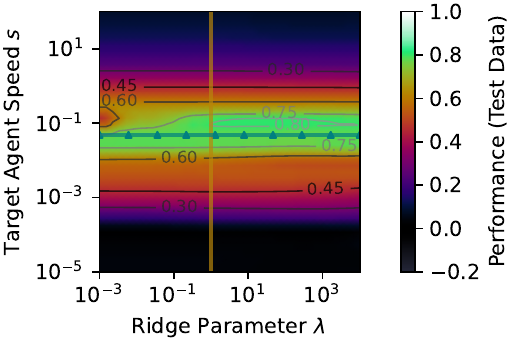}
    \caption{\textbf{Predictive performances for a scan over the speed-controller target agent speed $s$ and the Ridge coefficient $\lambda_{\text{Ridge}}$} with a fixed speed-controller strength $K_{sc} = 0.02069$ and a linearly attractive driver with $K_d = 100.0$. The blue horizontal line with pyramid symbols at $s = 0.04833$ corresponds to the near-critically damped speed-controller settings, which are marked with a pyramid symbol in \figref[a,b]{\ref{fig:driver_attraction}}. The orange vertical line indicates the default choice of $\lambda_{\text{Ridge}} = 1.0$. The choice of the Ridge coefficient impacts performance only significantly in a small region around $s \approx 0.1$ for the given $K_{sc}$.}
    \label{fig:driver_attraction_lambda_ridge_scan}
\end{figure}

\subsection{Agent-Agent Repulsion}

\begin{figure}[H]
    \centering
    \includegraphics{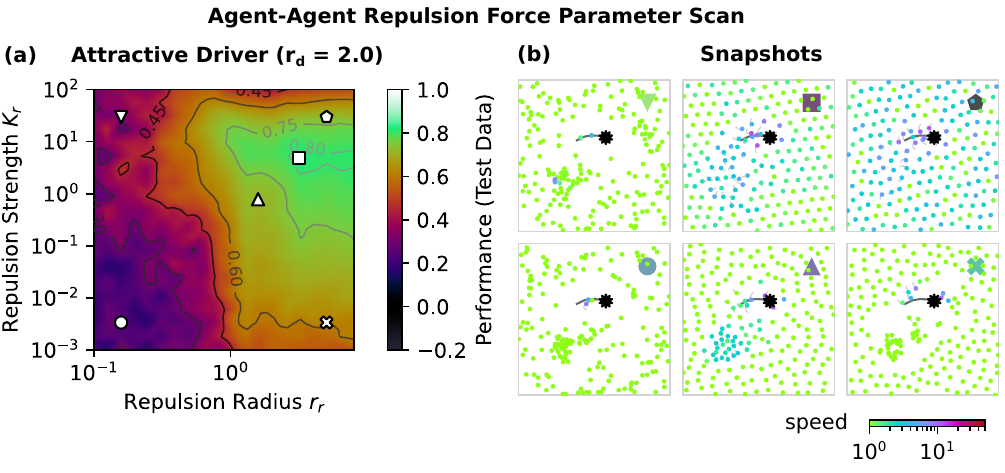}
    \caption{\textbf{Predictive performances for a scan over the agent-agent repulsion strength $K_r$ and agent-agent repulsion radius $r_r$ with an inversely attractive driver with strength $K_{d} = 100.0$ and range $r_d = 2.0$}, analogously to \figref[b]{\ref{fig:repulsion}} with a range $r_d = \infty$. }
    \label{fig:repulsion_attractive_driver_range_2.0}
\end{figure}

\begin{figure}[H]
    \centering
    \includegraphics{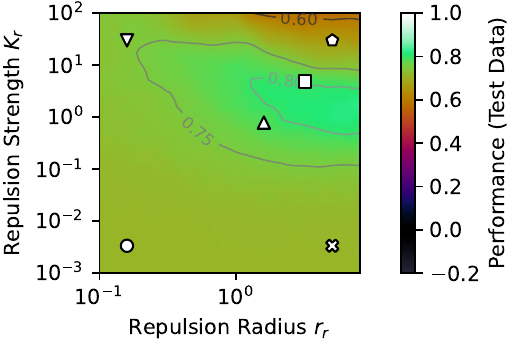}
    \caption{\textbf{Predictive performances for a scan over the agent-agent repulsion parameters for an attractive Lorenz-96 driver trajectory}, analogously to \figref[d]{\ref{fig:repulsion}}. The Lorenz-96 driver time series was taken from \refrefs{\cite{Gaimann2025, DARUS-4620_2025, DARUS-4619_2025}}.}
    \label{fig:repulsion_attractive_driver_Lorenz96}
\end{figure}

\subsection{Agent-Agent Repulsion and Number of Agents}

\begin{figure}[H]
    \centering
    \includegraphics{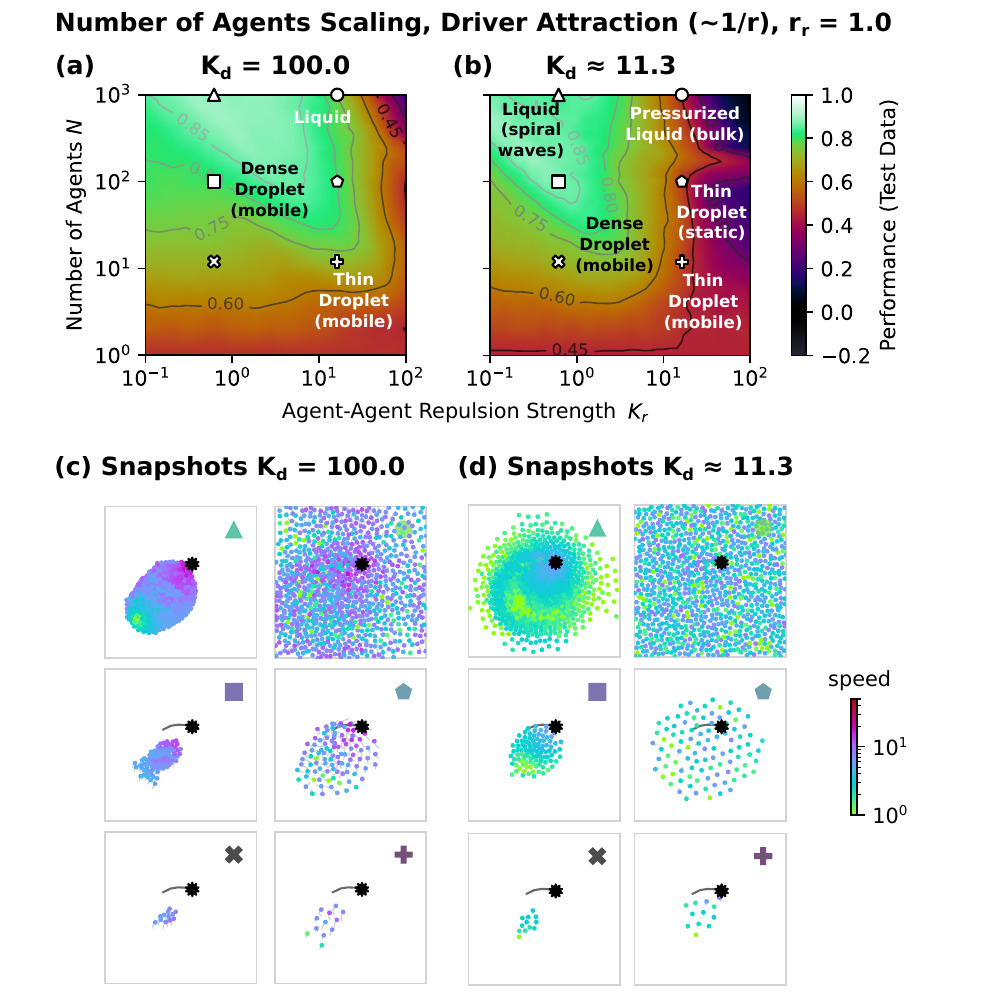}
    \caption{
        \textbf{Predictive performances versus employed number of agents $\bm{N_a}$ and agent-agent repulsion strengths $\bm{K_r}$ with an inversely attractive driver}, analogous to \figref{\ref{fig:n_agents_repulsion_strength}} but with a smaller agent-agent repulsion radius of $r_r = 1.0$ and different driver attraction strengths (a) $K_d = 100.0$ and  (b) $K_d = 11.2883789$. (c,d) Snapshots of the active matter systems at different parameter combinations that are marked as symbols in sub-figures (a,b); color indicates agent speed. Refer to \tabref{\ref{tab:supplementary_videos_repulsion_nagents_driver_attraction_rr_1e0_Kd_1e2}} and \tabref{\ref{tab:supplementary_videos_repulsion_nagents_driver_attraction_rr_1e0_Kd_1e1}} for corresponding videos.
    }
    \label{fig:n_agents_repulsion_strength_supplement}
\end{figure}

\newpage
\subsection{\label{ax:supp_driver_speeds}Variations of Mean Driver Speed}

\begin{figure}[H]
    \centering
    \subfloat[$\bar{v}_d = 1.0$]{\includegraphics[scale=1.0]{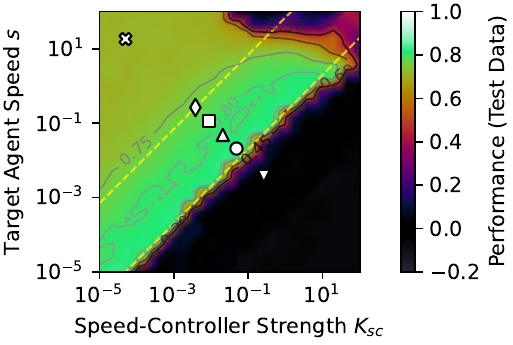}}\hfill
    \subfloat[$\bar{v}_d = 2.0$]{\includegraphics[scale=1.0]{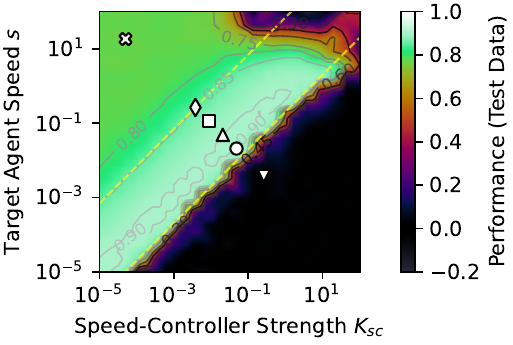}}\\
   \vspace{1em}
    \subfloat[$\bar{v}_d = 5.0$]{\includegraphics[scale=1.0]{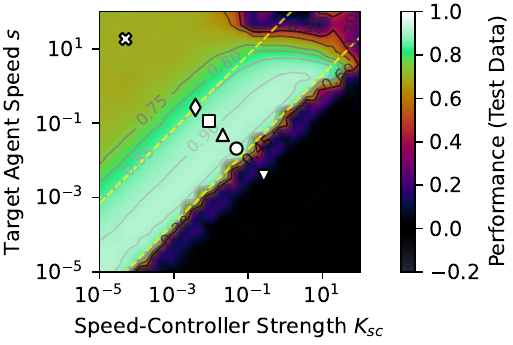}}\hfill
    \subfloat[$\bar{v}_d = 10.0$]{\includegraphics[scale=1.0]{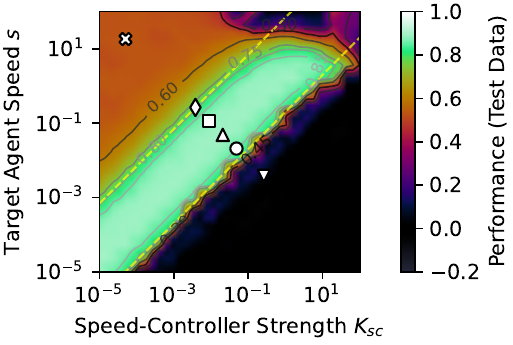}}\\
   \vspace{1em}
    \subfloat[$\bar{v}_d = 15.0$]{\includegraphics[scale=1.0]{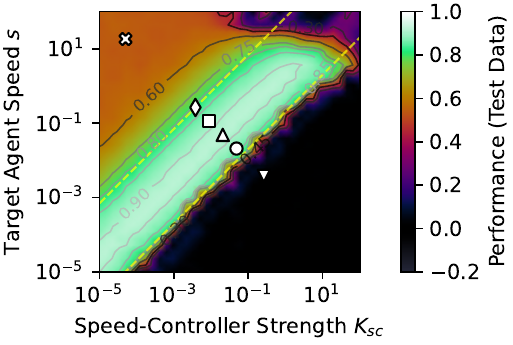}}\hfill
  \caption{\new{\textbf{Speed-controller parameter scans with different mean driver speeds $\bar{v}_d$ but same prediction horizons} for a repulsive driver with strength $K_d = 100.0$ as shown in \figref[a]{2} in \refref{\cite{Gaimann2025}}. To compare predictive performances across mean driver speeds, we choose a constant mean predicted distance ahead of $\bar{s}_{pred} = \bar{v}_d \Delta t_{pred} = 5.0$ length units. The Lorenz-63 trajectory here is generated using the \texttt{dysts} package as described in \refref{\cite{Gaimann2025}}, Appendix A. Sub-figure (d) has a similar driver mean speed to the one used in the main text ($\bar{v}_d = 9.91$). Refer to \tabref{\ref{tab:supplementary_videos_speed_controller_speed_1}} -- \tabref{\ref{tab:supplementary_videos_speed_controller_speed_15}} for corresponding videos.
  }}
    \label{fig:speed-controller-different-speeds-same-horizon}
\end{figure}

\newpage
\subsection{\label{ax:other}Other}

\begin{figure*}[htbp]
    \centering
    \includegraphics{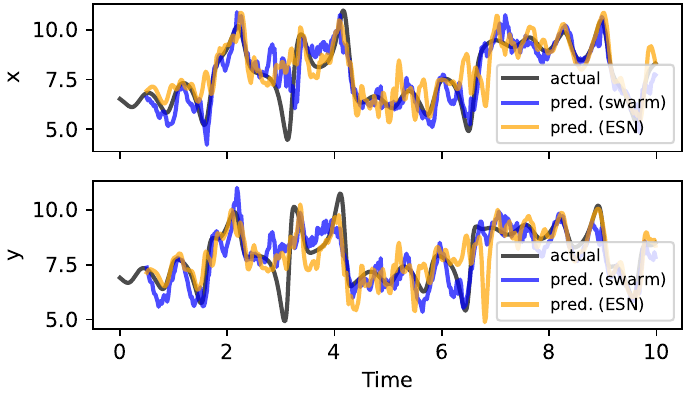}
    \caption{\new{\textbf{Predicted versus actual time series} for a near-critically damped active matter system with an inversely attractive driver (swarm, blue) (see also \figref[a]{\ref{fig:driver_attraction_speed_controller}}, pyramid symbol) and a simple Echo State Network with $N=600$ neurons and properties described in \refref{\cite{Gaimann2025}}, Appendix D (orange). }}
    \label{fig:time-series-comparison}
\end{figure*}
\section{Supplementary Tables}\label{ax:tables}
    \new{Columns:
\begin{itemize}
    \item PC ID: Parameter Combination Identifier (to access the raw dataset from the data repository \cite{DARUS-4805_2025}), and corresponding marker symbol used in the corresponding figure in the main text (see caption of respective table)
    \item $r_d$: driver interaction (cut-off) radius
    \item $K_d^{\mathrm{inv}}$: inverse driver repulsion (positive sign) or attraction (negative sign) strength
    \item $K_d^{\mathrm{lin}}$: linear driver attraction strength
    \item $K_{sc}$: speed-controller strength
    \item $s$: speed-controller target agent speed
    \item $K_h$: homing force strength
    \item $r_r$: agent–agent repulsion interaction (cut-off) radius
    \item $K_r$: agent–agent repulsion strength
    \item $N_a$: number of agents
    \item $\text{sMAPE}~(\%)$: Symmetric Mean Absolute Percentage Error
    \item $\text{NMSE}$: Normalized Mean Square Error
    \item $\text{NRMSE}$: Normalized Root Mean Square Error
    \item $P_y$: Predictive performance for the time series $y$ dimension (Pearson correlation coefficient)
    \item $P \equiv P_x$: Predictive performance for the time series $x$ dimension (Pearson correlation coefficient)\\
\end{itemize}
}


\begin{table}[ht]
  \centering
  \begin{tabular}{c c c c c c c c c c c c c c c}
    \toprule
    PC ID & $r_d$ & $K_{d}^{inv}$ & $K_{d}^{lin}$ & $K_{sc}$ & $s$ & $K_h$ & $r_r$ & $K_r$ & $N_a$ & $\text{sMAPE (\%)}$ & $\text{NMSE}$ & $\text{NRMSE}$ & $P_y$ & $P$ \\
    \midrule
    152 ($\nabla$) & 16 & -100 & --- & 0.264 & 0.00379 & --- & 1 & 2 & 200 & 15.2 & 1.14 & 1.06 & 0.0107 & 0.0314 \\
    190 (\Circle) & 16 & -100 & --- & 0.0483 & 0.0207 & --- & 1 & 2 & 200 & \textbf{6.32} & \textbf{0.286} & \textbf{0.535} & \textbf{0.798} & \textbf{0.882} \\
    209 ($\triangle$) & 16 & -100 & --- & 0.0207 & 0.0483 & --- & 1 & 2 & 200 & 6.72 & 0.306 & 0.553 & 0.795 & 0.878 \\
    228 ($\square$) & 16 & -100 & --- & 0.00886 & 0.113 & --- & 1 & 2 & 200 & 7.67 & 0.389 & 0.624 & 0.739 & 0.818 \\
    247 ($\lozenge$) & 16 & -100 & --- & 0.00379 & 0.264 & --- & 1 & 2 & 200 & 10.7 & 0.66 & 0.811 & 0.521 & 0.628 \\
    342 ($\times$) & 16 & -100 & --- & 5.46e-05 & 18.3 & --- & 1 & 2 & 200 & 14.7 & 1.01 & 1 & 0.0113 & 0.0585 \\
    \bottomrule
  \end{tabular}
  \caption{\new{Parameter combinations and performance metrics for the parameter scan shown in \figref[a]{\ref{fig:driver_attraction_speed_controller}}.}}
  \label{tab:2025-03-04-speed_controller_scan_negative_lymburn_driver_repulsion_100.0_inlay}
\end{table}


\begin{table}[ht]
  \centering
  \begin{tabular}{c c c c c c c c c c c c c c c}
    \toprule
    PC ID & $r_d$ & $K_{d}^{inv}$ & $K_{d}^{lin}$ & $K_{sc}$ & $s$ & $K_h$ & $r_r$ & $K_r$ & $N_a$ & $\text{sMAPE (\%)}$ & $\text{NMSE}$ & $\text{NRMSE}$ & $P_y$ & $P$ \\
    \midrule
    152 ($\nabla$) & --- & --- & 100 & 0.264 & 0.00379 & --- & 1 & 2 & 200 & 14 & 0.972 & 0.981 & 0.174 & 0.233 \\
    190 (\Circle) & --- & --- & 100 & 0.0483 & 0.0207 & --- & 1 & 2 & 200 & 11.6 & 0.984 & 0.988 & 0.511 & 0.546 \\
    209 ($\triangle$) & --- & --- & 100 & 0.0207 & 0.0483 & --- & 1 & 2 & 200 & 11 & \textbf{0.572} & \textbf{0.756} & \textbf{0.682} & \textbf{0.787} \\
    228 ($\square$) & --- & --- & 100 & 0.00886 & 0.113 & --- & 1 & 2 & 200 & \textbf{10.1} & 0.623 & 0.788 & 0.576 & 0.651 \\
    247 ($\lozenge$) & --- & --- & 100 & 0.00379 & 0.264 & --- & 1 & 2 & 200 & 13.2 & 0.893 & 0.941 & 0.288 & 0.364 \\
    342 ($\times$) & --- & --- & 100 & 5.46e-05 & 18.3 & --- & 1 & 2 & 200 & 14.7 & 1.01 & 1 & 0.00286 & 0.056 \\
    \bottomrule
  \end{tabular}
  \caption{\new{Parameter combinations and performance metrics for the parameter scan shown in \figref[b]{\ref{fig:driver_attraction_speed_controller}}.}}
  \label{tab:2025-03-04-speed_controller_scan_driver_attraction_100.0_inlay}
\end{table}


\begin{table}[ht]
  \centering
  \begin{tabular}{c c c c c c c c c c c c c c c}
    \toprule
    PC ID & $r_d$ & $K_{d}^{inv}$ & $K_{d}^{lin}$ & $K_{sc}$ & $s$ & $K_h$ & $r_r$ & $K_r$ & $N_a$ & $\text{sMAPE (\%)}$ & $\text{NMSE}$ & $\text{NRMSE}$ & $P_y$ & $P$ \\
    \midrule
    42 ($\times$) & 0.159 & -0.546 & --- & 0.0207 & 0.0483 & --- & 1 & 2 & 200 & 16.4 & 1.36 & 1.16 & 0.0191 & 0.0661 \\
    57 ($\square$) & 0.159 & -1.83e+05 & --- & 0.0207 & 0.0483 & --- & 1 & 2 & 200 & 15.6 & 1.24 & 1.11 & 0.0495 & 0.0513 \\
    268 ($+$) & 2 & -88.6 & --- & 0.0207 & 0.0483 & --- & 1 & 2 & 200 & 9.71 & 0.568 & 0.753 & 0.617 & 0.713 \\
    342 (\pentagon) & 5.04 & -0.546 & --- & 0.0207 & 0.0483 & --- & 1 & 2 & 200 & 13.2 & 0.938 & 0.967 & 0.301 & 0.443 \\
    357 (\Circle) & 5.04 & -1.83e+05 & --- & 0.0207 & 0.0483 & --- & 1 & 2 & 200 & 10.4 & 0.625 & 0.79 & 0.544 & 0.662 \\
    388 ($\triangle$) & 8 & -88.6 & --- & 0.0207 & 0.0483 & --- & 1 & 2 & 200 & \textbf{7.03} & \textbf{0.324} & \textbf{0.569} & \textbf{0.784} & \textbf{0.873} \\
    \bottomrule
  \end{tabular}
  \caption{\new{Parameter combinations and performance metrics for the parameter scan shown in \figref[a]{\ref{fig:driver_attraction}}.}}
  \label{tab:2025-08-18-nearcritdamped_200_agents_varied_inverse_driver_repulsion_inlay}
\end{table}


\begin{table}[ht]
  \centering
  \begin{tabular}{c c c c c c c c c c c c c c c}
    \toprule
    PC ID & $r_d$ & $K_{d}^{inv}$ & $K_{d}^{lin}$ & $K_{sc}$ & $s$ & $K_h$ & $r_r$ & $K_r$ & $N_a$ & $\text{sMAPE (\%)}$ & $\text{NMSE}$ & $\text{NRMSE}$ & $P_y$ & $P$ \\
    \midrule
    42 (\Circle) & 0.159 & 0.0695 & --- & 0.0207 & 0.0483 & 2 & 1 & 2 & 200 & 14.6 & 1 & 0.996 & 0.0136 & 0.0977 \\
    231 ($\triangle$) & 1.26 & 428 & --- & 0.0207 & 0.0483 & 2 & 1 & 2 & 200 & 6.75 & 0.314 & 0.56 & 0.783 & 0.862 \\
    315 ($\square$) & 3.18 & 2.07e+04 & --- & 0.0207 & 0.0483 & 2 & 1 & 2 & 200 & \textbf{6.52} & \textbf{0.287} & \textbf{0.535} & \textbf{0.791} & \textbf{0.887} \\
    357 (\pentagon) & 5.04 & 1.44e+05 & --- & 0.0207 & 0.0483 & 2 & 1 & 2 & 200 & 7 & 0.321 & 0.567 & 0.776 & 0.867 \\
    399 ($\times$) & 8 & 1e+06 & --- & 0.0207 & 0.0483 & 2 & 1 & 2 & 200 & 7.36 & 0.363 & 0.603 & 0.75 & 0.843 \\
    \bottomrule
  \end{tabular}
  \caption{\new{Parameter combinations and performance metrics for the parameter scan shown in \figref[a]{\ref{fig:nearcritdamped_driver_repulsion_scan_heatmap}}.}}
  \label{tab:2025-08-06-overdamped_varied_driver_repulsion_no_alignment_inlay}
\end{table}


\begin{table}[ht]
  \centering
  \begin{tabular}{c c c c c c c c c c c c c c c}
    \toprule
    PC ID & $r_d$ & $K_{d}^{inv}$ & $K_{d}^{lin}$ & $K_{sc}$ & $s$ & $K_h$ & $r_r$ & $K_r$ & $N_a$ & $\text{sMAPE (\%)}$ & $\text{NMSE}$ & $\text{NRMSE}$ & $P_y$ & $P$ \\
    \midrule
    42 (\Circle) & 2 & 100 & --- & 0.0207 & 0.0483 & 2 & 0.159 & 0.00336 & 200 & 10.2 & 0.638 & 0.797 & 0.577 & 0.652 \\
    57 ($\nabla$) & 2 & 100 & --- & 0.0207 & 0.0483 & 2 & 0.159 & 29.8 & 200 & 8.7 & 0.489 & 0.698 & 0.681 & 0.743 \\
    251 ($\triangle$) & 2 & 100 & --- & 0.0207 & 0.0483 & 2 & 1.59 & 0.785 & 200 & \textbf{5.88} & \textbf{0.248} & \textbf{0.497} & \textbf{0.818} & \textbf{0.904} \\
    314 ($\square$) & 2 & 100 & --- & 0.0207 & 0.0483 & 2 & 3.18 & 4.83 & 200 & 7.06 & 0.34 & 0.583 & 0.76 & 0.849 \\
    342 ($\times$) & 2 & 100 & --- & 0.0207 & 0.0483 & 2 & 5.04 & 0.00336 & 200 & 9.53 & 0.543 & 0.736 & 0.636 & 0.721 \\
    357 (\pentagon) & 2 & 100 & --- & 0.0207 & 0.0483 & 2 & 5.04 & 29.8 & 200 & 9.67 & 0.542 & 0.736 & 0.615 & 0.734 \\
    \bottomrule
  \end{tabular}
  \caption{\new{Parameter combinations and performance metrics for the parameter scan shown in \figref[a]{\ref{fig:repulsion}}.}}
  \label{tab:2025-08-15-nearcritdamped_varied_repulsion_no_alignment_inlay}
\end{table}


\begin{table}[ht]
  \centering
  \begin{tabular}{c c c c c c c c c c c c c c c}
    \toprule
    PC ID & $r_d$ & $K_{d}^{inv}$ & $K_{d}^{lin}$ & $K_{sc}$ & $s$ & $K_h$ & $r_r$ & $K_r$ & $N_a$ & $\text{sMAPE (\%)}$ & $\text{NMSE}$ & $\text{NRMSE}$ & $P_y$ & $P$ \\
    \midrule
    42 (\Circle) & 16 & -100 & --- & 0.0207 & 0.0483 & --- & 0.159 & 0.00336 & 200 & 8.4 & 0.447 & 0.669 & 0.692 & 0.805 \\
    57 ($\nabla$) & 16 & -100 & --- & 0.0207 & 0.0483 & --- & 0.159 & 29.8 & 200 & 9.13 & 0.524 & 0.723 & 0.638 & 0.731 \\
    251 ($\triangle$) & 16 & -100 & --- & 0.0207 & 0.0483 & --- & 1.59 & 0.785 & 200 & \textbf{6.83} & \textbf{0.323} & \textbf{0.568} & \textbf{0.785} & \textbf{0.863} \\
    314 ($\square$) & 16 & -100 & --- & 0.0207 & 0.0483 & --- & 3.18 & 4.83 & 200 & 8.09 & 0.415 & 0.644 & 0.728 & 0.822 \\
    342 ($\times$) & 16 & -100 & --- & 0.0207 & 0.0483 & --- & 5.04 & 0.00336 & 200 & 8.4 & 0.448 & 0.669 & 0.692 & 0.804 \\
    357 (\pentagon) & 16 & -100 & --- & 0.0207 & 0.0483 & --- & 5.04 & 29.8 & 200 & 10 & 0.597 & 0.772 & 0.584 & 0.676 \\
    \bottomrule
  \end{tabular}
  \caption{\new{Parameter combinations and performance metrics for the parameter scan shown in \figref[b]{\ref{fig:repulsion}}.}}
  \label{tab:2025-08-18-nearcritd_varied_rep_driver_attr_no_algnm_inf_range_inlay}
\end{table}


\begin{table}[ht]
  \centering
  \begin{tabular}{c c c c c c c c c c c c c c c}
    \toprule
    PC ID & $r_d$ & $K_{d}^{inv}$ & $K_{d}^{lin}$ & $K_{sc}$ & $s$ & $K_h$ & $r_r$ & $K_r$ & $N_a$ & $\text{sMAPE (\%)}$ & $\text{NMSE}$ & $\text{NRMSE}$ & $P_y$ & $P$ \\
    \midrule
    105 ($\times$) & 4 & 1e+03 & --- & 0.0207 & 0.0483 & 2 & 4 & 0.616 & 12 & 9.48 & 0.553 & 0.743 & 0.619 & 0.709 \\
    114 ($+$) & 4 & 1e+03 & --- & 0.0207 & 0.0483 & 2 & 4 & 16.2 & 12 & 9.11 & 0.516 & 0.718 & 0.64 & 0.738 \\
    225 ($\square$) & 4 & 1e+03 & --- & 0.0207 & 0.0483 & 2 & 4 & 0.616 & 100 & 6.78 & 0.309 & 0.556 & 0.781 & 0.875 \\
    234 (\pentagon) & 4 & 1e+03 & --- & 0.0207 & 0.0483 & 2 & 4 & 16.2 & 100 & 8.09 & 0.419 & 0.647 & 0.708 & 0.804 \\
    385 ($\triangle$) & 4 & 1e+03 & --- & 0.0207 & 0.0483 & 2 & 4 & 0.616 & 1000 & \textbf{5.48} & \textbf{0.227} & \textbf{0.476} & \textbf{0.838} & \textbf{0.916} \\
    394 (\Circle) & 4 & 1e+03 & --- & 0.0207 & 0.0483 & 2 & 4 & 16.2 & 1000 & 7.33 & 0.344 & 0.586 & 0.742 & 0.861 \\
    \bottomrule
  \end{tabular}
  \caption{\new{Parameter combinations and performance metrics for the parameter scan shown in \figref[a]{\ref{fig:n_agents_repulsion_strength}}.}}
  \label{tab:2025-03-03-ring_study_n_agents_repulsion_strength_inlay}
\end{table}


\begin{table}[ht]
  \centering
  \begin{tabular}{c c c c c c c c c c c c c c c}
    \toprule
    PC ID & $r_d$ & $K_{d}^{inv}$ & $K_{d}^{lin}$ & $K_{sc}$ & $s$ & $K_h$ & $r_r$ & $K_r$ & $N_a$ & $\text{sMAPE (\%)}$ & $\text{NMSE}$ & $\text{NRMSE}$ & $P_y$ & $P$ \\
    \midrule
    105 ($\times$) & 16 & -100 & --- & 0.0207 & 0.0483 & --- & 4 & 0.616 & 12 & 10 & 0.601 & 0.775 & 0.579 & 0.694 \\
    114 ($+$) & 16 & -100 & --- & 0.0207 & 0.0483 & --- & 4 & 16.2 & 12 & 9.59 & 0.562 & 0.749 & 0.6 & 0.709 \\
    225 ($\square$) & 16 & -100 & --- & 0.0207 & 0.0483 & --- & 4 & 0.616 & 100 & \textbf{7.51} & \textbf{0.382} & \textbf{0.618} & 0.742 & 0.826 \\
    234 (\pentagon) & 16 & -100 & --- & 0.0207 & 0.0483 & --- & 4 & 16.2 & 100 & 8.95 & 0.499 & 0.706 & 0.651 & 0.756 \\
    385 ($\triangle$) & 16 & -100 & --- & 0.0207 & 0.0483 & --- & 4 & 0.616 & 1000 & 8.45 & 0.432 & 0.657 & \textbf{0.745} & \textbf{0.84} \\
    394 (\Circle) & 16 & -100 & --- & 0.0207 & 0.0483 & --- & 4 & 16.2 & 1000 & 9.37 & 0.515 & 0.717 & 0.624 & 0.751 \\
    \bottomrule
  \end{tabular}
  \caption{\new{Parameter combinations and performance metrics for the parameter scan shown in \figref[b]{\ref{fig:n_agents_repulsion_strength}}.}}
  \label{tab:2025-08-22-inverse_ring_study_n_agents_repulsion_strength_r_r_4.0_Kd_100_inlay}
\end{table}

\section{\label{ax:supplementary_videos} Supplementary Videos}

All supplementary videos can be accessed in the Data Repository of the University of Stuttgart (DaRUS), in the dataverse Stuttgart Center for Simulation Science EXC 2075 / Project 6-15:\\
\href{https://doi.org/10.18419/DARUS-4806}{\texttt{https://doi.org/10.18419/DARUS-4806}} \cite{DARUS-4806_2025}\\
Snapshots show the simulated soft matter systems at $t = 3.3$ (after 165 integration time steps of $\Delta t = 0.02$) or at $t = 1.8$ (after 90 integration time steps; for driver attraction/repulsion parameter scans; for the bulk systems; for the undriven system only) if not mentioned otherwise.\\

\begin{table}[H]
    \small
    \begin{center}
    \begin{tabular}{|>{\centering\arraybackslash}p{\vidTableIdColumnWidth}|>{\centering\arraybackslash}p{\vidTableSnapshotColumnWidth}|>{\centering\arraybackslash}p{\vidTableDescriptionColumnWidth}|}
    \hline
    \textbf{ID} & \textbf{Snapshot} & \textbf{Description} \\ 
    \hline
    \hline
    
    {1} & \includegraphics{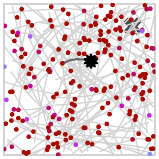} & Speed-controller scan, inverse driver attraction (cross symbol) \\ 
    \hline
    {2} & \includegraphics{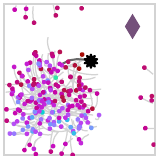} & Speed-controller scan, inverse driver attraction (diamond symbol) \\ 
    \hline
    {3} & \includegraphics{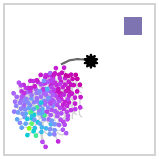} & Speed-controller scan, inverse driver attraction (square symbol) \\ 
    \hline
    {4} & \includegraphics{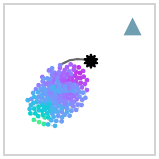} & Speed-controller scan, inverse driver attraction (pyramid symbol) \\ 
    \hline
    {5} & \includegraphics{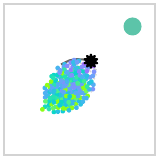} & Speed-controller scan, inverse driver attraction (circle symbol) \\ 
    \hline
    {6} & \includegraphics{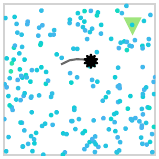} & Speed-controller scan, inverse driver attraction (nabla symbol) \\ 
    \hline
    \end{tabular}
    \end{center}
    
    \caption{ \centering Supplementary videos for the speed-controller parameter scan with an inverse driver attraction in \figref[a]{\ref{fig:driver_attraction_speed_controller}}. \newline Label in repository: \texttt{speed-controller-driver-attraction-inverse}.}
    \label{tab:supplementary_videos_driver_attraction_inverse}
\end{table}

\begin{table}[H]
    \small
    \begin{center}
    \begin{tabular}{|>{\centering\arraybackslash}p{\vidTableIdColumnWidth}|>{\centering\arraybackslash}p{\vidTableSnapshotColumnWidth}|>{\centering\arraybackslash}p{\vidTableDescriptionColumnWidth}|}
    \hline
    \textbf{ID} & \textbf{Snapshot} & \textbf{Description} \\ 
    \hline
    \hline
    
    {1} & \includegraphics{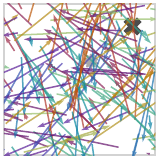} & Speed-controller scan, inverse driver attraction, velocity fluctuations (cross symbol) \\ 
    \hline
    {2} & \includegraphics{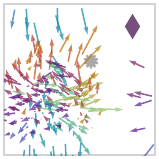} & Speed-controller scan, inverse driver attraction, velocity fluctuations (diamond symbol) \\ 
    \hline
    {3} & \includegraphics{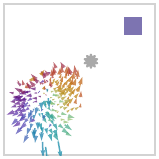} & Speed-controller scan, inverse driver attraction, velocity fluctuations (square symbol) \\ 
    \hline
    {4} & \includegraphics{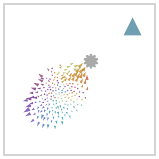} & Speed-controller scan, inverse driver attraction, velocity fluctuations (pyramid symbol) \\ 
    \hline
    {5} & \includegraphics{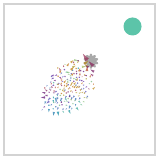} & Speed-controller scan, inverse driver attraction, velocity fluctuations (circle symbol) \\ 
    \hline
    {6} & \includegraphics{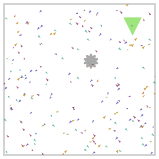} & Speed-controller scan, inverse driver attraction, velocity fluctuations (nabla symbol) \\ 
    \hline
    \end{tabular}
    \end{center}
    
    \caption{ \centering Supplementary videos for visualizing agent velocity fluctuations for the speed-controller parameter scan with inverse driver attraction in \figref[e]{\ref{fig:driver_attraction_correlations}}.  \newline Label in repository: \texttt{speed-controller-driver-attraction-inverse-fluctuations}.}
    \label{tab:supplementary_videos_driver_attraction_inverse_fluctuations}
\end{table}

\begin{table}[H]
    \small
    \begin{center}
    \begin{tabular}{|>{\centering\arraybackslash}p{\vidTableIdColumnWidth}|>{\centering\arraybackslash}p{\vidTableSnapshotColumnWidth}|>{\centering\arraybackslash}p{\vidTableDescriptionColumnWidth}|}
    \hline
    \textbf{ID} & \textbf{Snapshot} & \textbf{Description} \\ 
    \hline
    \hline
    
    {1} & \includegraphics{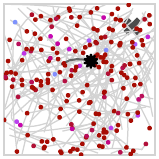} & Speed-controller scan, linear driver attraction (cross symbol) \\ 
    \hline
    {2} & \includegraphics{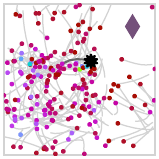} & Speed-controller scan, linear driver attraction (diamond symbol) \\ 
    \hline
    {3} & \includegraphics{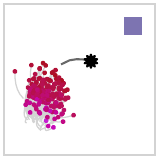} & Speed-controller scan, linear driver attraction (square symbol) \\ 
    \hline
    {4} & \includegraphics{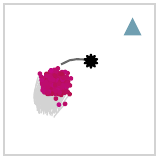} & Speed-controller scan, linear driver attraction (pyramid symbol) \\ 
    \hline
    {5} & \includegraphics{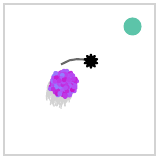} & Speed-controller scan, linear driver attraction (circle symbol) \\ 
    \hline
    {6} & \includegraphics{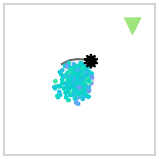} & Speed-controller scan, linear driver attraction (nabla symbol) \\ 
    \hline
    \end{tabular}
    \end{center}
    
    \caption{ \centering Supplementary videos for the speed-controller scan with a linear driver attraction in \figref[b]{\ref{fig:driver_attraction_speed_controller}}. \newline Label in repository: \texttt{speed-controller-driver-attraction-linear}.}
    \label{tab:supplementary_videos_driver_attraction_linear}
\end{table}

\begin{table}[H]
    \small
    \begin{center}
    \begin{tabular}{|>{\centering\arraybackslash}p{\vidTableIdColumnWidth}|>{\centering\arraybackslash}p{\vidTableSnapshotColumnWidth}|>{\centering\arraybackslash}p{\vidTableDescriptionColumnWidth}|}
    \hline
    \textbf{ID} & \textbf{Snapshot} & \textbf{Description} \\ 
    \hline
    \hline
    
    {1} & \includegraphics{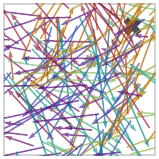} & Speed-controller scan, linear driver attraction, velocity fluctuations (cross symbol) \\ 
    \hline
    {2} & \includegraphics{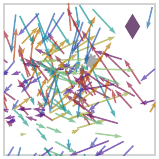} & Speed-controller scan, linear driver attraction, velocity fluctuations (diamond symbol) \\ 
    \hline
    {3} & \includegraphics{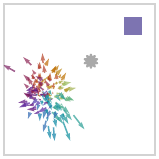} & Speed-controller scan, linear driver attraction, velocity fluctuations (square symbol) \\ 
    \hline
    {4} & \includegraphics{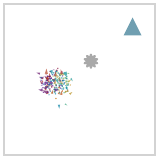} & Speed-controller scan, linear driver attraction, velocity fluctuations (pyramid symbol) \\ 
    \hline
    {5} & \includegraphics{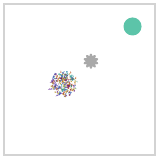} & Speed-controller scan, linear driver attraction, velocity fluctuations (circle symbol) \\ 
    \hline
    {6} & \includegraphics{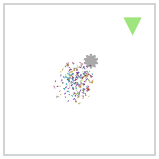} & Speed-controller scan, linear driver attraction, velocity fluctuations (nabla symbol) \\ 
    \hline
    \end{tabular}
    \end{center}
    
    \caption{ \centering Supplementary videos for visualizing agent velocity fluctuations for the speed-controller parameter scan with linear driver attraction in \figref[f]{\ref{fig:driver_attraction_correlations}}. \newline Label in repository: \texttt{speed-controller-driver-attraction-linear-fluctuations}.}
    \label{tab:supplementary_videos_driver_attraction_linear_flucts}
\end{table}

\begin{table}[H]
    \small
    \begin{center}
    \begin{tabular}{|>{\centering\arraybackslash}p{\vidTableIdColumnWidth}|>{\centering\arraybackslash}p{\vidTableSnapshotColumnWidth}|>{\centering\arraybackslash}p{\vidTableDescriptionColumnWidth}|}
    \hline
    \textbf{ID} & \textbf{Snapshot} & \textbf{Description} \\ 
    \hline
    \hline
    
    {1} & \includegraphics{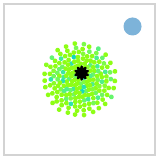} & Driver repulsion scan (circle symbol)\\
    \hline
    {2} & \includegraphics{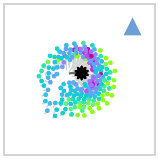} & Driver repulsion scan (pyramid symbol)\\ 
    \hline
    {3} & \includegraphics{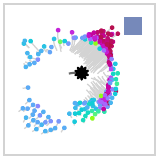} & Driver repulsion scan (square symbol)\\ 
    \hline
    {4} & \includegraphics{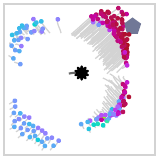} & Driver repulsion scan (pentagon symbol)\\ 
    \hline
    {5} & \includegraphics{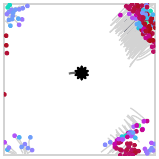} & Driver repulsion scan (cross symbol)\\ 
    \hline
    \end{tabular}
    \end{center}
    
    \caption{
    \centering Supplementary videos for default driver repulsion parameter scans using near-critically damped speed-controller settings in \figref{\ref{fig:Lymburn_critical_driver_repulsion_scan_heatmap}}. \newline Label in repository: \texttt{driver-repulsion-nearcritdamped}.
    }
   \label{tab:supplementary_videos_driver_repulsion_nearcritdamped}
\end{table}

\vspace{1cm}

\begin{table}[H]
    \small
    \begin{center}
    \begin{tabular}{|>{\centering\arraybackslash}p{\vidTableIdColumnWidth}|>{\centering\arraybackslash}p{\vidTableSnapshotColumnWidth}|>{\centering\arraybackslash}p{\vidTableDescriptionColumnWidth}|}
    \hline
    \textbf{ID} & \textbf{Snapshot} & \textbf{Description} \\ 
    \hline
    \hline
    
    {1} & \includegraphics{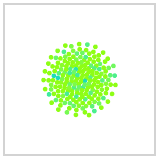} & Undriven system, near-critically damped\\

    \hline
    \end{tabular}
    \end{center}
    
    \caption{
    \centering Supplementary video for the undriven ``ground state'' using a near-critically damped speed-controller setting, serving as baseline for \figref{\ref{fig:Lymburn_critical_driver_repulsion_scan_heatmap}}. \newline Label in repository: \texttt{undriven-nearcritdamped}.
    }
   \label{tab:supplementary_videos_undriven}
\end{table}

\begin{table}[H]
    \small
    \begin{center}
    \begin{tabular}{|>{\centering\arraybackslash}p{\vidTableIdColumnWidth}|>{\centering\arraybackslash}p{\vidTableSnapshotColumnWidth}|>{\centering\arraybackslash}p{\vidTableDescriptionColumnWidth}|}
    \hline
    \textbf{ID} & \textbf{Snapshot} & \textbf{Description} \\ 
    \hline
    \hline
    
    {1} & \includegraphics{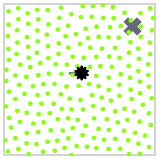} & Inverse driver attraction scan (cross symbol)\\
    \hline
    {2} & \includegraphics{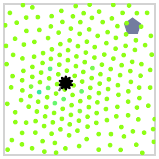} & Inverse driver attraction scan (pentagon symbol)\\ 
    \hline
    {3} & \includegraphics{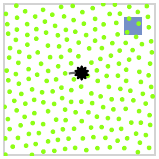} & Inverse driver attraction scan (square symbol)\\ 
    \hline
    {4} & \includegraphics{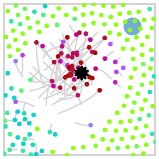} & Inverse driver attraction scan (circle symbol)\\ 
    \hline
    {5} & \includegraphics{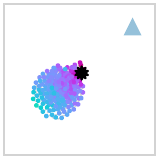} & Inverse driver attraction scan (pyramid symbol)\\ 
    \hline
    {6} & \includegraphics{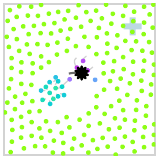} & Inverse driver attraction scan (plus symbol)\\ 
    \hline
    \end{tabular}
    \end{center}
    
    \caption{
    \centering Supplementary videos for parameter scans for the inverse driver attraction force using near-critically damped speed-controller settings in \figref{\ref{fig:driver_attraction}}. \newline Label in repository: \texttt{driver-attraction-inverse-nearcritdamped}.
    }
\label{tab:supplementary_videos_driver_attraction_nearcritdamped}
\end{table}

\begin{table}[H]
    \small
    \begin{center}
    \begin{tabular}{|>{\centering\arraybackslash}p{\vidTableIdColumnWidth}|>{\centering\arraybackslash}p{\vidTableSnapshotColumnWidth}|>{\centering\arraybackslash}p{\vidTableDescriptionColumnWidth}|}
    \hline
    \textbf{ID} & \textbf{Snapshot} & \textbf{Description} \\ 
    \hline
    \hline
    
    {1} & \includegraphics{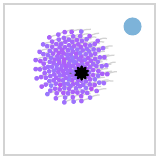} & Driver rep.\ scan, Lymburn-damped (circle symbol)\\
    \hline
    {2} & \includegraphics{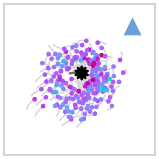} & Driver rep.\ scan, Lymburn-damped (pyramid symbol)\\ 
    \hline
    {3} & \includegraphics{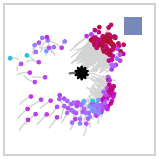} & Driver rep.\ scan, Lymburn-damped (square symbol)\\ 
    \hline
    {4} & \includegraphics{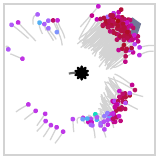} & Driver rep.\ scan, Lymburn-damped (pentagon symbol)\\ 
    \hline
    {5} & \includegraphics{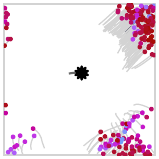} & Driver rep.\ scan, Lymburn-damped (cross symbol)\\ 
    \hline
    {6} & \includegraphics{img_dump/2024-09-30-Lymburn_critical_varied_driver_repulsion_large_box_pc_399_movie_test_frame_90.pdf
    } & Driver rep.\ scan, Lymburn-damped, $l_{\text{box}}^{\text{sim}} = 32.0$, $l_{\text{box}}^{\text{obs}} = 16.0$ (cross symbol)\\ 
    \hline
    {7} & \includegraphics{img_dump/2024-09-30-Lymburn_critical_varied_driver_repulsion_large_box_large_kernels_pc_399_movie_test_frame_90.pdf} & Driver rep.\ scan, Lymburn-damped, $l_{\text{box}}^{\text{sim}} = 32.0$, $l_{\text{box}}^{\text{obs}} = 32.0$ (cross symbol)\\ 
    \hline
    {8} & \includegraphics{img_dump/2024-10-21-Lymburn_critical_varied_driver_repulsion_LARGE_box_LARGE_kernels_pc_399_movie_test_frame_90.pdf} & Driver rep.\ scan, Lymburn-damped, $l_{\text{box}}^{\text{sim}} = 64.0$, $l_{\text{box}}^{\text{obs}} = 64.0$ (cross symbol)\\ 
    \hline
    \end{tabular}
    \end{center}
    
    \caption{
    \centering Supplementary videos for default driver repulsion parameter scans using ``critical'' ($K_{sc} = 2.0; s = 10.0$) speed-controller settings of Lymburn \etal \cite{Lymburn2021} in \figref{\ref{fig:Lymburn_critical_driver_repulsion_scan_heatmap}}. Label in repository: \texttt{driver-repulsion-lymburndamped}.
    }
   \label{tab:supplementary_videos_driver_repulsion}
\end{table}

\begin{table}[H]
    \small
    \begin{center}
    \begin{tabular}{|>{\centering\arraybackslash}p{\vidTableIdColumnWidth}|>{\centering\arraybackslash}p{\vidTableSnapshotColumnWidth}|>{\centering\arraybackslash}p{\vidTableDescriptionColumnWidth}|}
    \hline
    \textbf{ID} & \textbf{Snapshot} & \textbf{Description} \\ 
    \hline
    \hline
    {1} & \includegraphics{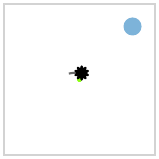} & Inverse driver attraction scan, single agent (circle symbol)\\
    \hline
    {2} & \includegraphics{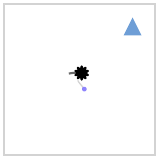} & Inverse driver attraction scan, single agent (pyramid symbol)\\ 
    \hline
    {3} & \includegraphics{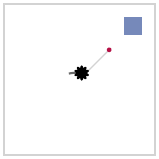} & Inverse driver attraction scan, single agent (square symbol)\\ 
    \hline
    {4} & \includegraphics{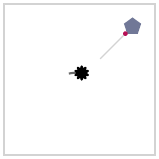} & Inverse driver attraction scan, single agent (pentagon symbol)\\ 
    \hline
    {5} & \includegraphics{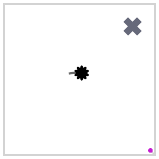} & Inverse driver attraction scan, single agent (cross symbol)\\ 
    \hline
    {6} & \includegraphics{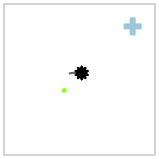} & Inverse driver attraction scan, single agent (plus symbol)\\
    \hline
    {7} & \includegraphics{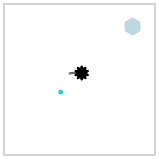} & Inverse driver attraction scan, single agent (hexagon symbol)\\ 
    \hline
    \end{tabular}
    \end{center}
    
    \caption{
    \centering Supplementary videos for the inverse driver attraction parameter scan with a single agent using near-critically damped speed-controller settings in \figref[c]{\ref{fig:few_particles}}. \newline Label in repository: \texttt{driver-attraction-inverse-nearcritdamped-single-agent}.
    }
   \label{tab:supplementary_videos_driver_attraction_inverse_nearcritdamped_1_agent}
\end{table}

\begin{table}[H]
    \small
    \begin{center}
    \begin{tabular}{|>{\centering\arraybackslash}p{\vidTableIdColumnWidth}|>{\centering\arraybackslash}p{\vidTableSnapshotColumnWidth}|>{\centering\arraybackslash}p{\vidTableDescriptionColumnWidth}|}
    \hline
    \textbf{ID} & \textbf{Snapshot} & \textbf{Description} \\ 
    \hline
    \hline
    
    {1} & \includegraphics{img_dump/2024-10-21-single_agent_overdamped_varied_driver_repulsion_pc_42_movie_test_frame_90.pdf} & Driver repulsion scan, single agent (circle symbol)\\
    \hline
    {2} & \includegraphics{img_dump/2024-10-21-single_agent_overdamped_varied_driver_repulsion_pc_231_movie_test_frame_90.pdf} & Driver repulsion scan, single agent (pyramid symbol)\\ 
    \hline
    {3} & \includegraphics{img_dump/2024-10-21-single_agent_overdamped_varied_driver_repulsion_pc_315_movie_test_frame_90.pdf} & Driver repulsion scan, single agent (square symbol)\\ 
    \hline
    {4} & \includegraphics{img_dump/2024-10-21-single_agent_overdamped_varied_driver_repulsion_pc_357_movie_test_frame_90.pdf} & Driver repulsion scan, single agent (pentagon symbol)\\ 
    \hline
    {5} & \includegraphics{img_dump/2024-10-21-single_agent_overdamped_varied_driver_repulsion_pc_399_movie_test_frame_90.pdf} & Driver repulsion scan, single agent (cross symbol)\\ 
    \hline
    {6} & \includegraphics{img_dump/2025-05-14-nearcritdamped_single_agent_varied_inverse_driver_repulsion_pc_263_movie_test_frame_90.pdf} & Driver repulsion scan, single agent (plus symbol)\\
    \hline
    {7} & \includegraphics{img_dump/2025-05-14-nearcritdamped_single_agent_varied_inverse_driver_repulsion_pc_325_movie_test_frame_90.pdf} & Driver repulsion scan, single agent (hexagon symbol)\\ 
    \hline
    \end{tabular}
    \end{center}
    
    \caption{
    \centering Supplementary videos for default driver repulsion parameter scans with a single agent using near-critically damped speed-controller settings in \figref{\ref{fig:Lymburn_critical_driver_repulsion_scan_heatmap}}. \newline Label in repository: \texttt{driver-repulsion-nearcritdamped-single-agent}.
    }
   \label{tab:supplementary_videos_driver_repulsion_nearcritdamped_1_agent}
\end{table}

\begin{table}[H]
    \small
    \begin{center}
    \begin{tabular}{|>{\centering\arraybackslash}p{\vidTableIdColumnWidth}|>{\centering\arraybackslash}p{\vidTableSnapshotColumnWidth}|>{\centering\arraybackslash}p{\vidTableDescriptionColumnWidth}|}
    \hline
    \textbf{ID} & \textbf{Snapshot} & \textbf{Description} \\ 
    \hline
    \hline
    
    {1} & \includegraphics{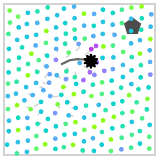} & Agent-agent rep., repulsive driver (pentagon symbol)\\ 
    \hline
    {2} & \includegraphics{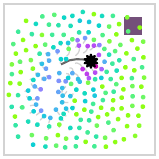} & Agent-agent rep., repulsive driver (square symbol)\\ 
    \hline
    {3} & \includegraphics{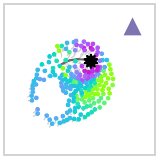} & Agent-agent rep., repulsive driver (pyramid symbol)\\ 
    \hline
    {4} & \includegraphics{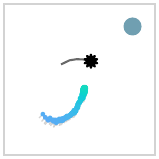} & Agent-agent rep., repulsive driver (circle symbol)\\ 
    \hline
    {5} & \includegraphics{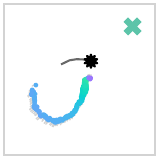} & Agent-agent rep., repulsive driver (cross symbol)\\ 
    \hline
    {6} & \includegraphics{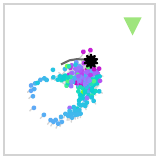} & Agent-agent rep., repulsive driver (nabla symbol)\\ 
    \hline
    \end{tabular}
    \end{center}
    
    \caption{ \centering Supplementary videos for agent-agent repulsion parameter scans for a repulsive driver in \figref{\ref{fig:repulsion}}. \newline Label in repository: \texttt{agent-repulsion-driver-repulsion}.}
    \label{tab:supplementary_videos_repulsion_driver_rep}
\end{table}

\begin{table}[H]
    \small
    \begin{center}
    \begin{tabular}{|>{\centering\arraybackslash}p{\vidTableIdColumnWidth}|>{\centering\arraybackslash}p{\vidTableSnapshotColumnWidth}|>{\centering\arraybackslash}p{\vidTableDescriptionColumnWidth}|}
    \hline
    \textbf{ID} & \textbf{Snapshot} & \textbf{Description} \\ 
    \hline
    \hline
    
      {1} & \includegraphics{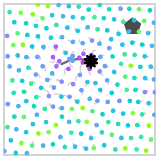} & Agent-agent rep., attractive driver (circle symbol)\\ 
    \hline
    {2} & \includegraphics{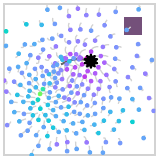} & Agent-agent rep., attractive driver (pyramid symbol)\\ 
    \hline
    {3} & \includegraphics{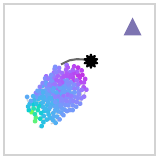} & Agent-agent rep., attractive driver (square symbol)\\ 
    \hline
    {4} & \includegraphics{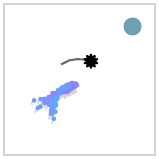} & Agent-agent rep., attractive driver (pentagon symbol)\\ 
    \hline
    {5} & \includegraphics{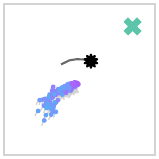} & Agent-agent rep., attractive driver (cross symbol)\\ 
    \hline
    {6} & \includegraphics{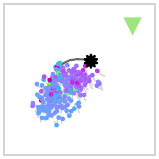} & Agent-agent rep., attractive driver (nabla symbol)\\ 
    \hline
    \end{tabular}
    \end{center}

    \caption{ \centering Supplementary videos for agent-agent repulsion parameter scans for an inversely attractive driver in \figref{\ref{fig:repulsion}}. \newline Label in repository: \texttt{agent-repulsion-driver-attraction}.}
    \label{tab:supplementary_videos_repulsion_driver_attr}
\end{table}

\begin{table}[H]
    \small
    \begin{center}
    \begin{tabular}{|>{\centering\arraybackslash}p{\vidTableIdColumnWidth}|>{\centering\arraybackslash}p{\vidTableSnapshotColumnWidth}|>{\centering\arraybackslash}p{\vidTableDescriptionColumnWidth}|}
    \hline
    \textbf{ID} & \textbf{Snapshot} & \textbf{Description} \\ 
    \hline
    \hline
    {1} & \includegraphics{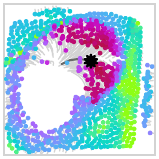} & $N_{\text{agents}}$ versus agent repulsion strength, repulsive driver (pyramid symbol)\\
    \hline
    {2} & \includegraphics{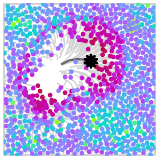} & $N_{\text{agents}}$ versus agent repulsion strength, repulsive driver (circle symbol)\\ 
    \hline
    {3} & \includegraphics{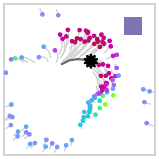} & $N_{\text{agents}}$ versus agent repulsion strength, repulsive driver  (square symbol)\\ 
    \hline
    {4} & \includegraphics{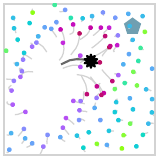} & $N_{\text{agents}}$ versus agent repulsion strength, repulsive driver (pentagon symbol)\\ 
    \hline
    {5} & \includegraphics{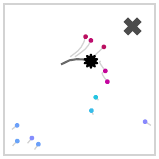} & $N_{\text{agents}}$ versus agent repulsion strength, repulsive driver (cross symbol)\\ 
    \hline
    {6} & \includegraphics{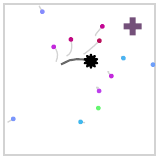} & $N_{\text{agents}}$ versus agent repulsion strength, repulsive driver (plus symbol)\\
    \hline
    \end{tabular}
    \end{center}
    
    \caption{
    \centering Supplementary videos for the number of agents versus agent-agent repulsion strength scan ($r_r = 4.0$) for a repulsive driver ($K_d = 100.0$) in \figref[a,c]{\ref{fig:n_agents_repulsion_strength}}. \newline Label in repository: \texttt{agent-repulsion-nagents-driver-repulsion}.
    }
   \label{tab:supplementary_videos_repulsion-nagents-driver-repulsion}
\end{table}

\begin{table}[H]
    \small
    \begin{center}
    \begin{tabular}{|>{\centering\arraybackslash}p{\vidTableIdColumnWidth}|>{\centering\arraybackslash}p{\vidTableSnapshotColumnWidth}|>{\centering\arraybackslash}p{\vidTableDescriptionColumnWidth}|}
    \hline
    \textbf{ID} & \textbf{Snapshot} & \textbf{Description} \\ 
    \hline
    \hline

    {1} & \includegraphics{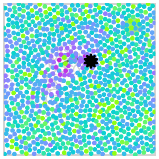}  & $N_{\text{agents}}$ versus $K_r$, attractive driver ($r_r = 4.0$, $K_d = 100.0$) (pyramid symbol)\\ 
    \hline
    {2} & \includegraphics{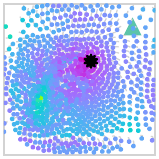}  & $N_{\text{agents}}$ versus $K_r$, attractive driver ($r_r = 4.0$, $K_d = 100.0$) (circle symbol)\\ 
    \hline
    {3} & \includegraphics{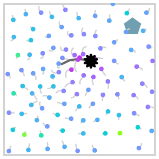}& $N_{\text{agents}}$ versus $K_r$, attractive driver ($r_r = 4.0$, $K_d = 100.0$) (pentagon symbol)\\ 
    \hline
    {4} & \includegraphics{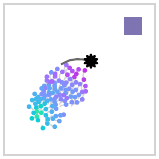}  & $N_{\text{agents}}$ versus $K_r$, attractive driver ($r_r = 4.0$, $K_d = 100.0$) (square symbol)\\ 
    \hline
    {5} & \includegraphics{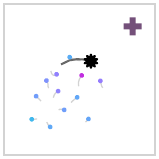} & $N_{\text{agents}}$ versus $K_r$, attractive driver ($r_r = 4.0$, $K_d = 100.0$) (cross symbol)\\ 
    \hline
    {6} & \includegraphics{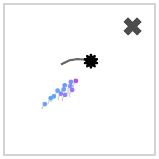} & $N_{\text{agents}}$ versus $K_r$, attractive driver ($r_r = 4.0$, $K_d = 100.0$) (plus symbol)\\ 
    \hline
    \end{tabular}
    \end{center}
    
    \caption{
    \centering Supplementary videos for the number of agents versus agent-agent repulsion strength scan ($r_r = 4.0$) for an inversely attractive driver ($K_d = 100.0$) in \figref[b,d]{\ref{fig:n_agents_repulsion_strength}}. \newline Label in repository: \texttt{agent-repulsion-nagents-driver-attraction}.
    }
   \label{tab:supplementary_videos_repulsion-nagents-driver-attraction}
\end{table}

\begin{table}[H]
    \small
    \begin{center}
    \begin{tabular}{|>{\centering\arraybackslash}p{\vidTableIdColumnWidth}|>{\centering\arraybackslash}p{\vidTableSnapshotColumnWidth}|>{\centering\arraybackslash}p{\vidTableDescriptionColumnWidth}|}
    \hline
    \textbf{ID} & \textbf{Snapshot} & \textbf{Description} \\ 
    \hline
    \hline
    
    {1} & \includegraphics{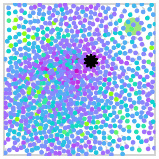}  & $N_{\text{agents}}$ versus $K_r$, attractive driver ($r_r = 1.0$, $K_d = 100.0$) (pyramid symbol)\\ 
    \hline
    {2} & \includegraphics{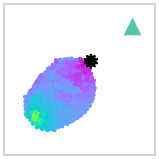}  & $N_{\text{agents}}$ versus $K_r$, attractive driver ($r_r = 1.0$, $K_d = 100.0$) (circle symbol)\\ 
    \hline
    {3} & \includegraphics{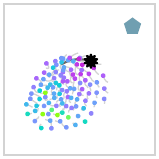}& $N_{\text{agents}}$ versus $K_r$, attractive driver ($r_r = 1.0$, $K_d = 100.0$) (pentagon symbol)\\ 
    \hline
    {4} & \includegraphics{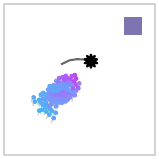}  & $N_{\text{agents}}$ versus $K_r$, attractive driver ($r_r = 1.0$, $K_d = 100.0$) (square symbol)\\ 
    \hline
    {5} & \includegraphics{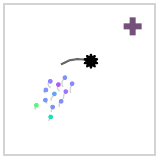} & $N_{\text{agents}}$ versus $K_r$, attractive driver ($r_r = 1.0$, $K_d = 100.0$) (cross symbol)\\ 
    \hline
    {6} & \includegraphics{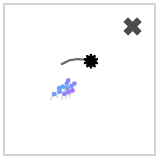} & $N_{\text{agents}}$ versus $K_r$, attractive driver ($r_r = 1.0$, $K_d = 100.0$) (plus symbol)\\ 
    \hline
    \end{tabular}
    \end{center}
    
    \caption{ \centering Supplementary videos for agent-agent repulsion parameter scans ($r_r = 1.0$) for an inversely attractive driver with a driver attraction strength of $K_d = 100.0$ in \figref[a,c]{\ref{fig:n_agents_repulsion_strength_supplement}}. \newline Label in repository: \texttt{agent-repulsion-nagents-driver-attraction-r\_r-1e0-K\_d-1e2}.}
    \label{tab:supplementary_videos_repulsion_nagents_driver_attraction_rr_1e0_Kd_1e2}
\end{table}

\begin{table}[H]
    \small
    \begin{center}
    \begin{tabular}{|>{\centering\arraybackslash}p{\vidTableIdColumnWidth}|>{\centering\arraybackslash}p{\vidTableSnapshotColumnWidth}|>{\centering\arraybackslash}p{\vidTableDescriptionColumnWidth}|}
    \hline
    \textbf{ID} & \textbf{Snapshot} & \textbf{Description} \\ 
    \hline
    \hline

    {1} & \includegraphics{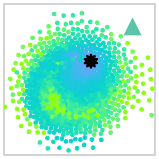} & $N_{\text{agents}}$ versus $K_r$, attractive driver ($r_r = 1.0$, $K_d \approx 11.3$) (pyramid symbol)\\
    \hline
    {2} & \includegraphics{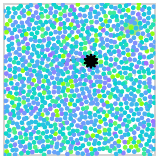} & $N_{\text{agents}}$ versus $K_r$, attractive driver ($r_r = 1.0$, $K_d \approx 11.3$) (circle symbol)\\ 
    \hline
    {3} & \includegraphics{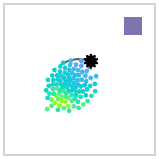} & $N_{\text{agents}}$ versus $K_r$, attractive driver ($r_r = 1.0$, $K_d \approx 11.3$) (square symbol)\\ 
    \hline
    {4} & \includegraphics{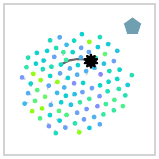} & $N_{\text{agents}}$ versus $K_r$, attractive driver ($r_r = 1.0$, $K_d \approx 11.3$) (pentagon symbol)\\ 
    \hline
    {5} & \includegraphics{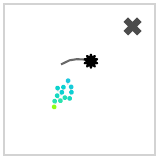} & $N_{\text{agents}}$ versus $K_r$, attractive driver ($r_r = 1.0$, $K_d \approx 11.3$) (cross symbol)\\ 
    \hline
    {6} & \includegraphics{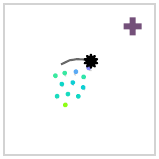} & $N_{\text{agents}}$ versus $K_r$, attractive driver ($r_r = 1.0$, $K_d \approx 11.3$) (plus symbol)\\
    \hline
    \end{tabular}
    \end{center}
    
    \caption{ \centering Supplementary videos for agent-agent repulsion parameter scans ($r_r = 1.0$) for an inversely attractive driver with a driver attraction strength of $K_d = 11.288378$ in \figref[b,d]{\ref{fig:n_agents_repulsion_strength_supplement}}. \newline Label in repository: \texttt{agent-repulsion-nagents-driver-attraction-r\_r-1e0-K\_d-1e1}.}
    \label{tab:supplementary_videos_repulsion_nagents_driver_attraction_rr_1e0_Kd_1e1}
\end{table}

\begin{table}[H]
    \small
    \begin{center}
    \begin{tabular}{|>{\centering\arraybackslash}p{\vidTableIdColumnWidth}|>{\centering\arraybackslash}p{\vidTableSnapshotColumnWidth}|>{\centering\arraybackslash}p{\vidTableDescriptionColumnWidth}|}
    \hline
    \textbf{ID} & \textbf{Snapshot} & \textbf{Description} \\ 
    \hline
    \hline

    {1} & \includegraphics{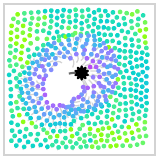} & Viscoelastic fluid, weaker driver repulsion ($K_d = 10^2$)\\ 
    \hline
    {2} & \includegraphics{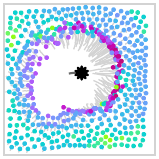} & Viscoelastic fluid, stronger driver repulsion ($K_d = 10^3$) \\ 
    \hline
    \end{tabular}
    \end{center}

    \caption{ \centering Supplementary videos for the viscoelastic fluids with different driver repulsion strengths shown in \figref{\ref{fig:viscoelastic_fluids}}.\newline Label in repository: \texttt{viscoelastic-fluids}.}
    \label{tab:supplementary_videos_bulk}
\end{table}

\begin{table}[H]
    \small
    \begin{center}
    \begin{tabular}{|>{\centering\arraybackslash}p{\vidTableIdColumnWidth}|>{\centering\arraybackslash}p{\vidTableSnapshotColumnWidth}|>{\centering\arraybackslash}p{\vidTableDescriptionColumnWidth}|}
    \hline
    \textbf{ID} & \textbf{Snapshot} & \textbf{Description} \\ 
    \hline
    \hline
    
    {1} & \includegraphics{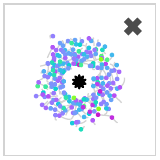} & Speed controller scan ($\bar{v}_{d} = 1.0$, cross symbol)\\ 
    \hline
    {2} & \includegraphics{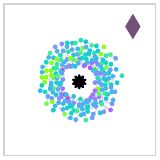} & Speed controller scan ($\bar{v}_{d} = 1.0$, diamond symbol) \\ 
    \hline
    {3} & \includegraphics{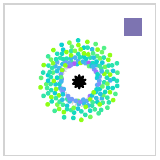} & Speed controller scan ($\bar{v}_{d} = 1.0$, square symbol) \\ 
    \hline
    {4} & \includegraphics{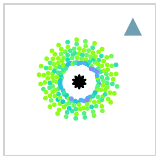} & Speed controller scan ($\bar{v}_{d} = 1.0$, pyramid symbol) \\ 
    \hline
    {5} & \includegraphics{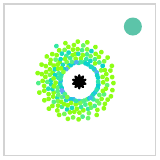} & Speed controller scan ($\bar{v}_{d} = 1.0$, circle symbol) \\ 
    \hline
    {6} & \includegraphics{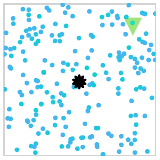} & Speed controller scan ($\bar{v}_{d} = 1.0$, nabla symbol)\\ 
    \hline
    \end{tabular}
    \end{center}
    
    \caption{\new{\textbf{Supplementary videos for the speed-controller parameter scan with a mean driver speed of $\bar{v}_d = 1.0$ and a constant mean predicted distance ahead of $\bar{s}_{pred} = 5.0$} in \figref[a]{\ref{fig:speed-controller-different-speeds-same-horizon}}. \newline Label in repository: \texttt{speed-controller-mean-driver-speed-1}. }}
    \label{tab:supplementary_videos_speed_controller_speed_1}
\end{table}

\begin{table}[H]
    \small
    \begin{center}
    \begin{tabular}{|>{\centering\arraybackslash}p{\vidTableIdColumnWidth}|>{\centering\arraybackslash}p{\vidTableSnapshotColumnWidth}|>{\centering\arraybackslash}p{\vidTableDescriptionColumnWidth}|}
    \hline
    \textbf{ID} & \textbf{Snapshot} & \textbf{Description} \\ 
    \hline
    \hline
    
    {1} & \includegraphics{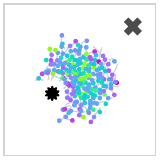} & Speed controller scan ($\bar{v}_{d} = 2.0$, cross symbol)\\ 
    \hline
    {2} & \includegraphics{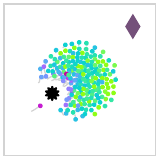} & Speed controller scan ($\bar{v}_{d} = 2.0$, diamond symbol) \\ 
    \hline
    {3} & \includegraphics{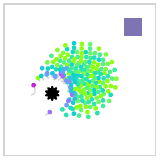} & Speed controller scan ($\bar{v}_{d} = 2.0$, square symbol) \\ 
    \hline
    {4} & \includegraphics{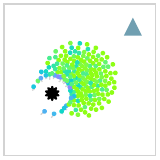} & Speed controller scan ($\bar{v}_{d} = 2.0$, pyramid symbol) \\ 
    \hline
    {5} & \includegraphics{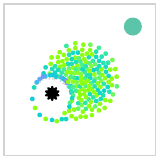} & Speed controller scan ($\bar{v}_{d} = 2.0$, circle symbol) \\ 
    \hline
    {6} & \includegraphics{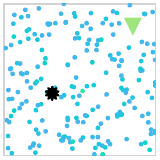} & Speed controller scan ($\bar{v}_{d} = 2.0$, nabla symbol)\\ 
    \hline
    \end{tabular}
    \end{center}
    
    \caption{\new{\textbf{Supplementary videos for the speed-controller parameter scan with a mean driver speed of $\bar{v}_d = 2.0$ and a constant mean predicted distance ahead of $\bar{s}_{pred} = 5.0$} in \figref[b]{\ref{fig:speed-controller-different-speeds-same-horizon}}. \newline Label in repository: \texttt{speed-controller-mean-driver-speed-2}. }}
    \label{tab:supplementary_videos_speed_controller_speed_2}
\end{table}

\begin{table}[H]
    \small
    \begin{center}
    \begin{tabular}{|>{\centering\arraybackslash}p{\vidTableIdColumnWidth}|>{\centering\arraybackslash}p{\vidTableSnapshotColumnWidth}|>{\centering\arraybackslash}p{\vidTableDescriptionColumnWidth}|}
    \hline
    \textbf{ID} & \textbf{Snapshot} & \textbf{Description} \\ 
    \hline
    \hline
    
    {1} & \includegraphics{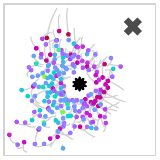} & Speed controller scan ($\bar{v}_{d} = 5.0$, cross symbol)\\ 
    \hline
    {2} & \includegraphics{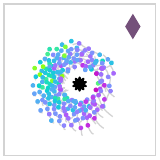} & Speed controller scan ($\bar{v}_{d} = 5.0$, diamond symbol) \\ 
    \hline
    {3} & \includegraphics{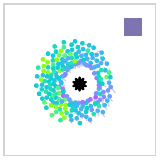} & Speed controller scan ($\bar{v}_{d} = 5.0$, square symbol) \\ 
    \hline
    {4} & \includegraphics{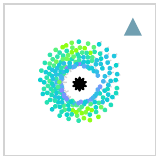} & Speed controller scan ($\bar{v}_{d} = 5.0$, pyramid symbol) \\ 
    \hline
    {5} & \includegraphics{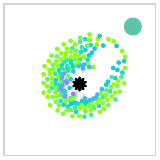} & Speed controller scan ($\bar{v}_{d} = 5.0$, circle symbol) \\ 
    \hline
    {6} & \includegraphics{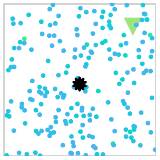} & Speed controller scan ($\bar{v}_{d} = 5.0$, nabla symbol)\\ 
    \hline
    \end{tabular}
    \end{center}
    
    \caption{\new{\textbf{Supplementary videos for the speed-controller parameter scan with a mean driver speed of $\bar{v}_d = 5.0$ and a constant mean predicted distance ahead of $\bar{s}_{pred} = 5.0$} in \figref[c]{\ref{fig:speed-controller-different-speeds-same-horizon}}. \newline Label in repository: \texttt{speed-controller-mean-driver-speed-5}. }}
    \label{tab:supplementary_videos_speed_controller_speed_5}
\end{table}

\begin{table}[H]
    \small
    \begin{center}
    \begin{tabular}{|>{\centering\arraybackslash}p{\vidTableIdColumnWidth}|>{\centering\arraybackslash}p{\vidTableSnapshotColumnWidth}|>{\centering\arraybackslash}p{\vidTableDescriptionColumnWidth}|}
    \hline
    \textbf{ID} & \textbf{Snapshot} & \textbf{Description} \\ 
    \hline
    \hline
    
    {1} & \includegraphics{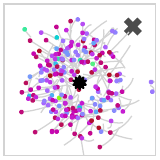} & Speed controller scan ($\bar{v}_{d} = 10.0$, cross symbol)\\ 
    \hline
    {2} & \includegraphics{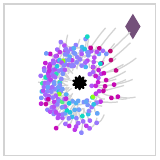} & Speed controller scan ($\bar{v}_{d} = 10.0$, diamond symbol) \\ 
    \hline
    {3} & \includegraphics{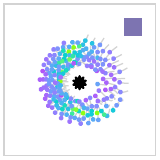} & Speed controller scan ($\bar{v}_{d} = 10.0$, square symbol) \\ 
    \hline
    {4} & \includegraphics{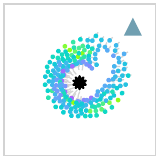} & Speed controller scan ($\bar{v}_{d} = 10.0$, pyramid symbol) \\ 
    \hline
    {5} & \includegraphics{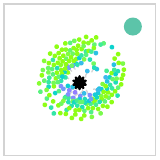} & Speed controller scan ($\bar{v}_{d} = 10.0$, circle symbol) \\ 
    \hline
    {6} & \includegraphics{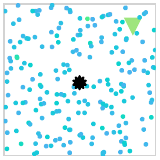} & Speed controller scan ($\bar{v}_{d} = 10.0$, nabla symbol)\\ 
    \hline
    \end{tabular}
    \end{center}
    
    \caption{\new{\textbf{Supplementary videos for the speed-controller parameter scan with a mean driver speed of $\bar{v}_d = 10.0$ and a constant mean predicted distance ahead of $\bar{s}_{pred} = 5.0$} in \figref[d]{\ref{fig:speed-controller-different-speeds-same-horizon}}. \newline Label in repository: \texttt{speed-controller-mean-driver-speed-10}. }}
    \label{tab:supplementary_videos_speed_controller_speed_10}
\end{table}

\begin{table}[H]
    \small
    \begin{center}
    \begin{tabular}{|>{\centering\arraybackslash}p{\vidTableIdColumnWidth}|>{\centering\arraybackslash}p{\vidTableSnapshotColumnWidth}|>{\centering\arraybackslash}p{\vidTableDescriptionColumnWidth}|}
    \hline
    \textbf{ID} & \textbf{Snapshot} & \textbf{Description} \\ 
    \hline
    \hline
    
    {1} & \includegraphics{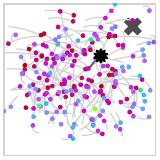} & Speed controller scan ($\bar{v}_{d} = 15.0$, cross symbol)\\ 
    \hline
    {2} & \includegraphics{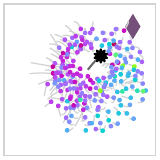} & Speed controller scan ($\bar{v}_{d} = 15.0$, diamond symbol) \\ 
    \hline
    {3} & \includegraphics{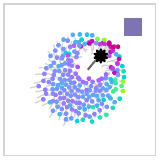} & Speed controller scan ($\bar{v}_{d} = 15.0$, square symbol) \\ 
    \hline
    {4} & \includegraphics{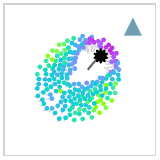} & Speed controller scan ($\bar{v}_{d} = 15.0$, pyramid symbol) \\ 
    \hline
    {5} & \includegraphics{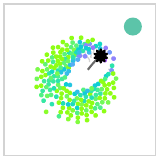} & Speed controller scan ($\bar{v}_{d} = 15.0$, circle symbol) \\ 
    \hline
    {6} & \includegraphics{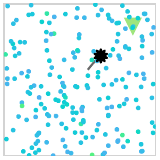} & Speed controller scan ($\bar{v}_{d} = 15.0$, nabla symbol)\\ 
    \hline
    \end{tabular}
    \end{center}
    
    \caption{\new{\textbf{Supplementary videos for the speed-controller parameter scan with a mean driver speed of $\bar{v}_d = 15.0$ and a constant mean predicted distance ahead of $\bar{s}_{pred} = 5.0$} in \figref[e]{\ref{fig:speed-controller-different-speeds-same-horizon}}. \newline Label in repository: \texttt{speed-controller-mean-driver-speed-15}. }}
    \label{tab:supplementary_videos_speed_controller_speed_15}
\end{table}
\end{widetext}

\end{document}